\documentclass[10pt,journal,compsoc]{IEEEtran}

\PassOptionsToPackage{dvipsnames,svgnames}{xcolor}
\usepackage{amsmath,amsfonts}
\usepackage{algorithmic}
\usepackage{array}
\usepackage{stfloats}
\usepackage{lipsum}                    % used to generate placeholder text
\usepackage{mathptmx}                  % use matching math font
\usepackage{mwe}                       % used to generate placeholder figures
\usepackage{url}
\usepackage{verbatim}
\usepackage{fontawesome5}
\usepackage{colortbl}
\usepackage{booktabs}                  % only used for the table example
\usepackage{mathptmx}                  % use matching math font
\usepackage{footnote}
\usepackage[hang,flushmargin]{footmisc}
\usepackage{listings}
\usepackage{multirow}
\usepackage{enumitem}
\usepackage{soul}
\usepackage{arydshln}
\usepackage{color}
\usepackage{xcolor}
\usepackage{sparklines}
\usepackage{mdframed} % For creating framed environments with custom backgrounds
\usepackage{helvet}
\usepackage{pgf}
\usepackage{svg}
\usepackage{cite}
\usepackage[export]{adjustbox}          % for IEEE profile
\usepackage{tikz}

\definecolor{backcolour}{rgb}{0.95,0.95,0.92}
\definecolor{darkblue}{RGB}{0,0,139} % A color between blue and navy
\definecolor{darkgreen}{RGB}{0,100,0} % Dark green color
\definecolor{brownishorange}{RGB}{204,85,0} % Brownish-orange color
\definecolor{new-green}{rgb}{0.104,0.667,0.229}
\newcommand{\rev}[1]{{#1}} % print text black (option 1)
\definecolor{darkestblue}{HTML}{000052} % 0% (start)
\definecolor{midblue1}{HTML}{54548B}    % 33%
\definecolor{midblue2}{HTML}{AAAAC6}    % 67%
\definecolor{lightestblue}{HTML}{FFFFFF}% 100% (end)

\newcommand{\SquareCell}[2]{%
  \fcolorbox{black}{#1}{%
    \begin{minipage}[c][0.1cm][c]{0.1cm}%
      \centering \color{white}#2
    \end{minipage}%
  }%
}

\newcommand{\ZeroCell}[1]{%
  \fcolorbox{black}{white}{%
    \begin{minipage}[c][0.1cm][c]{0.1cm}%
      \centering \color{black}#1
    \end{minipage}%
  }%
}

\newcommand{\hmCell}[1]{%
  \begingroup
    \ifdim #1 pt < 2pt
      \cellcolor{blue!10}\color{black}#1%
    \else\ifdim #1 pt < 3pt
      \cellcolor{blue!20}\color{black}#1%
    \else\ifdim #1 pt < 3.5pt
      \cellcolor{blue!30}\color{black}#1%
    \else\ifdim #1 pt < 4pt
      \cellcolor{blue!40}\color{black}#1%
    \else
      \cellcolor{blue!50}\color{white}#1%
    \fi\fi\fi\fi
  \endgroup
}

\newcommand{\kmCell}[1]{%
  \begingroup
    \ifdim #1 pt < 2pt
      \cellcolor{purple!10}\color{black}#1%
    \else\ifdim #1 pt < 3pt
      \cellcolor{purple!20}\color{black}#1%
    \else\ifdim #1 pt < 3.5pt
      \cellcolor{purple!30}\color{black}#1%
    \else\ifdim #1 pt < 4pt
      \cellcolor{purple!40}\color{black}#1%
    \else
      \cellcolor{purple!50}\color{white}#1%
    \fi\fi\fi\fi
  \endgroup
}

\definecolor{appleredlight}{RGB}{255, 105, 97}
\definecolor{appleorangelight}{RGB}{255, 179, 64}
\definecolor{appleyellowlight}{RGB}{255, 212, 38}
\definecolor{applegreenlight}{RGB}{48, 219, 91}
\definecolor{applemintlight}{RGB}{102, 212, 207}
\definecolor{appleteallight}{RGB}{93, 230, 255}
\definecolor{applecyanlight}{RGB}{112, 215, 255}
\definecolor{applebluelight}{RGB}{64, 156, 255}
\definecolor{appleindigolight}{RGB}{125, 122, 255}
\definecolor{applepurplelight}{RGB}{218, 143, 255}
\definecolor{applepinklight}{RGB}{255, 100, 130}
\definecolor{applebrownlight}{RGB}{181, 148, 105}

\definecolor{applerednormal}{RGB}{255, 69, 58}
\definecolor{appleorangenormal}{RGB}{255, 159, 10}
\definecolor{appleyellownormal}{RGB}{255, 214, 10}
\definecolor{applegreennormal}{RGB}{48, 209, 88}
\definecolor{applemintnormal}{RGB}{99, 230, 226}
\definecolor{appletealnormal}{RGB}{64, 200, 224}
\definecolor{applecyannormal}{RGB}{100, 210, 255}
\definecolor{applebluenormal}{RGB}{10, 132, 255}
\definecolor{appleindigonormal}{RGB}{94, 92, 230}
\definecolor{applepurplenormal}{RGB}{191, 90, 242}
\definecolor{applepinknormal}{RGB}{255, 55, 95}
\definecolor{applebrownnormal}{RGB}{172, 142, 104}

\usepackage{longfbox} % for drawing frames on stakeholder aliases
\usepackage{makecell}

\makeatletter
\newdimen\@tempdimd
\makeatother

\newfboxstyle{patternparamrad}{padding-top=1pt, padding-bottom=0.5pt, padding-left=2pt, padding-right=2pt, border-style=solid, height=6pt, border-radius=2.5pt}

\newfboxstyle{patternparam}{padding-top=1pt, padding-bottom=0.5pt, padding-left=2pt, padding-right=2pt, border-style=none, height=6pt}

\def\Sal{\lfbox[patternparamrad, background-color=white, border-color=applebluenormal]{{\color{applebluenormal}{{\normalfont \textsc{Sal}}}}}\xspace}

\def\Col{\lfbox[patternparamrad, background-color=white, border-color=applerednormal]{{\color{applerednormal}{{\normalfont \textsc{Col}}}}}\xspace}

\def\Chart{\lfbox[patternparamrad, background-color=white, border-color=appleorangenormal]{{\color{appleorangenormal}{{\normalfont \textsc{Chart}}}}}\xspace}

\def\Text{\lfbox[patternparamrad, background-color=white, border-color=applepurplenormal]{{\color{applepurplenormal}{{\normalfont \textsc{Text}}}}}\xspace}

\def\CVD{\lfbox[patternparamrad, background-color=white, border-color=applegreennormal]{{\color{applegreennormal}{{\normalfont \textsc{CVD}}}}}\xspace}

\newcommand{\techname}{Visualizationary}
\newcommand{\topic}[1]{\noindent\textbf{#1}}

\begin{document}

%% ---------------------------------------------------------------------
%% Title, author(s), and affiliations
%% ---------------------------------------------------------------------

%% Paper title
\title{\techname: Automating Design Feedback for Visualization Designers Using LLMs}

%% Authors
\author{Sungbok Shin, Sanghyun Hong, and Niklas Elmqvist,~\IEEEmembership{Fellow,~IEEE}
\IEEEcompsocitemizethanks{
    \IEEEcompsocthanksitem Sungbok Shin is with Aviz at Inria and Universit\'e Paris-Saclay, Saclay, France. 
    The work was done while the author was affiliated with the University of Maryland, College Park. 
    E-mail: sungbok.shin@inria.fr
    \IEEEcompsocthanksitem Sanghyun Hong is with the Department of Computer Science at Oregon State University, Corvallis, OR, USA.
    E-mail: sanghyun.hong@oregonstate.edu
    \IEEEcompsocthanksitem Niklas Elmqvist is with the Department of Computer Science at Aarhus University, Aarhus, Denmark.
    E-mail: elm@cs.au.dk}
    \thanks{Manuscript received XXX XX, 2024; revised XXX XX, 2024.}
}

% The paper headers
\markboth{IEEE Transactions on Visualization and Computer Graphics}{Shin \MakeLowercase{\textit{et al.}}: \techname}

%% -------------------------------------------------------------
%% Abstract
%% -------------------------------------------------------------
\IEEEtitleabstractindextext{%
\begin{abstract}
Interactive visualization editors empower users to author visualizations without writing code, but do not provide guidance on the art and craft of effective visual communication.
In this paper, we explore the potential of using an off-the-shelf large language models (LLMs) to provide actionable and customized feedback to visualization designers.
Our implementation, %called 
\textsc{\techname}, demonstrates how ChatGPT can be used for this purpose through two key components: a preamble of visualization design guidelines and a suite of perceptual filters that extract salient metrics from a visualization image.
We present findings from a longitudinal user study involving 13 visualization designers---6 novices, 4 intermediates, and 3 experts---who authored a new visualization from scratch over %the course of 
several days.
Our results indicate that providing guidance in natural language via an LLM can aid even seasoned designers in refining their visualizations.
All our supplemental materials %accompanying this paper 
are available at \url{https://osf.io/v7hu8}.
\end{abstract}

%% Keywords
}

%% -------------------------------------------------------------
%% Paper Content
%% -------------------------------------------------------------
\maketitle
\IEEEdisplaynontitleabstractindextext
\IEEEpeerreviewmaketitle

%% ---------------------------------------------------------------------
%% INTRODUCTION
%% ---------------------------------------------------------------------
\IEEEraisesectionheading{\section{Introduction}\label{sec:intro}}
\IEEEPARstart{T}{o} democratize the transformative power of data in today's information society, it is not sufficient that people get access to visualizations created by others; they must also be empowered to author their \textit{own} visualizations.
In response, recent years have seen an influx of interactive tools  that enable users to create visualizations without writing code, such as iVisDesigner~\cite{Ren2014}, Data Illustrator~\cite{Liu2018}, and Lyra~\cite{DBLP:journals/cgf/SatyanarayanH14}.
However, no matter how advanced these tools become, a fundamental barrier remains: authoring effective visualizations is a complex task requiring visual design skills, an understanding of aesthetics, and experience---expertise that not all would-be designers possess.

Fortunately, visualization designers have nothing to lose but their chains. 
We introduce \textsc{\techname}, a system to seize the means of visualization design by using off-the-shelf large-language models (LLMs).
Our system enhances the design of communicative visualizations by providing perceptual feedback.  
While it may not fully replace feedback from peers or colleagues, it serves as a valuable alternative when it is impossible to gain human feedback and can benefit designers of all expertise levels, from novices to seasoned ones.

Although competing approaches for improving visualization designs, such as heuristics (like small, \rev{smart} defaults in Lyra) and Grammar of Graphics (GoG)-style specifications have their own advantages, they also carry limitations due to their deterministic nature. 
Heuristic-driven methods lack human intervention, which can constrain creativity and limit the potential for novel ideas. 
GoG-style specifications, despite offering precise chart descriptions, lead to producing standardized solutions, reducing the diversity of possible visualization outcomes.

In contrast, \techname{} automates perceptual feedback using a suite of automated techniques that simulate how a viewer would perceive a visualization.
By providing actionable insights without the need to standardize changes, 
\techname{} preserves the designer's creativity while enhancing effectiveness.

The basic idea behind \techname{} is to investigate how an LLM can be used to aid the iterative design process of a visualization artifact in a workflow that we call \textit{analyze-clarify-guide-track} (ACGT):

\begin{enumerate}
    \item \textbf{\underline{A}nalyze} the state of a visualization using automated perceptual filters, yielding a corresponding analysis report; 
    \item \textbf{\underline{C}larify} the results from the filters in the report such that it is understandable to even novice visualization designers;
    \item \textbf{\underline{G}uide} the designer in making changes that address identified concerns in their visualization; and 
    \item \textbf{\underline{T}rack} the trajectory of the visualization artifact over time during its iterative design process.
\end{enumerate}

We present a web-based implementation of \techname{} that leverages a server-side LLM, demonstrating how a ``vanilla'' LLM can be used for this highly complex visualization design task.
\techname{} operates as follows:
We first translate chart images into text using a vision-language model (VLM)~\cite{liu2022deplot} and then generate automatic design guidance using an LLM as a template and text processor.
Users can also upload their existing designs as an image file, leaving them free to use a tool of their own choice.
The system presents the guidance in an interactive report organized hierarchically, enabling designer to drill into the findings at their desired level of detail.

In evaluation, we present findings from a longitudinal study conducted with 6 novice, 4 intermediate, and 3 expert visualization designers over a period spanning a few days.
During the study, each designer worked with \techname{} to create visualizations and improve their designs. 
We report on four successive revisions of their visualization as well as their quantitative and qualitative feedback from 13 visualization designers of varying expertise, from a novice with only 1 year of experience in the field to an expert with 11 years of experience.
The designs created by these designers are then assessed by 3 senior visualization researchers.
Our findings show the effectiveness as well as shortcomings of our concept from various perspectives, including level of expertise, type of feedback, and design optimization. 
These insights contribute to a deeper understanding of what constitutes effective design feedback for visualizations.

%% -------------------------------------------------------------
%% BACKGROUND
%% -------------------------------------------------------------
\section{Background}
\label{sec:background}

% Here we present the state of the art in three topics: 
Here we provide a brief overview of three key areas,
required for understanding our work: 
design feedback, automating visualization design feedback, and large-language models (LLMs).

\subsection{Design Feedback}
\label{subsec:design-feedback}

Feedback is essential in all design disciplines, and it typically comes from two sources: user and usage feedback versus peer and supervisor feedback.
In academic contexts, the former, which emphasizes validation through empirical evaluation, is more prevalent~\cite{munzner2009, xu14voyant, yen20decipher, oppenlaender20crowdui, letschas21effectivecrowdsourcedfeedback}.
Munzner's nested model for visual design and validation~\cite{munzner2009} primarily incorporates this type of feedback.
\rev{Conversely, peer feedback holds greater prevalence in practice, particularly for designs aimed at widespread use~\cite{alabood23systematic, oppenlaender21crowdsoucingvspeer, shin24simulation}.
Nevertheless, this methodology has also permeated academic circles, notably within interaction design~\cite{DBLP:journals/interactions/BardzellBL10, DBLP:journals/iwc/Bardzell11, cheng20critiqueme}.}

For feedback to be effective, it must not only be understandable by the intended recipient, but it must also provide enough information so that the recipient can act upon the feedback (if necessary).
What Munzner calls low-level feedback~\cite{munzner14visualization} is focused on fine-grained insights and improvements, and is often actionable because of its low abstraction level: e.g., ``\textit{this red color is a poor choice for a color-blind person},'' ``\textit{this bar must be made larger},'' and ``\textit{this text uses a font with poor legibility}.''
However, higher-level design feedback in the form of critique, qualitative evaluation~\cite{DBLP:journals/cgf/ChandrasegaranB17}, and abstract metrics can be much harder to understand, let alone act upon.
For example, learning that your visualization suffers from ``poor contrast,'' ``poor color choices,'' or is ``not aesthetically pleasing'' can be challenging to address.

\subsection{Automatic Design Feedback on Visualizations}

At its core, data visualization is very much a design discipline and still relies heavily on empirical methods to iteratively refine an artifact over time~\cite{munzner14visualization}.
This has led to a significant focus on design guidelines and rules of thumb~\cite{franconeri21visualdatacommunication, Tufte1983}, textbooks with a design focus~\cite{munzner14visualization}, and numerous resources derived from real-world practice~\cite{Few2006}.
However, to personalize such design guidelines and complement personal tutoring, recent work in the research area has begun to provide various mechanisms for automating design feedback for the designer.

One feedback approach is focused on \textit{linting}, a technique drawn from compiler design where source code is statically analyzed to flag poor coding practices (e.g., no indentation, poor identifier naming, and lack of variable initialization) that often cause errors.
Analogously, visualization linting can detect chart construction errors~\cite{hopkins20visualint}, visualization mirages~\cite{mcnutt20mirage}, and deceptively-designed line charts~\cite{fan22linechartdeception}.
For example, VizLinter~\cite{chen22vizlinter} helps rectify problems in Vega-Lite specifications.

Another method uses a form of recommendation to automatically generate charts based on the specific dataset to visualize.
Mackinlay's Ph.D.\ work on automatic visualization~\cite{Mackinlay1986} was one of the first endeavors in this area, and it was later realized in production as Tableau's Show Me feature~\cite{mackinlay07showme}. 
More recent visualization recommendation methods have been proposed that generate charts using data properties~\cite{key12vizdeck, wongsuphasawat16voyager, DBLP:conf/chi/WongsuphasawatQ17}, perceptual principles~\cite{wongsuphasawat16voyager}, expert feedback~\cite{luo18deepeye}, large-scale dataset-visualization pairs~\cite{hu19vizml}, and design knowledge~\cite{moritz19draco}.

Accessibility is seeing emerging interest in the visualization field~\cite{Elmqvist2023, DBLP:journals/interactions/MarriottLBCEGHM21}. 
Heuristic frameworks on general accessibility such as Chartability~\cite{DBLP:journals/cgf/ElavskyBM22} could conceivably be operationalized to support automatic accessibility auditing. 
For color vision deficiency, Angerbauer et al.~\cite{DBLP:conf/chi/AngerbauerRCOPM22} present results from a large-scale crowdsourced assessment that could be similarly automated.

Finally, quantitative feedback can also help a designer improve their visualization.
VisLab~\cite{choi23vislab} provides a web-based system facilitating the collection of such quantitative feedback from crowdworkers that can be used to evaluate and refine a visualization in online experiments.
Perceptual Pat~\cite{shin23perceptualpat} tracks the evolution of a visualization artifact over time, assembling a report consisting of several perceptual metrics for each iteration using computer vision and related methods.
In comparison to this system, our work goes beyond merely displaying perceptual metrics about a visualization artifact to explaining the findings and then suggesting how the user may address shortcomings.

\subsection{Large Language Models}
\label{subsec:llms}

A \textit{language model} is a statistical model that predicts the likelihood of word sequences in natural language based on a large corpus of textual training data.
Recent work~\cite{brown2020language, wei2022emergent, ganguli2022predictability} shows that the primary discriminator in accomplishing complex tasks using language models is their \emph{scale}.
In light of this trend, there has been a growing number of works on so-called ``large'' language models (LLMs). %~\cite{brown2020language, OpenAI2023GPT4TR, chiang2023vicuna, touvron2023llama, anil2023palm}.
LLMs have billions of model parameters, and thus, training these models is computationally demanding.
%It needs a large training corpus, hardware accelerators, and machines equipped with them.

A standard paradigm for addressing the training costs is to pre-train LLMs on a large text corpus and then allow users to fine-tune these models on a set of desired tasks, e.g., question answering, language translation, and text summarization. %~\cite{brown2020language, chowdhery2022palm, rae2021scaling, anil2023palm, liang2022holistic, goyal2022news}.
In pre-training, a model is typically trained to do next-word prediction, i.e., given a prompt, the model returns a sequence of word tokens likely to complete the prompt.
This process is performed by third-parties such as OpenAI, Google, or Meta. %, and such \emph{pre-trained models} are offered to users. %as a form of internet service, e.g., ChatGPT~\cite{OpenAI2023GPT4TR}, or as-is, e.g., open-source models such as Vicuna~\cite{chiang2023vicuna} or LLaMA~\cite{touvron2023llama}.
In particular, services such as ChatGPT offer a personalized experience by separating sessions for each user and fine-tuning the pre-trained model based on the user's inputs.
Recent models further support \emph{multi-modality}, with images and text being combined, and enable them to answer questions, e.g., in data visualizations: ``How much does X outperform Y?''~\cite{liu2022deplot, wang2019matcha}

Our work leverages both types of pre-trained LLMs to generate feedback: (1) Multi-modal LLMs that take a user's data visualization and questions on the visualization as input and return the answers to the questions, and (2) uni-modal LLMs (or standard ones) that receive text inputs and return the corresponding outputs.
We use multi-model LLMs to generate textual descriptions of a user's visualization. 
These descriptions, along with user queries, are then fed into uni-modal LLMs, which generate design feedback based on the provided input.
To minimize hallucination, we provide strict guidance as a preamble so that the language model does not provide random and inconsistent feedback~\cite{mialon23augmentedLM}.

%% -------------------------------------------------------------
%% FRAMEWORK
%% -------------------------------------------------------------
% ----------------------------------------------------------------
% Framework
% ----------------------------------------------------------------
\section{Automated Visualization Feedback}
\label{sec:framework}

We propose an automated visualization design feedback mechanism for supporting iterative design of a visualization artifact.
The designer uploads consecutive versions of a visualization artifact as they iterate on the design.
The automated system then provides natural language feedback to facilitate improvement.

\subsection{Target Audience}

Our work in this paper primarily targets \textbf{novice} or \textbf{intermediate designers} who do not have the required experience or expertise to critique their own work with an objective eye, or the ability to make the required changes to address problems.
In other words, our automated design feedback mechanism is not intended to be a replacement for a trained designer's critical eye.
% However, 
Because design critique is critical during any design process, 
we even find that 
automated design feedback is useful even for \textbf{experts} %~\cite{shin23perceptualpat}.
(see \S\ref{subsec:feedback-patterns}).
% Similarly, automation is no replacement for a human peer or mentor to give feedback and suggestions to the novice designer.
% However, mentors or peers may not always be available.

\subsection{Design Rationale}

Based on Munzner's visualization design and nested evaluation frameworks~\cite{munzner2009, munzner14visualization}, interaction design critique~\cite{DBLP:journals/iwc/Bardzell11}, and guidelines for visualization design~\cite{franconeri21visualdatacommunication, moritz19draco, DBLP:journals/tvcg/RenLB19}, we identify several benefits for automating visualization design feedback. We explain our rationale by comparing our system’s conceptual approach with the type of feedback a human designer would provide:

\begin{itemize}[noitemsep, topsep=0.2cm, leftmargin=1.7em]

    \rev{\item[R1]\textbf{Ubiquitous:}
    An automated system can provide immediate feedback, enabling rapid design iterations anytime~\cite{Elmqvist2013, DBLP:journals/cacm/Elmqvist23} the designer needs support. 
    By contrast, human feedback may not always be readily-available.
    Furthermore, the system should not depend on the authoring tool in use.}
    
    \item[R2]\textbf{Real-time:}
    Hence, an automated visualization feedback system can provide data in a real-time, enabling designers to make immediate improvements during the design process.

    \item[R3]\textbf{Consistency:}
    Automated feedback systems apply consistent evaluation criteria, reducing subjectivity and bias as well as human error in assessments.
    Human feedback may be less consistent and sometimes even conflicting.

    \item[R4]\textbf{Scalability:}
    Compared to a human evaluator, an automated design feedback mechanism \rev{can handle a large volume of evaluations for multiple users simultaneously.}

    \item[R5]\textbf{Inexpensive:}
    Automating feedback can reduce (but not eliminate) human evaluators, which can be costly.

\end{itemize}

\subsection{Analyze-Clarify-Guide-Track (ACGT) Workflow}
\label{subsec:acgt-workflow}
We propose an automated visualization design workflow called \textit{analyze-clarify-guide-track} (ACGT) with four components:

\begin{enumerate}[noitemsep, topsep=0.2cm, leftmargin=0.5cm]

    \item\textbf{\underline{A}nalyze}:
    To provide written feedback to the designer, we first need to collect textual data from the visualization screenshot; i.e., we must make the transition from images to text.
    We do this using automated vision filters, such as color contrast, visual hierarchy, and layout consistency (see \S~\ref{subsec:filters}).
    The outcome of this analysis is a report on the visualization artifact.
    
    \item\textbf{\underline{C}larify}:
    Interpreting visualization metrics on their own can be challenging~\cite{shin23perceptualpat}, let alone to use as a basis for revisions. 
    The \textit{clarify} phase translates results from the perceptual filters into understandable terms (R3), thus bridging the gap between technical analysis and practical design improvements.
    It breaks down complex issues identified in the analysis into layman's terms, explaining which and why certain changes are needed.
    
    \item\textbf{\underline{G}uide}:
    It is not enough to know the problem; we must also know how to fix it.
    Our mechanism provides such practical guidance to the designer as natural language on how to address the identified concerns. 
    This guidance may include recommendations, actionable insights, or best practices.
    
    \item \textbf{\underline{T}rack}:
    Understanding the overall trajectory of the design process is important to avoid local optima.
    The tracking component monitors the progress of the design over time during iterative refinement.
    It maintains a historical record of design iterations, changes implemented, and corresponding improvements in visualization quality.
    This tracking allows designers to see the evolution of their work, understand the impact of changes made based on earlier recommendations, and make informed decisions.
    
\end{enumerate}

\section{How \techname{} Works}
\label{sec:overview}

% \techname{} is a web-based visualization design feedback system that automates natural language critique for visualization designers.
\techname{} is an automated design feedback system 
that provides design critiques for visualization designers in natural language.
The system is implemented as a web-based platform,
where the front-end offers an interactive interface for users 
to explore various visual feedback, 
while the back-end supports these processes using off-the-shelf language models.
Here we describe its architecture and core components in detail.

\begin{figure}[tbh]
    \centering
    \includegraphics[width=\columnwidth]{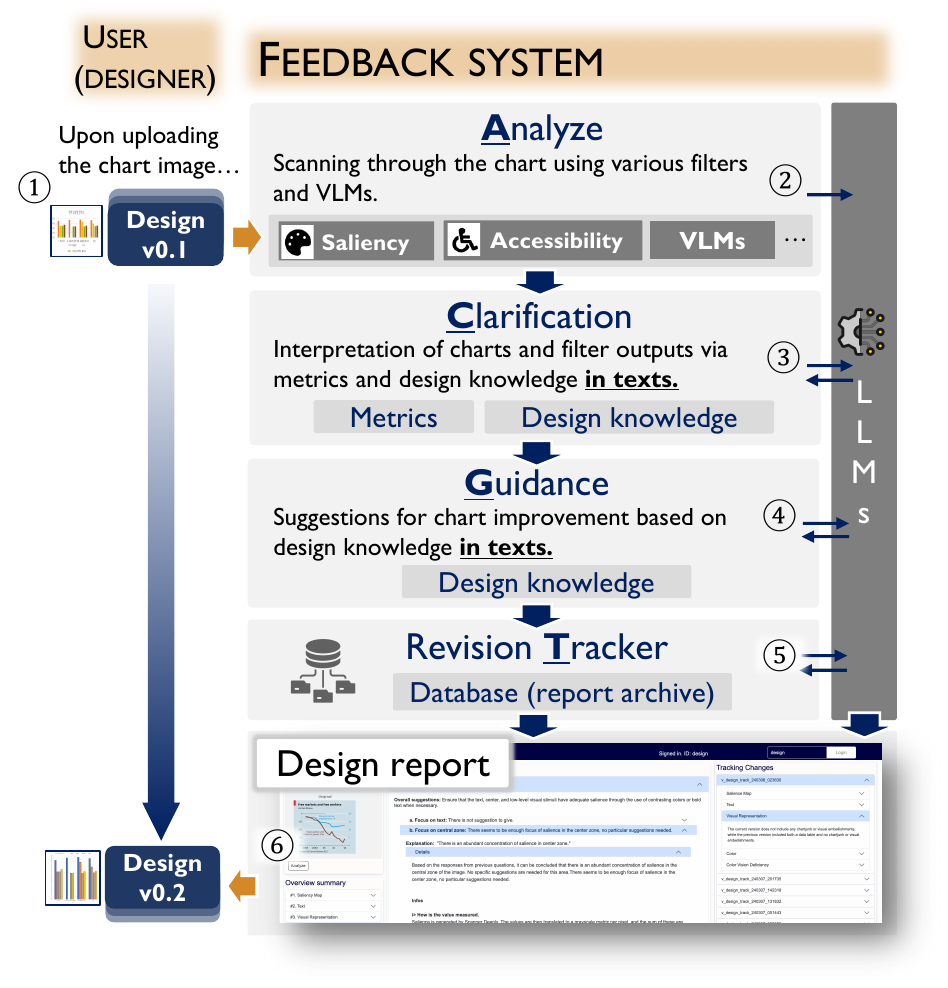}
    \vspace{-0.9cm}
    \caption{\textbf{System overview.}
    \techname{} is composed of a visualization design management interface and a feedback system; the white circles refer to the analysis and report steps, respectively.
    The system first receives a data visualization (chart) from the project container interface \textcircled{1} and generates design feedback (explanations \textcircled{3} and suggestions \textcircled{4}) based on the filters used to analyze \textcircled{2} the visualization.
    The feedback system is designed to support the \emph{analysis-clarify-guide-track} workflow and leverages a large language model (LLM) to automate the steps.
    We archive the feedback generated by \techname{} \textcircled{5} with the revision tracker.
    The design report component \textcircled{6} summarizes the feedback and the revision history for the iterative design improvement process.}
    \label{fig:system-overview}
\end{figure}

\subsection{System Overview}
\label{subsec:overview}

\begin{figure*}[tbh]
    \centering
    \includegraphics[width=\textwidth]{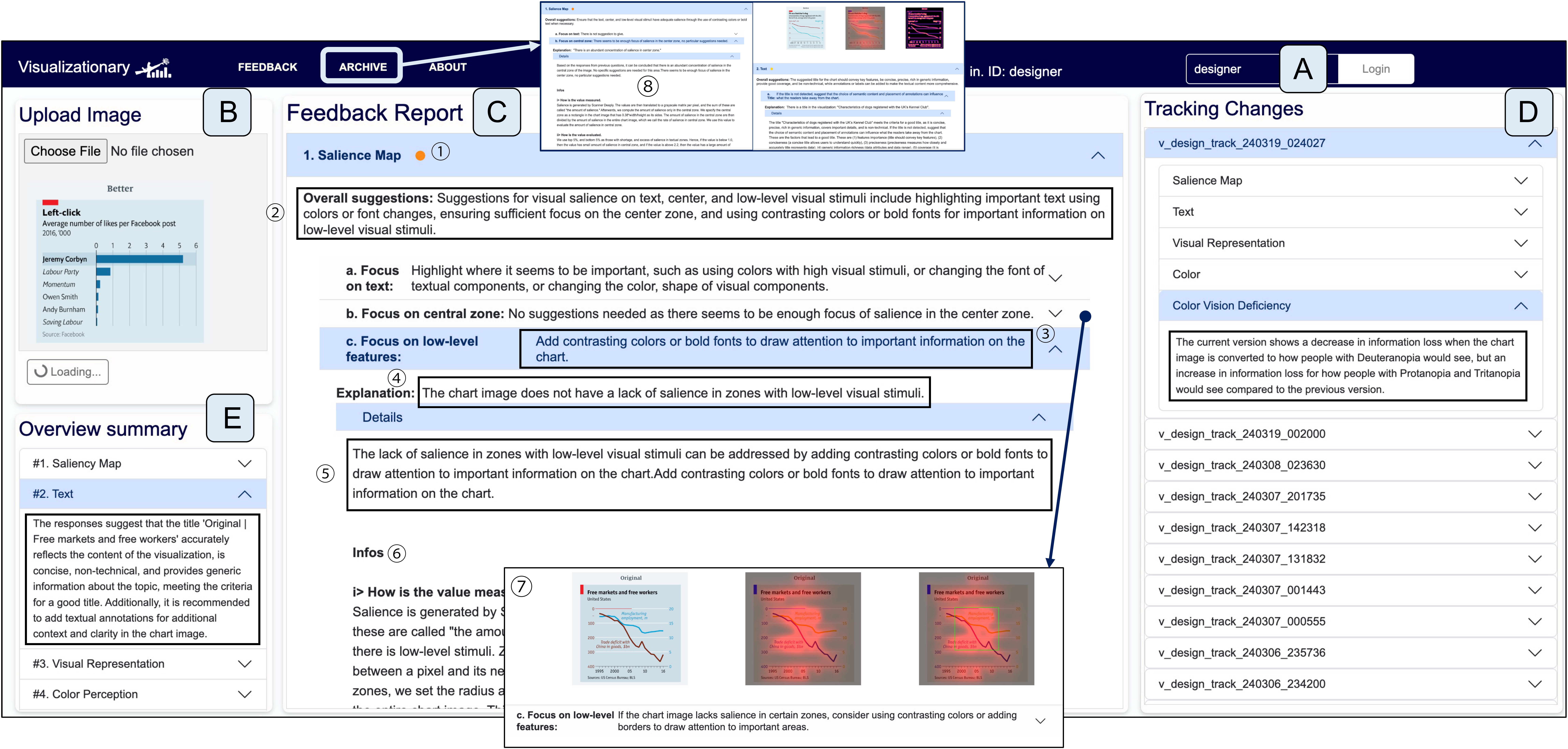}
    \caption{\textbf{\techname{} interface.}
    Example of an interactive report on a visualization artifact generated using our novel analyze-clarify-guide-track (ACGT) framework. 
    The image is a depiction of our system, \techname{} where a new chart image (Figure 1(B)) is being updated while the current report is about the past visualization image.
    The report is generated by automatically analyzing (“analyze”) the artifact using an expandable collection of image filters grouped into categories: salience, text, visual representation, color, and accessibility.
    After reporting results from each image filter (center column), the report then interprets (“clarify”) the findings using natural language understandable to a novice visualization designer.
    This is followed by natural language suggestions (“guide”) on how to address weaknesses in the design.
    Finally, the right side of the report shows changes over time (“track”) as the designer iteratively refines the artifact.}
    \label{fig:teaser}
\end{figure*}

\techname{} is a web-based client-server system (R4) that tracks a visualization artifact throughout its design process.
Our system can support multiple users at once, so we placed no time constraints on image uploads (R4). 
However, no instances occurred where multiple users accessed it simultaneously.
The workflow (Fig.~\ref{fig:system-overview}) is based on the user---a visualization designer---iteratively uploading a screenshot of the current state of their visualization artifact and receiving automated feedback from the system.
The feedback then presumably helps the designer make improvements to the artifact in the next design iteration. 
However, unlike existing visualization design feedback systems such as Perceptual Pat~\cite{shin23perceptualpat}, which merely report metrics, \techname{} implements the ACGT workflow from \S\ref{sec:framework}: beyond analyzing the image, it also explains (``clarify'') the feedback using natural language (R3), suggests ways for addressing shortcomings (``guide''), and tracks the changes to the design over multiple iterations.

This workflow is realized through \rev{three} components:
(1) a web-based visualization design interface (R1, R4, R5) (\S~\ref{subsec:visually-dmi}),
(2) interactive design reports providing feedback at user-controlled levels of detail (\S~\ref{subsec:reports}), and
(3) an automated feedback system for extracting visualization data and transforming into textual feedback (R5) (\S~\ref{subsec:filters}). 
The former two belong to \techname{}'s frontend, whereas the third is a part of the backend.
Afterwards, we detail our prompting methodologies for providing visualization feedback (\S~\ref{sec:generating-feedback}), and our implementation notes (\S~\ref{subsec:implementation}). 

\subsection{\techname{} Design Management Interface}
\label{subsec:visually-dmi}
The frontend for \techname{} is a web-based single-page application for managing the iterative design process of a visualization artifact. %(Fig.~\ref{fig:design-mgmt}).
A \textit{project} is a container for a design process, including all of the visualization revisions and their corresponding \textit{design reports} (see \S~\ref{subsec:reports}).
The frontend has operations for creating, editing, and deleting projects associated with a user.

% frontend -- design management interface

% \begin{figure}[tbh]
%     \centering
%     \frame{\includegraphics[width=\linewidth]{figures/image_visualizationary.png}}
%     \caption{\textbf{\techname{} design management interface.}
%     The system shows an overview of the timeline of revisions for a visualization artifact.
%     Each revision consists of a screenshot of the artifact as well as an interactive design report.
%     The right column shows the revision tracker, which provides a high-level view of the overall design trajectory.
%     }
%     \label{fig:design-mgmt}
% \end{figure}

Viewing a project shows a timeline of the revisions and allows the user to bring up the details for each.
Projects can be extended by uploading a new screenshot of the visualization artifact in \texttt{.jpg} or \texttt{.png} file (see Fig.~\ref{fig:teaser}) (R1) and then launching the analyze-clarify-guide-track workflow on the new revision.
This process typically takes a couple of minutes to complete (R2), after which a new design report is added to the timeline for the designer's perusal.
In this way, the \techname{} interface is designed to help the design process of a visualization artifact.

\subsection{Visualization Design Reports}
\label{subsec:reports}

Results from the feedback backend are presented as interactive \textit{visualization design reports}.
A design report is an interactive document consisting of a hierarchy of expandable sections designed in a bottom-up fashion. 
Sections are organized based on the main categories of visualization filters in \S~\ref{subsec:filters}.
The contents of a section in the report include both natural language and illustrative images (such as heatmaps, extracted shapes, or gaze maps) (see Fig.~\ref{fig:teaser}~(C)). 
Each section provides a short overall summary of suggestions (Fig.~\ref{fig:teaser}~(C) \textcircled{2}) of the underlying data for that metric followed by two standardized paragraphs:

\begin{itemize}[noitemsep, topsep=0.em, leftmargin=0.3cm]
    \item\textbf{Explanations:} Explanation of visualization feedback in natural language (``clarify'' in the ACGT workflow) (e.g., Fig.~\ref{fig:teaser}~(C) \textcircled{4}).
    
    \item\textbf{Suggestions:} Recommendations addressing shortcomings in the design for that specific metric (``guide'') (e.g., Fig.~\ref{fig:teaser}~(C) \textcircled{3}).
\end{itemize}

Expanding the subheadings yields additional explanations (see Fig.~\ref{fig:teaser}~(C) \textcircled{5}) and suggestions.
Sections can be expanded down to raw visual metrics (see Fig.~\ref{fig:teaser}~(C) \textcircled{6}).
We also provide additional information and warnings about each filter, to help the designer make decisions while reading the report. 
The rationale for this design is to present feedback in a summarized form that can be easily navigated.
It also allows experienced designers to see high-level themes without having to drill down into low-level details.

\techname{} provides additional functionality.  
First, to help navigating the report, we indicate sections that require modification using a circular visual mark, either yellow or red next to the title of each section (see Fig.~\ref{fig:teaser}~(C) \textcircled{1}).
If there exist at most one issue in a section, then the corresponding circle is yellow.
If there is more than one, then the circle is orange.
Second, to help designers quickly skim section summaries, we provide a short overview summary of responses per category in Fig.~\ref{fig:teaser}~(E).

We also aim to maintain transparency in the feedback process, given that both filters and LLMs can yield misleading feedback. 
By being transparent, designers can better determine whether to trust the model.
For instance, the report shows how the system extracts chart data and calculates its metrics.

\subsection{Feedback System}
\label{subsec:filters}

Here we present our feedback system (see Fig.~\ref{fig:system-overview}).
The feedback is provided using the ACGT Workflow, as explained in  \S~\ref{subsec:acgt-workflow}.
We use automated models to extract perceptual characteristics from visualizations, with the goal of providing pragmatic and actionable design feedback.
\rev{These functions are a suite of computer vision and vision-language models (VLMs) that we collectively call \textit{visualization filters}. }
\rev{We present a total of 10 filters from 5 topics. }
We first detail how these visualization filters are translated to design feedback (what corresponds to ACG of our ACGT workflow). 
Afterwards, we describe how the revision \underline{T}racker works. 

\smallskip
\noindent\textbf{Visualization Filters. } We explain the filters according to the ACGT framework.
To each filter, we provide the intended role, \underline{A}nalysis process, \underline{C}larification criteria, and \underline{G}uidance to users. 
In a nutshell, the filters automatically \underline{a}nalyze the visualizations. 
Then, using the information from these filters, \underline{c}larification is made based on predefined heuristics about the metric and also on the judgment of LLMs, backed by the design knowledge added as a preamble.
Finally, based on the clarification, the LLM provides actionable \underline{g}uidance to users in the form of textual feedback. 
\rev{We instill the clarifications along with the guidance relevant to the decisions into ChatGPT as a prompt (see \S~\ref{sec:generating-feedback}). 
Then, based on a series of template questions, ChatGPT provides the appropriate answer. }

However, despite our intents, not all filters we provide are translated into actionable feedback.
This heuristic as well as examples of the feedback are described in our supplementary material.
Note that the list is not exhaustive and can be extended.

% \begin{itemize}

%     \item[\faSun] \textbf{Virtual eyetracker} refers to the extent to which areas within a scene capture the attention of an observer~\cite{riche13metricssaliency}.

%     \item[\faFont]\textbf{Textual characters} are visual objects that are used for labels, titles, and legends in visualization~\cite{stokes23tvcgtextscharts}.
    
%     \item[\faChartBar]\textbf{Visual representations} can generate feedback specific to the technique, such as visualization linting~\cite{chen22vizlinter, hopkins20visualint}.
    
%     \item[\faPalette]\textbf{Color perception} is a basic building block of data visualization~\cite{munzner14visualization, Rhyne2016}.
%     Visualization practice stipulates using a limited number of distinguishable and easily named colors~\cite{DBLP:conf/visualization/Healey96}, potentially as a function of the mark used~\cite{szafir18modelingcolordifference}.

%     \item[\faAccessibleIcon]\textbf{Accessibility} \rev{To reach a broader audience, visualizations must be made inclusive. 
%     Accessibility is emerging as an important new topic in the visualization community. 
%     Various issues, such as color vision deficiency and low vision, exist alongside numerous challenges that still need to be addressed.~\cite{elavsky24datanav, DBLP:journals/cgf/ChoiJPCE19, DBLP:journals/tvcg/ChunduryPRTLE22, DBLP:journals/interactions/MarriottLBCEGHM21, Elmqvist2023}.}
%     %Automated feedback can be highly useful even for seasoned designers.
%     % improving the effectiveness and reach of the visualization for a broader audience
% \end{itemize}

\begin{itemize}[noitemsep, topsep=0.3cm, leftmargin=0.5cm]

    \item[\faSun] \textbf{Virtual Eyetracker.}
    Virtual eyetracker refers to the extent to which areas within a scene capture the attention of an observer~\cite{riche13metricssaliency}.
    \rev{We utilize a virtual eyetracker~\cite{shin23scannerdeeply} that predicts gaze on an overview task of a visualization image, and provide quantitative guidance. 
    From the eyetracker, users can find where the audience would look at a visualization image as an overview. 
    We have two reasons for providing metrics as guidance.}
    First, we notify designers if heavy focus of salience at an area is indeed what they intended. 
    As a case in point, if salience is primarily directed toward text, the metrics can prompt designers to reflect on whether they should shift more attention to graphical elements. 
    Second, we alert users to potential biases by providing metrics that indicate when salience is heavily concentrated in a particular area. 
    Because the eyetracker's performance is not perfect, it is important to inform users about possible biases, particularly when salience rates are unusually high or low. 
    For example, the center of an image may show unusually high salience even if the area contains no graphical elements.
    
    \begin{itemize}[noitemsep, topsep=0.2cm, leftmargin=0.3cm]
        
        \item\textit{A virtual eyetracker.} 
        \textbf{Role:} \rev{Predicts visual salience in a visualization to estimate what readers will notice during an overview task.}  
        \textbf{Analysis process:} \rev{A Scanner Deeply~\cite{shin23scannerdeeply} generates an expected salience map about the inputted chart image on an overview task.}
        \textbf{Clarification:} \rev{The salience map is overlaid as a heatmap on the chart image.} 
        The most salient areas appear in red or yellow, while less salient areas are shown in darker tones.
        \textbf{Guidance to users: } The salience map indicates whether the highlighted areas match the designer’s intended focus~\cite{bylinskii17visualimportance}. 
        It also suggests methods to improve salience in specific zones.
        In order to help designers focus on their areas of interests, we introduce the concept of minimizing the data-ink ratio~\cite{Tufte1983}, methods to highlight zones of interest (e.g., using colors, making bold, etc.), and impact of grids in charts~\cite{wattenbarg03multiscale}. 
    
        \item\textit{Focus on text.} 
        \textbf{Role:} \rev{Warns designers when the salience may focus overly on text. }
        \textbf{Analysis process:} \rev{OCR software is used to detect textual zones, and texts are boxed by each letter. 
        Then, we calculate how much salience is focused on these boxes. }
        \textbf{Clarification:} Mark as \textit{overly salient to text} if the salience in text boxes is above the $10^{th}$ percentile of the chart saliency in the dataset by Shin et al.~\cite{shin23scannerdeeply}.
        \textbf{Guidance to users: } We warn designers that the salience of text is higher than other visualizations.  
        We caution that the model’s salience measurements may be biased or inaccurate because it tends to highlight text too readily~\cite{shin23scannerdeeply}.
        The designer has to decide whether to believe the results or not.
        
        \item\textit{Focus on center.}
        \textbf{Role:} \rev{Warns designers when} the salience focuses overly on the center of the chart image.
        \textbf{Analysis process:} We measure the concentration of salience in the center. 
        The center is defined as a rectangle in the middle of the image, with its height and width as $1/3\times$ of the chart.
        \textbf{Clarification: } \rev{A visualization is flagged as \textit{overly salient in the center} if saliency in this region is above $10^{th}$ percentile.}
        \textbf{Guidance to users: }  We warn designers that the salience on the central zone~\cite{bylinskii19differentmodels} is higher than other visualizations.
        We also inform the possibility that the model’s salience measurements may be biased or inaccurate.

        \item\textit{Focus on visual attention.} 
        \textbf{Role:} \rev{Warns designers when a visualization may not effectively direct visual attention to important regions.}
        \textbf{Analysis process:} \rev{Identify areas likely to attract attention by locating zones near RGB color transitions (within a 5-pixel radius), and measure how much of these zones are covered by high saliency values.}
        \textbf{Clarification:} 
        %Mark as \textit{scarcely salient} if the salience in zones with visual attention is less than $90^{th}$ percentile.
        \rev{A visualization is flagged as \textit{``scarcely salient''} if saliency in these regions falls below the 90$^{th}$ percentile of visualizations in the dataset.}
        \textbf{Guidance to users:} Designers are alerted when attention-relevant zones receive less saliency than is typical. 
        This matters because strong visualizations generally guide attention toward meaningful visual structures. Low saliency in such areas may reduce effectiveness.
        
        %We warn designers when the salience on areas with visual attention is lower than other visualizations. 
        %We present two possibilities. 
        %First, visual salience in visualizations generally focuses on areas with visual attention, where charts are typically located~\cite{itti98saliency, harel06gbvs}. 
        %Hence, if the salience is minimal, it is worth noting.

% Focus on Visual Attention/ 

% Role: Warns designers when a visualization may not effectively direct visual attention to important regions.

% Analysis Process: Identify areas likely to attract attention by locating zones near RGB color transitions (within a 5-pixel radius), and measure how much of these zones are covered by high saliency values.

% Clarification: A visualization is flagged as "scarcely salient" if saliency in these regions falls below the 90th percentile of visualizations in the dataset.

% Guidance to Users: Designers are alerted when attention-relevant zones receive less saliency than is typical. This matters because strong visualizations generally guide attention toward meaningful visual structures. Low saliency in such areas may reduce effectiveness.

    \end{itemize}

    \item[\faFont]\textbf{Textual characters.}
    Textual characters are visual objects that are used for labels, titles, and legends in visualization. 
    Using text alongside charts can improve comprehension of visualizations \cite{stokes23tvcgtextscharts}. 
    \rev{To support this process, the system aims to detect title, and textual information in the chart.}
    %We evaluate the existence of textual components in the visualization image~\cite{stokes23tvcgtextscharts}. 

    \begin{itemize}[noitemsep, topsep=0.1cm, leftmargin=0.2cm]

        \item\textit{Title. }
        \textbf{Role:} Detects a title and provides recommendations. 
        \textbf{Analysis process:} Title is detected via DePlot~\cite{liu23deplot}. 
        \rev{DePlot is a VLM that translates visualizations into tables and textual components.}
        Gathering all the information, ChatGPT provides title suggestions for the chart.
        \textbf{Clarification:} DePlot reads the chart image, and decides if there is a title in the image.
        \textbf{Guidance to users:} If a title is not detected, then recommend adding a one and provide suggestions.

        \item\textit{Textual content. } 
        \textbf{Role:} \rev{Detects} text in visualization image. 
        \textbf{Analysis process:} \rev{The OCR filter scans the chart image, and detects textual information. 
        We use the PyTesseract-OCR as our OCR filter~\cite{PyTesseract_perf}. 
        PyTesseract-OCR is a Python library that detects textual information from the visualization image.
        }
        \textbf{Clarification:} If the OCR filter detects at least one letter in the text, then it contains the text. 
        \textbf{Guidance to users:}
        When text is detected the system explains the benefits of texts in charts. 
        If no text is detected, it explains in detail the benefits of having textual explanations in visualizations~\cite{stokes23tvcgtextscharts}.
    \end{itemize}

    \item[\faChartBar]\textbf{Visual representations.}
    There exist various functionalities by knowing the chart's data and its visual representations, such as visualization linting~\cite{chen22vizlinter, hopkins20visualint}. 
    %\textbf{\rev{Implementation overview:}}
    Our focus is on providing chart recommendations based on the data table and design preamble, as well as detecting unnecessary visual embellishments (chartjunk). 
    By leveraging these suggestions, we aim to improve the visual representation of the data.
    
    \begin{itemize}[noitemsep, topsep=0.1cm, leftmargin=0.2cm]

        \item\textit{Optimal chart type:} 
        \textbf{Role:} \rev{Suggests the designer an optimal chart, based on the data table translated from the chart. }
        \textbf{Analysis process:} \rev{The data table is extracted from DePlot.
        The suggestion is generated based on a preamble of design knowledge and ChatGPT, which is described below.}
        \textbf{Clarification:} 
        The suggestions are determined based on two types of information: (1) the guideline as a preamble of design knowledge (perceptual rankings of visual cues, described in Mackinlay's APT~\cite{Mackinlay1986}) and (2) the data table provided by DePlot. 
        Using these two information, ChatGPT makes the final suggestion. 
        We recognize that this is a relatively simple model with limitations, such as the inability to recommend treemaps and a restricted knowledge base for certain visualization choices.
        \rev{No suggestion is provided if data table, or data is not extractable from the chart image.}
        \textbf{Guidance to users:} A chart is suggested, and ChatGPT provides both the data table and the reasoning behind its choice. 
        This helps the designer decide whether to follow the advice.
        
        \item\textit{Visual embellishment (``chart junk''):} 
        \textbf{Role:} Detects any chartjunk~\cite{DBLP:conf/chi/BatemanMGGMB10} in the visualization, and if present, warns about its potential benefits and drawbacks. 
        \textbf{Analysis process}: A computer vision model, such as YoloR~\cite{wang21yolor}  detects objects. 
        \textbf{Clarification:} If the object detection model detects objects within the chart image, then we say that the model has \textit{chartjunk}.
        \textbf{Guidance to users:} If chartjunk is detected, then warn of its benefits as well as dangers~\cite{borkin13memoryvis, DBLP:conf/chi/BatemanMGGMB10}.
        
    \end{itemize}

    \item[\faPalette]\textbf{Color perception.} Color perception is a basic building block of data visualization~\cite{munzner14visualization, Rhyne2016}.
    Visualization practice stipulates using a limited number of distinguishable and easily named colors~\cite{DBLP:conf/visualization/Healey96}, potentially as a function of the mark~\cite{szafir18modelingcolordifference}. 
    We detect the number of different and similar colors in a chart image. 
    We then remind users about appropriate color usage.
    For instance, similar hues should represent continuous values, while distinct colors should signify categories~\cite{franconeri21visualdatacommunication,munzner14visualization}.
    
    \begin{itemize}[noitemsep, topsep=0.1cm, leftmargin=0.2cm]

        \item\textit{Variability of colors:} 
        \textbf{Role:} \rev{Detects different colors and reminds the designer to check if different colors are used effectively (representing categorical values).}
        \textbf{Analysis process:} \rev{A custom function processes the chart image to identify distinct colors. 
        Colors are grouped as \textit{similar} if their maximum Euclidean distance in RGB space is less than 10, with each group represented by its centroid. 
        Each group is treated as a single distinct color.}
        \textbf{Clarification:} \rev{If the function identifies more than 2 distinct colors from the chart image, then the image is identified as having multiple colors.}
        \textbf{Guidance to users:} \rev{After showing all distinct colors used, the filter recommends that different colors represent categories. 
        However, it is not capable of detecting mismatch of colors. }
        
        \item\textit{Similarity of colors:} 
        \textbf{Role:} \rev{Detects similar colors and suggests the designer to identify if they are used effectively (representing continuous values). }
        \textbf{Analysis process:} A customized function created in Python detects groups of similar colors \rev{(the definition is identical to similar colors described in variability of colors)}. 
        A group of similar colors are determined using differences in RGB values between detected colors. 
        \textbf{Clarification:} If the function identifies more than 2 similar groups of colors then the image is identified as having similar colors.
        \textbf{Guidance to users:} The filter recommends that similar colors represent continuous, related values.
        \rev{This filter too, is not capable of detecting color mismatch. }

    \end{itemize}

    \item[\faAccessibleIcon]\textbf{Accessibility.}
    Accessibility is emerging as an important new topic in the visualization field~\cite{Elmqvist2023}, including issues such as color vision deficiency, blindness, and low vision.~\cite{elavsky24datanav, DBLP:journals/cgf/ChoiJPCE19, DBLP:journals/tvcg/ChunduryPRTLE22, DBLP:journals/interactions/MarriottLBCEGHM21, Elmqvist2023}.
    %\textbf{\rev{Implementation overview:}} 
    Although accessibility encompasses many considerations, we specifically addresses color vision deficiency (CVD).
    We focus on CVD because it affects a significant portion of the population (about 1 in 12 males and 1 in 200 females are affected by CVD) and directly impacts the effectiveness of color-encoded data~\cite{donoharmguide}.
    
    \begin{itemize}[noitemsep, topsep=0.1cm, leftmargin=0.2cm]
        \item \textit{CVD:} 
        \textbf{Role:} Helps detect visualizations that are significantly affected for people with color vision deficiency (deuteranopia, protanopia, tritanopia).
        \textbf{Analysis process:} \rev{We evaluate the chart image from two perspectives. 
        First, we quantitatively measure the information loss in the simulated visualizations using the entropy difference from the original image. 
        Second, we simulate the chart image as if it were viewed with color vision deficiency. 
        }
        \textbf{Clarification:} If the entropy loss in the simulated visualization image exceeds a certain threshold, then the chart image is marked as \textit{significantly affected.}
        \textbf{Guidance to users:}
        \rev{We caution that individuals with color vision deficiency may miss information. 
        To observe the exact differences, one should view the simulated chart image.
        We then provide recommended CVD-safe palettes.
        However, it is up to the designer to select the palette.}
    \end{itemize}
    
\end{itemize}

When the system detects no quantitative issue in \textit{clarification}, it indicates that no problem was found and then presents the filter’s intent.
For example, when a title is detected, the system says, ``A title is detected. The intention of this filter is to determine whether a title exists and, if it does not, recommend adding one, because a title helps users better understand the visualization.''
Another example is when the system detects no chartjunk. 
In that case, the system replies as, ``\textit{We did not detect any chartjunk.}''

% We instill these metrics along with the suggestions relevant to the decisions into ChatGPT as a prompt (see \S~\ref{sec:generating-feedback}). 
% Then, based on a series of template questions, ChatGPT provides the appropriate answer. 

\smallskip
\noindent\textbf{Revision Tracker. } The \textit{revision tracker} in Fig.~\ref{fig:teaser}~(D) gives a high-level overview of changes from one revision to another for the entire design process.
This component realizes the ``track'' in the ACGT workflow. 
The intention is to give the designer a bird's eye view of the design process, thereby seeing the improvements over time and also avoiding local optima. 
The natural language in the tracker is derived using the LLM with the reports to be compared as prompts; see the next section.
The revision tracker primarily monitors changes in quantifiable metrics for each topic. 
It displays any increases or decreases and provides commentary on whether those changes are beneficial. 
Because it focuses on metrics, it does not capture qualitative comparisons (e.g., saliency maps or simulated CVD images). 
To address that gap, we also offer a detailed tracking capability.
For detailed tracking, \techname{} allows designers to check all versions of their past reports by going to `Archives' in the navigation bar (see Fig.~\ref{fig:teaser} \textcircled{8}).
The interface provides two separate screens, each displaying a report of past version selected by the user, to facilitate the comparison process.

\subsection{Prompt Engineering for Visualization Feedback}
\label{sec:generating-feedback}

We use off-the-shelf LLMs, such as ChatGPT, to generate the interpretations and suggestions in the visualization design reports. 
However, formulating effective prompts requires careful \emph{prompt engineering}~\cite{reynolds2021prompt, wei2022emergent, wei2022chain, zhou2022least, arora2022ask}. 
Prompt engineering is typically based on empirical data specific to each model, and our approach using LLMs for visualization feedback is no different.
An LLM is not deterministic, and may occasionally produce unexpected output.
It also tends to be quite verbose, producing long responses to even short and well-contained queries.
We thus refer to guidelines from prior work~\cite{liu2022design} in designing prompts to avoid such non-deterministic and long-winded behavior:

\begin{itemize}%[noitemsep, topsep=0.em, leftmargin=1.2em]
    \item \textit{Use exact keywords:} use concepts as reserved words;
    \item \textit{No synonyms:} varying labels yields nondeterminism;
    \item \textit{Avoid unnecessary data:} irrelevant data expands scope;
    \item \textit{Constrain feedback length:} avoid unfocused answers;
    \item \textit{Ask specific queries:} limit the scope of the query; and
    \item \textit{Avoid open-endedness:} encourage short/direct responses.
\end{itemize}

To further minimize hallucinations, we incorporate relevant grounding contexts (e.g., design knowledge about data visualizations and filters)~\cite{mialon23augmentedLM} in the prompt, preventing the LLM from generating text outside its learned scope.

We construct two types of prompts: \textit{analyze-clarify-guide} prompts and \textit{track} prompts.
To create those prompts, we define the following templates $T$ (full prompts are included in OSF~\footnote{\url{https://osf.io/v7hu8}}):

\textbf{ACG-template.}
We call the analyze-clarify-guide (ACG) prompt template $T_{acg}$.
\verb|[Q]| is the question we define to create interpretations and suggestions.
For instance, we use
``\textit{Analyze the visual salience on text.
Provide interpretations in 2 sentences.}''
or ``\textit{Provide suggestions about the result of the previous question in 2 sentences.}''
We define 9--12 questions for each visualization metric.
We replace the metric name in different questions, e.g., from visualization salience to color blindness.
We call the prompt conditions \verb|[cond]|:
``\textit{Please interpret exactly in the following way, as if you are an assistant to a visualization designer, explaining to novice visualization designers.
If no visual salience is detected, then just interpret it as No salience is detected in the chart image.
If (the rate of salience in the textual zone over the rate of the textual zone in chart image) from the measured result is less than 0.6, then interpret it as a lack of salience in textual elements.}''
At the end of each prompt, we append \verb|[filter-suggestions]|, a compact representation of visualization design knowledge in natural language; see the supplementary material. 

We mainly use ``if, then'' tone when providing design knowledge. 
While this may not be the most natural style of discussion, we want to convey that our suggestions are optional and to prevent designers from applying feedback inaccurately.

\textbf{T-template.}
$T_{t}$ is used for generating track (T) prompts.
We define the question \verb|[Q]| for extracting differences between two different versions of data visualization as follows:
``\textit{Given the information, in one sentence, concisely, what are the changes made between the current and previous versions about visual salience?
Although the changes may seem minor, try to describe even the small, acute changes made between the current and previous versions, in terms of the visual salience.}''
\verb|[{curr/prev}-output]| is the collection of visualization design knowledge, expressed as natural language, calculated by the visualization filter components.
\verb|[{curr/prev}-interpretations]| is the interpretations \techname{} generated for the two visualizations.

\begin{mdframed}[backgroundcolor=backcolour, linecolor=white, innerleftmargin=4pt, innerrightmargin=4pt, innertopmargin=-3pt, innerbottommargin=4pt]
\begin{align*}
    T_{acg} &= \textcolor{darkblue}{\textbf{\texttt{[Q]}}} +\\
    & \text{\textnormal{\textcolor{darkgreen}{\textbf{``Solve the problem based on this guideline:''}}}} + \\
            & \quad \textbf{\texttt{[cond]}} + \textcolor{brownishorange}{\textbf{\texttt{[filter-suggestions]}}}. \\
    T_{t} &= \textcolor{darkblue}{\textbf{\texttt{[Q]}}} + \\
    & \text{\textnormal{\textcolor{darkgreen}{\textbf{``Here are details about the current version:''}}}}~+ \\
          & \quad \textbf{\texttt{[curr-output]}} + \textbf{\texttt{[curr-interpretations]}} \nonumber \\
          & \quad + \text{\textnormal{\textcolor{darkgreen}{\textbf{``Here are details about the previous version:'' }}}} + \\
          & \quad \textbf{\texttt{[prev-output]}} + \textbf{\texttt{[prev-interpretations]}}.
\end{align*}
\end{mdframed}

\subsection{Implementation Notes}
\label{subsec:implementation}

The \techname{} system is implemented as a web application using HTML, CSS, JavaScript, and JQuery.
We use the Python Flask web framework\footnote{\url{https://flask.palletsprojects.com/}}; the analysis components were implemented server-side in Python 3.
We store data into MongoDB.\footnote{\url{https://www.mongodb.com/}}
During the user study, \techname{} was hosted on Amazon Web Services (AWS).
We use \verb|gpt-3.5-turbo| as LLM.

%% -------------------------------------------------------------
%% USER STUDY
%% -------------------------------------------------------------
\section{User Study}
\label{sec:user-study}

We performed a user study to assess the impact of LLM-generated feedback in the visualization design pipeline using the \techname{} system.
We opted for a qualitative study where designers of different skill were asked to use \techname{} to support the design process of a new visualization over several days.
We then let a small group of visualization experts assess the design process without seeing the \techname{} reports.
We also interviewed our participants to collect their perception of the impact of \techname{} on their design work.
The study received approval from our university's ethics review board (IRB).

\begin{table}[t!]
    \centering
    \caption{\textbf{User study demographics.}
    A total of 13 visualization designers of different seniority participated in the user study. 
    These participants were all above the age of 18, had at least a bachelor's degree in visualization or related field, and had more than a year of experience in designing visualizations.
    We categorize those with less than or equal to 3 years of experience in visualization as novice (6 participants, N1-N6), those with 4 to 7 years as intermediate (4 participants, I1-I4), and those with more than 7 years as expert groups (3 participants, E1-E3), respectively.
    We present participants' gender, age, education (Edu), job title, and years of experience in Vis (Exp.). }
    \scalebox{0.99}{
        \begin{tabular}{llllll}
        % \rowcolor{SteelBlue}
        \toprule
        \textbf{ID} &
        \textbf{Gender} &
        \textbf{Age} &
        \textbf{Edu.} &
        \textbf{Job Title} &
        \textbf{Exp.} 
        \\
        % \textbf{Group} \\
        \midrule
        % \rowcolor{LightSteelBlue!30}
        N1 & Male & 33 & B.S. & Software engineer & 1 yr \\
        \rowcolor{gray!20}
        N2 & Female & 33 & B.S. & Primary school teacher & 1 yr \\
        N3 & Male & 32 & Ph.D. & Postdoc in CS & 1 yr \\
        \rowcolor{gray!20}
        N4 & Male & 26 & B.S. & Ph.D.\ student in CS & 3 yrs \\
        % \rowcolor{LightSteelBlue!30}
        N5 & Male & 29 & M.S. & Software engineer & 3 yrs \\
        \rowcolor{gray!20}
        N6 & Female & 24 & B.S. & M.S. student in Data Science & 3 yrs \\
        \midrule
        % \rowcolor{LightSteelBlue!30}
        I1 & Male & 30 & B.S. & Ph.D.\ student in CS & 5 yrs \\
        \rowcolor{gray!20}
        I2 & Male & 34 & Ph.D. & Research scientist & 5 yrs \\
        % \rowcolor{LightSteelBlue!30}
        I3 & Male & 28 & M.S. & Ph.D.\ student in CS & 6 yrs \\
        \rowcolor{gray!20}
        I4 & Male & 31 & M.S. & Ph.D.\ student in CS & 6 yrs \\
        \midrule
        % \rowcolor{LightSteelBlue!30}
        E1 & Male & 31 & M.S. & Data scientist & 9 yrs \\
        \rowcolor{gray!20}
        E2 & Male & 33 & Ph.D. & Assistant professor in Vis & 10 yrs \\
        % \rowcolor{LightSteelBlue!30}
        E3 & Male & 35 & Ph.D. & Postdoc in Vis & 11 yrs \\
        \bottomrule
    \end{tabular}
    }
    \label{tab:participant}
\end{table}

\begin{table*}[ht]
    \centering
    \caption{\textbf{Participants' study setup. }
    Here we list the settings the participants used to conduct our study. We present the tools, datasets and chart types used by the participants.
    We also show the total time (\texttt{hh}:\texttt{mm}) taken for these participants to finish the work (all participants updated five designs within a single continuous time frame).}
    \begin{tabular}{llllc}
        % \rowcolor{SteelBlue}
        \toprule 
        \textbf{ID} &
        \textbf{Tool Used} &
        \textbf{Dataset Topics} &
        \textbf{Chart Type Explored} &
        \rev{\textbf{Time Between V1 and V5}} \\
        \midrule
        
        N1 & Microsoft Excel & Time management & Stacked area/bar charts, line chart & 01:47 \\
        \rowcolor{gray!20}
        N2 & Microsoft Excel & Survey analysis & Bar chart & 01:23 \\
        % \rowcolor{LightSteelBlue!30}
        N3 & Python (Matplotlib) & Topic document analysis & Bar chart & 01:14 \\
        \rowcolor{gray!20}
        N4 & Python (Matplotlib) & Scientific result analysis & Bar chart, line chart & 02:16 \\
        % \rowcolor{LightSteelBlue!30}
        N5 & Python (Matplotlib) & Dataset analysis & Line chart, bar chart & 01:48\\
        \rowcolor{gray!20}
        N6 & Tableau & Scientific result analysis & Dot plot, line chart & 02:31 \\
        \midrule
        % \rowcolor{LightSteelBlue!30}
        I1 & Javascript (D3.js) & Cluster analysis & Matrix scatterplot & 01:54\\
        \rowcolor{gray!20}
        I2 & Tableau & Credit card usage  & Bar chart, Stacked bar chart & 02:28\\
        % \rowcolor{LightSteelBlue!30}
        I3 & Microsoft Excel & User experience on UI & Bar chart & 03:18\\
        \rowcolor{gray!20}
        I4 & Microsoft Excel & Scientific result analysis & Bar chart, stacked bar chart  & 02:04 \\
        \midrule
        % \rowcolor{LightSteelBlue!30}
        E1 & Tableau & Sales data analysis & Geographic map, bubble chart & 02:35\\
        \rowcolor{gray!20}
        E2 & Spotfire & Market analysis & Bar chart & 02:15\\
        % \rowcolor{LightSteelBlue!30}
        E3 & Microsoft Excel & Scientific result analysis & Boxplot & 01:47\\
        \bottomrule 
    \end{tabular}
    \label{tab:participant-setup}
\end{table*}

\subsection{Participants}
\label{subsec:participants}

We advertised our study through messages to the Data Visualization Society, to alumni of our institution, and to other academic institutions. 
Out of many applicants, we selected only those who are over the age of 18, are capable of reading and speaking English fluently, and have experience in visualization for more than \rev{a year.} 
As a result, we were able to recruit 13 visualization designers of different seniority.
Table~\ref{tab:participant} shows the demographics of the participants. 
These participants were paid a \$40 Amazon gift card upon concluding the study. 
While \techname{} was originally designed for novice and intermediate visualization designers, our participants also included expert ones.
We included such participants because we were interested in seeing whether our system could benefit even an expert. 
%Furthermore, we also wanted to eliminate the confound of inexperienced designers having to first learn how to author visualizations. 

Beyond the designer participants, we also recruited three expert participants with at least seven years of experience in visualization: an associate professor (or equivalent), an assistant professor, and a Ph.D.\ candidate (within months of graduation) within the research field.
These experts only took part in the final independent design assessment step.

\subsection{Task}
\label{subsec:tasks}

Our study asked participants to design a new visualization over a period of 3--5 days while using the \techname{} interface to support their design process. 
Participants were free to choose a dataset and to craft any type of visualization using any tool or combination of tools.
We asked each participant to upload at least 5 versions (including the first and last versions) of their designs into \techname{}.
We imposed the following requirements:

\begin{itemize}
    \item They must spend at least two hours on the design process;
    \item They should work independently on the design task;
    \item Image size between $100 \times 100$ and $2000 \times 2000$ px;
    \item Images in either \texttt{.png} or \texttt{.jpg} formats;
    \item No photographs allowed; and
    \item No confidential information or data was used.
\end{itemize}

We also secured permission to use the designs for publication.

\subsection{Procedure}
\label{subsec:procedure}

The study was performed exclusively online using video conferencing and consisted of four steps.
We obtained participant demographic information and scheduled the dates for each step during recruitment. 
Here we describe the steps in detail.

\textbf{Step 1: Pre-study Briefing.}
We started each session with an overview of our study, answered any questions from the participant, and collected informed consent using an online consent form, as per the directives of our IRB.
Then the experiment administrator explained the longitudinal chart design task (\S~\ref{subsec:tasks}).
Next, the administrator showed how to operate the \techname{} system: creating an account, logging in, uploading screenshots, and accessing design reports.
The administrator answered any additional questions from the participant.
The study was concluded by finding an appointment for the post-interview.
On average, this briefing process took 30 minutes.

\textbf{Step 2: Longitudinal Chart Design.} 
Participants were given at least three days to design their visualization, during which time they were asked to upload at least five versions of their visualization. 
We imposed no restrictions on how participants allocated their time \rev{(R2)} but asked that they dedicate at least a total of two hours during the period on the design and that they use the \techname{} interface \rev{(R1)}. 
During this period, the experiment administrator was available to answer technical questions.

\textbf{Step 3: Post-study Interview.}
We conducted a post-interview interview with each participant at the end of the design period.
We started by asking each participant to explain how their designs evolved over time. 
We always selected the first and last version of the visualization artifact. 
If more than five versions were uploaded by a participant, we asked the participant to identify the three most significant versions other than the first and last ones.
We defined ``significant'' as versions where feedback had caused them to make radical changes to the design.
Following that, we posed questions about their experiences using \techname{} during the design process. 
Afterward, we asked about their concepts of an ideal feedback provider after the study. 
The questions we asked are shown in the next section. 
While we allocated 30 minutes for this session, some of the participants exceeded this time limit.

\textbf{Step 4: Expert Design Assessment.}
In the final step, our independent expert participants (all with experience teaching data visualization and grading visualization assignments) were given all five designs for each of the designer participants.
They were asked to provide their assessment of the quality of changes during the design process from the first to the final design for each participant using a 1--5 Likert scale (1 = significant decline in quality, 3 = neutral, 5 = significant quality improvement).
They were also asked to motivate their assessment using 1 to 2 sentences.
\rev{This meant that each expert gave 1 Likert scale rating and 1-2 assessment sentences per designer participant's each iteration (that is, 5 ratings and 5 assessment sentences in total).}

\subsection{Data Collection}

Interviews were video and audio recorded. 
We transcribed audio for further analysis.
We collected demographics using online forms before the study. 
All uploaded chart images and design reports were collected automatically by \techname{}.

% [Note] design evolution figure is here b/c of the latex's weird placement rules.
\begin{figure*}[ht]
    \setlength{\tabcolsep}{0pt}
    \begin{tabular}{lccccc}
        \hspace{0.4cm} &
        \colorbox{blue!10}{\parbox[t][0.3cm]{0.18\textwidth}{\centering Version 1}} &
        \colorbox{blue!20}{\parbox[t][0.3cm]{0.18\textwidth}{\centering Version 2}} &
        \colorbox{blue!30}{\parbox[t][0.3cm]{0.18\textwidth}{\centering Version 3}} &
        \colorbox{blue!40}{\parbox[t][0.3cm]{0.18\textwidth}{\centering Version 4}} &
        \colorbox{blue!50}{\parbox[t][0.3cm]{0.18\textwidth}{\centering Version 5}}
        \\
        
        \textbf{N1} & 
        \includegraphics[width=0.18\textwidth]{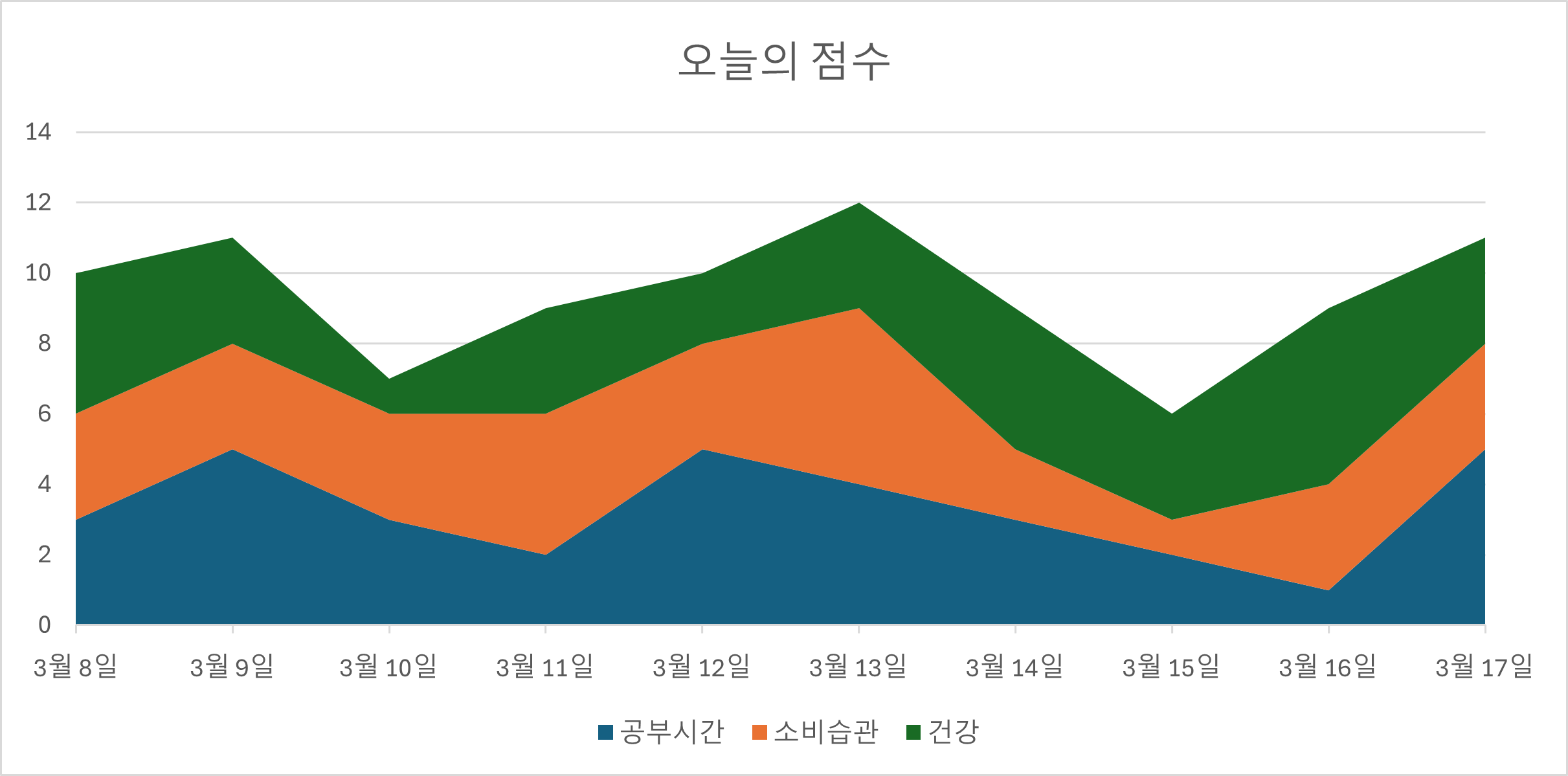} &
        \includegraphics[width=0.18\textwidth]{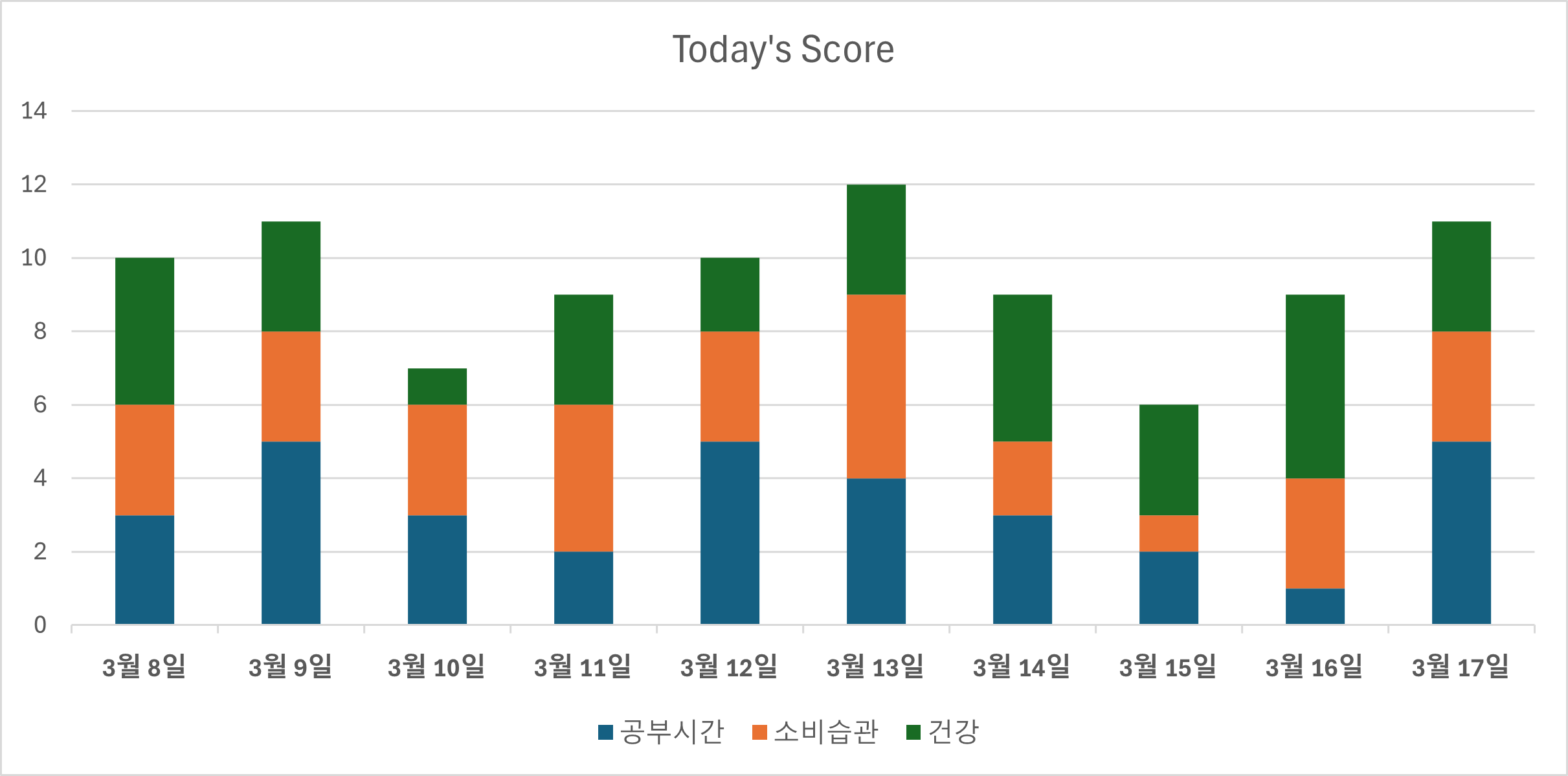} &
        \includegraphics[width=0.18\textwidth]{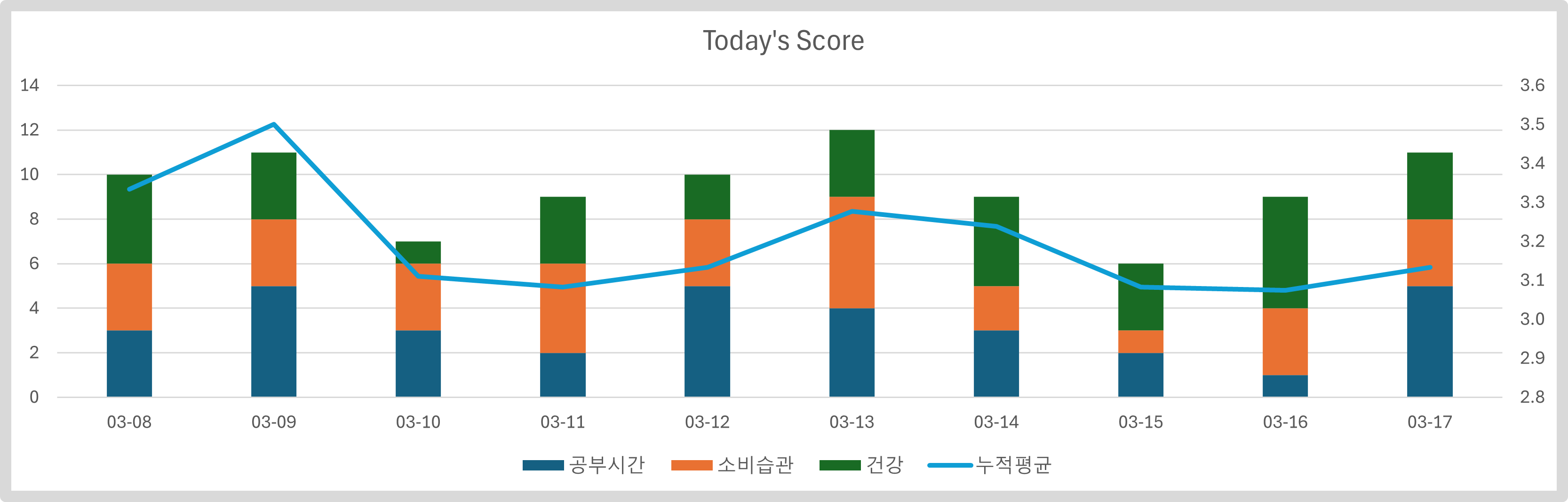} &
        \includegraphics[width=0.18\textwidth]{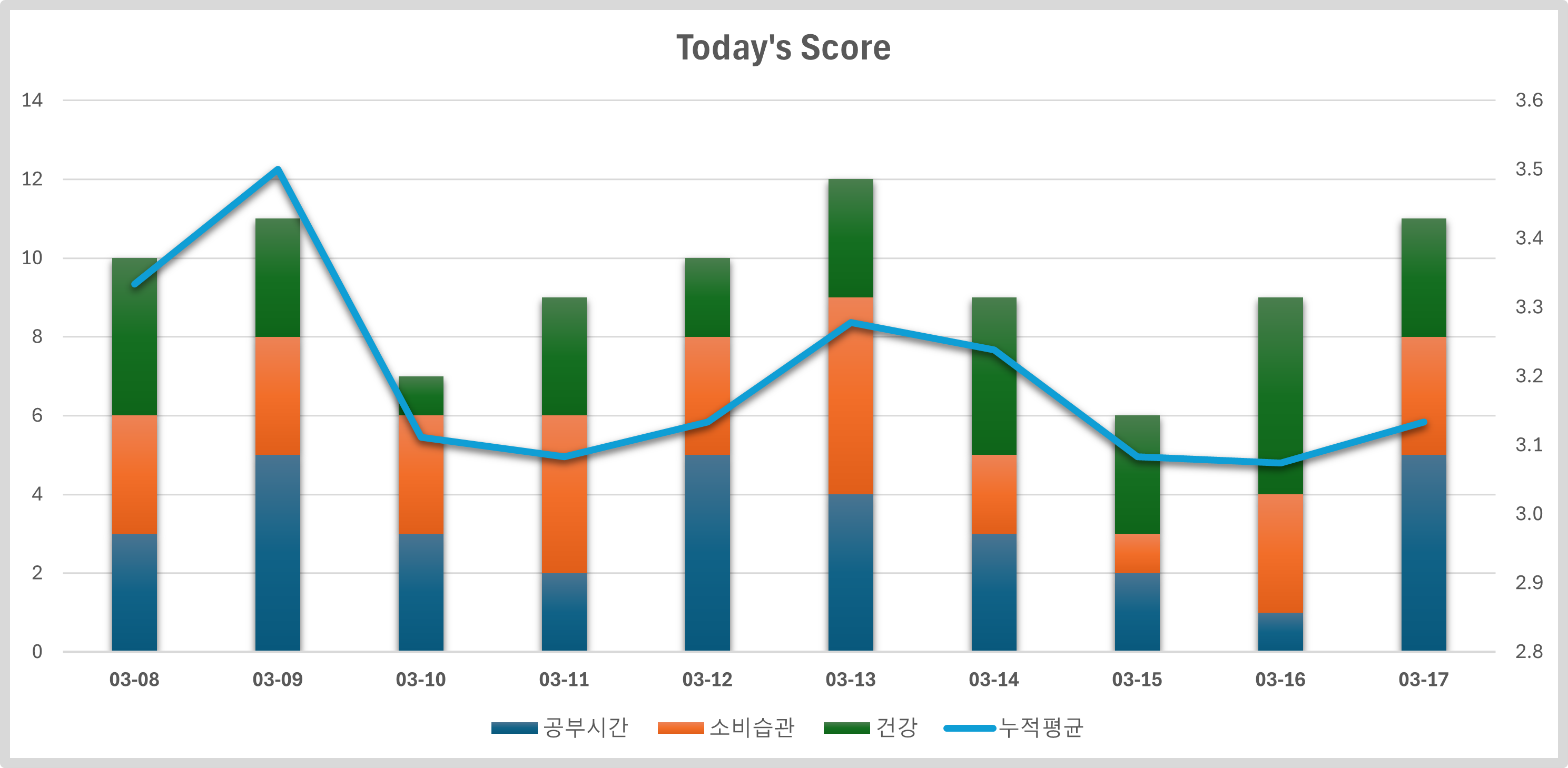} &
        \includegraphics[width=0.18\textwidth]{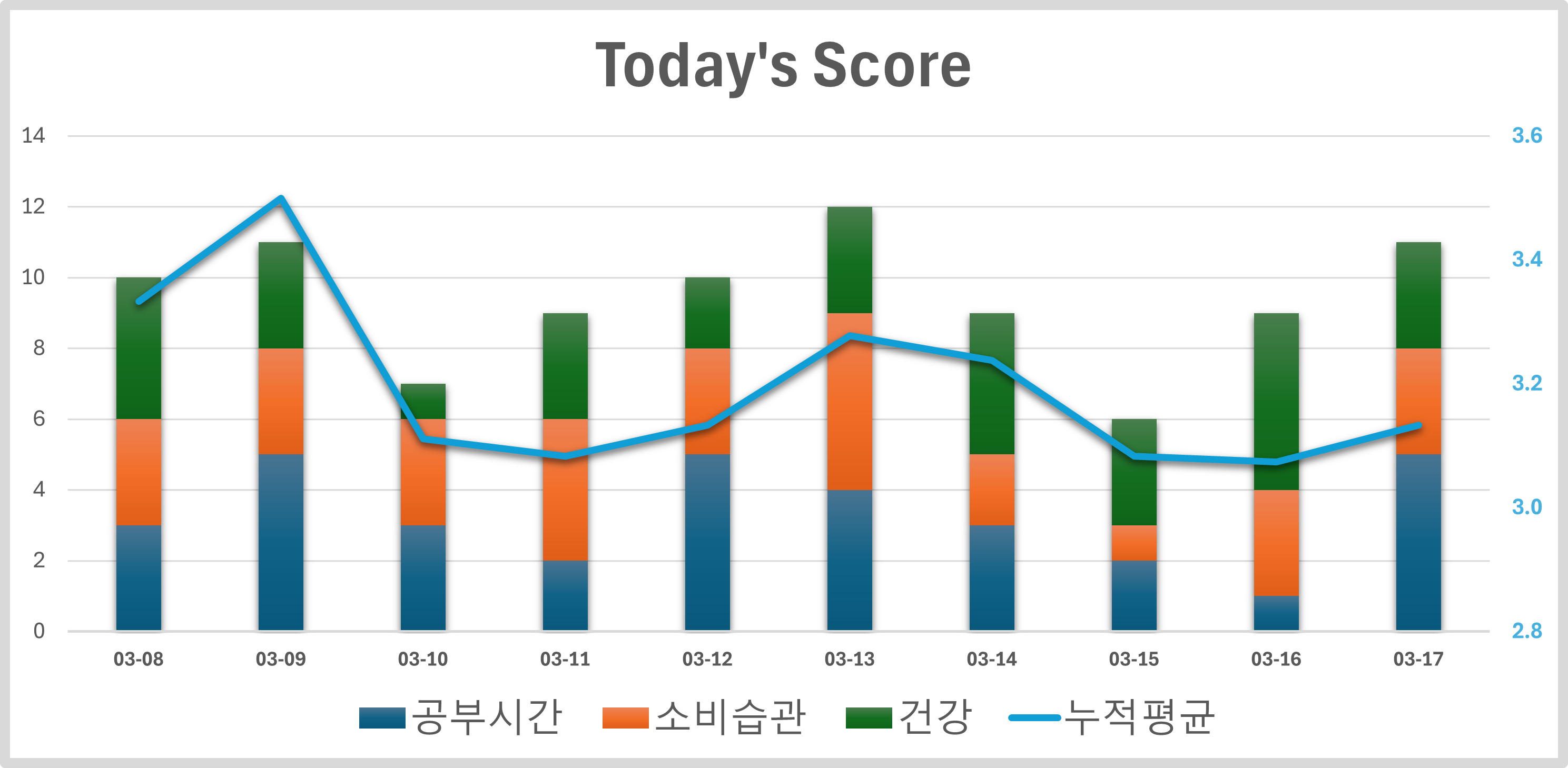}
        %\includegraphics[width=0.17\textwidth]{figures/P1_1} &
        %\resizebox{0.18\textwidth}{!}{\framebox{Vis}} &
        %\includegraphics[width=0.17\textwidth]{figures/P1_2} &
        %\resizebox{0.18\textwidth}{!}{\framebox{Vis}} &
        %\includegraphics[width=0.17\textwidth]{figures/P1_3} &
        %\resizebox{0.18\textwidth}{!}{\framebox{Vis}} &
        %\includegraphics[width=0.17\textwidth]{figures/P1_5} &
        %\resizebox{0.18\textwidth}{!}{\framebox{Vis}} &
        %\includegraphics[width=0.17\textwidth]{figures/P1_6} 
        %\resizebox{0.18\textwidth}{!}{\framebox{Vis}}
        \\
        \midrule

        \textbf{N2} & 
        \includegraphics[width=0.18\textwidth]{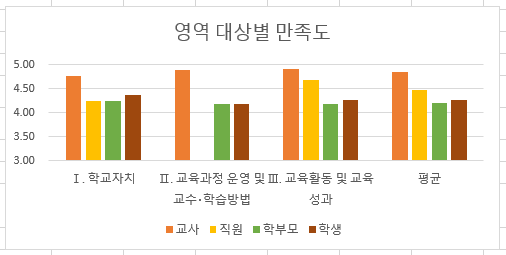} &
        \includegraphics[width=0.13\textwidth]{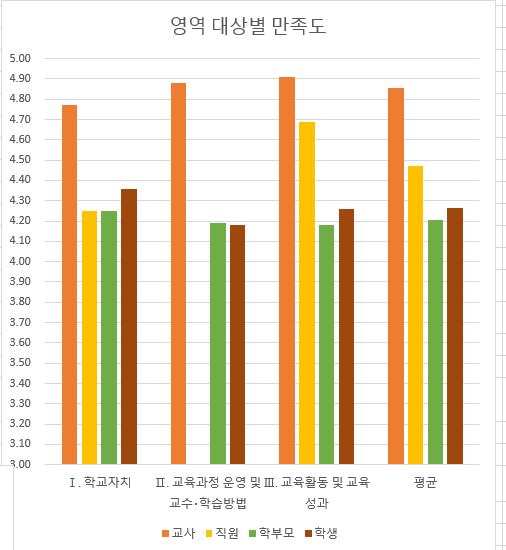} &
        \includegraphics[width=0.11\textwidth]{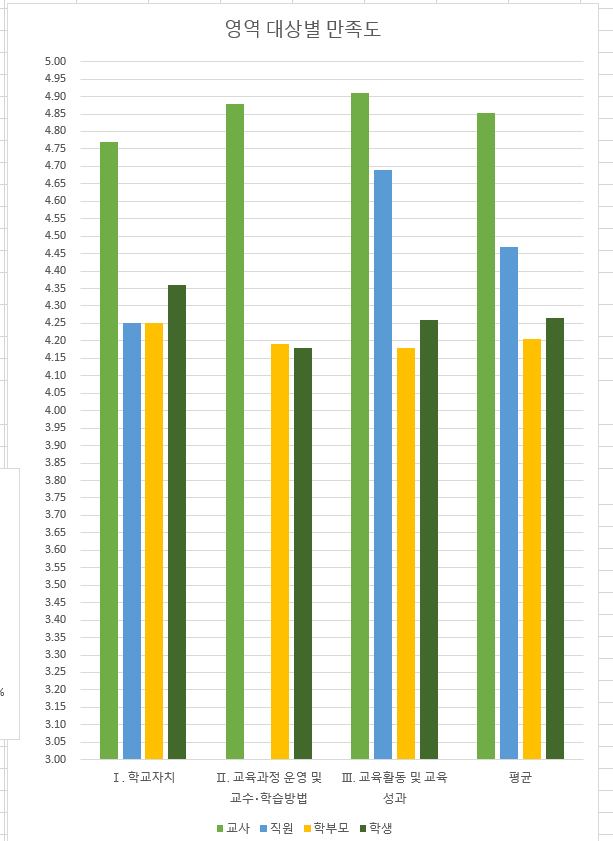} &
        \includegraphics[width=0.15\textwidth]{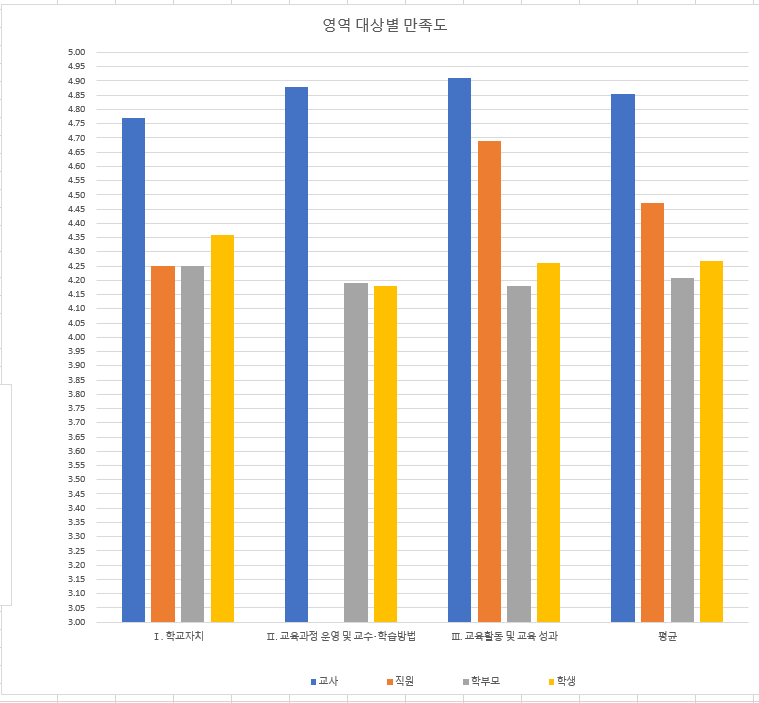} &
        \includegraphics[width=0.15\textwidth]{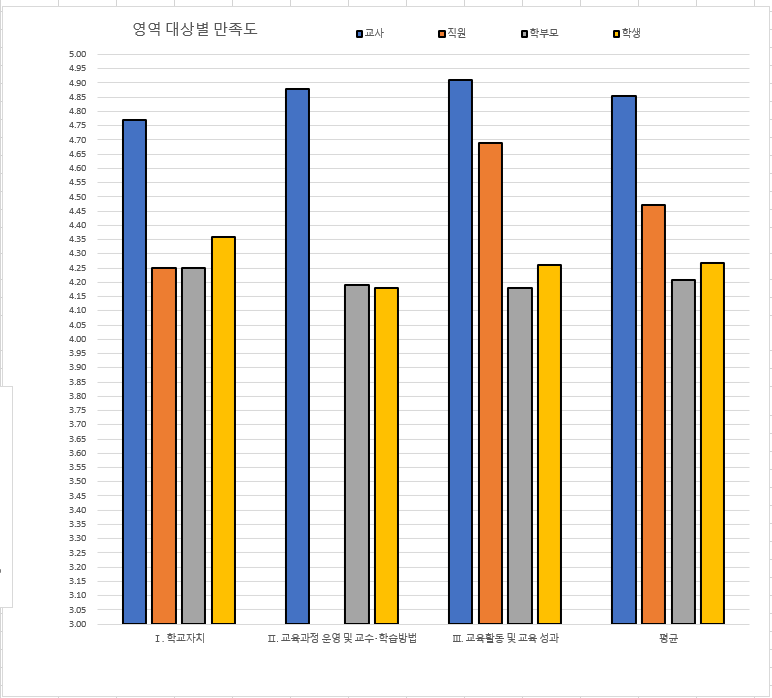}
        \\
        \midrule
        
        \textbf{N3} & 
        \includegraphics[width=0.18\textwidth]{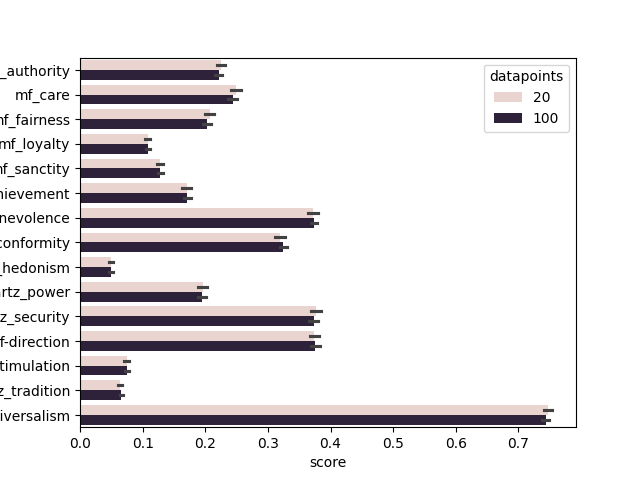} &
        \includegraphics[width=0.18\textwidth]{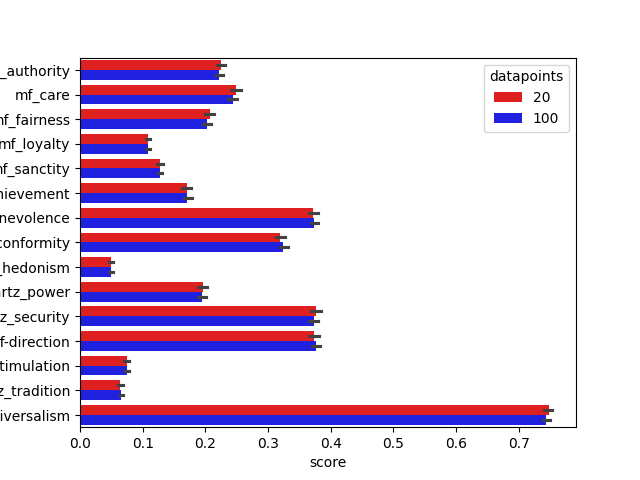} &
        \includegraphics[width=0.07\textwidth]{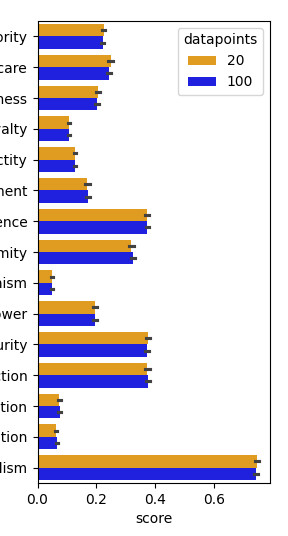} &
        \includegraphics[width=0.10\textwidth]{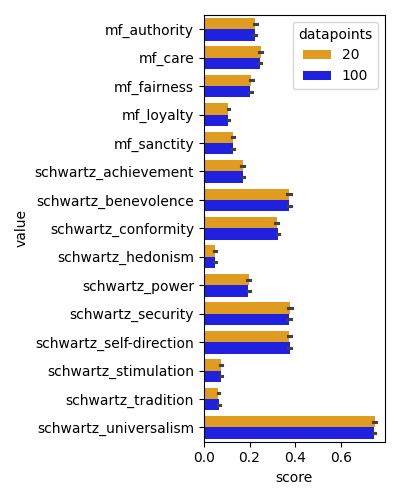} &
        \includegraphics[width=0.08\textwidth]{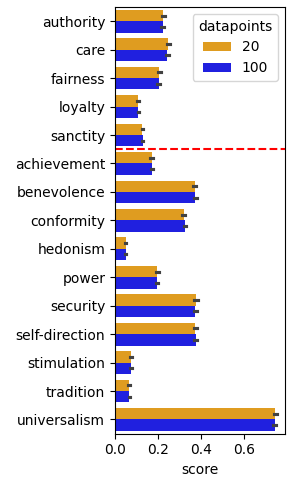}
        \\
        \midrule

        \textbf{N4} & 
        \includegraphics[width=0.18\textwidth]{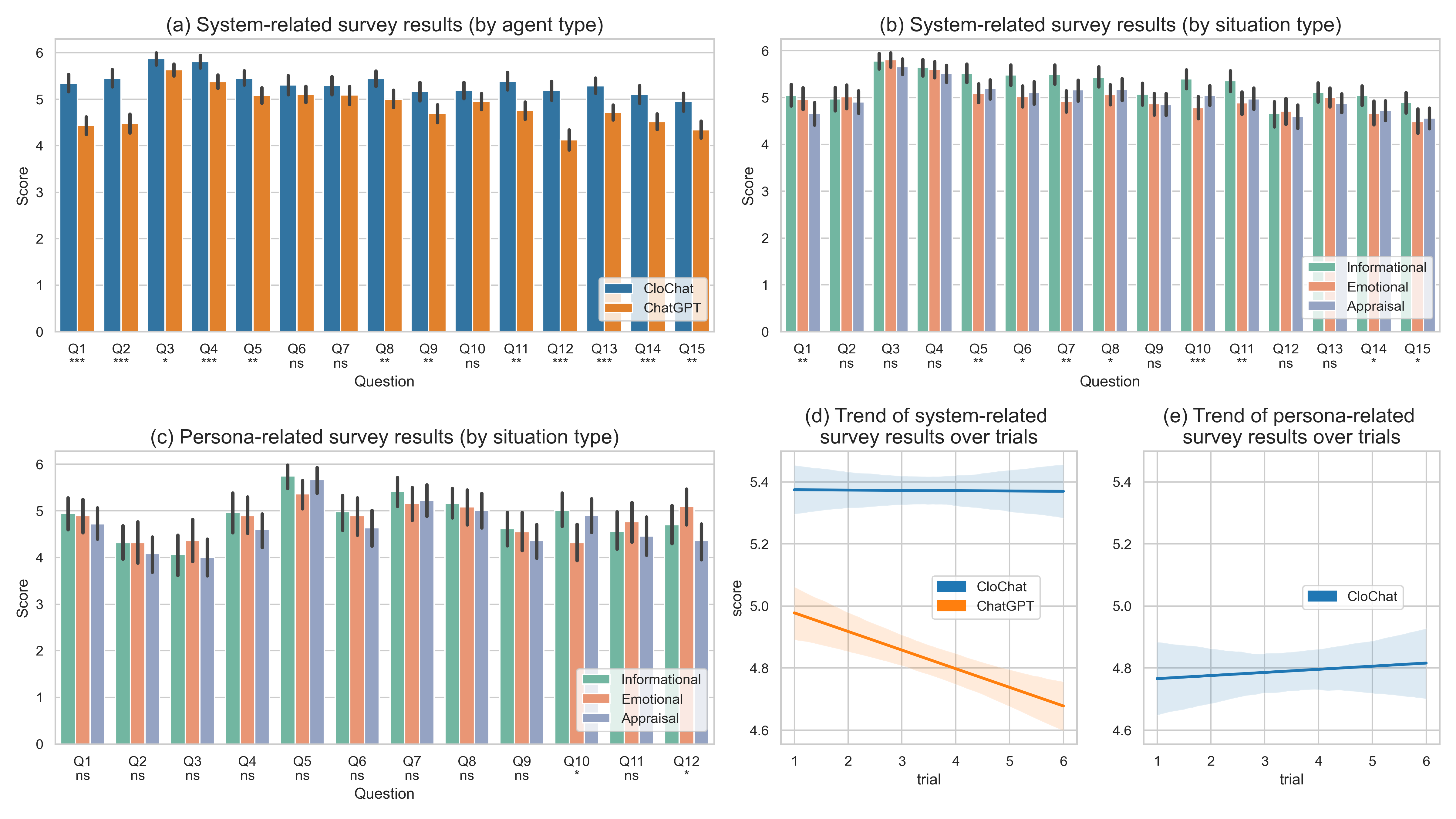} &
        \includegraphics[width=0.18\textwidth]{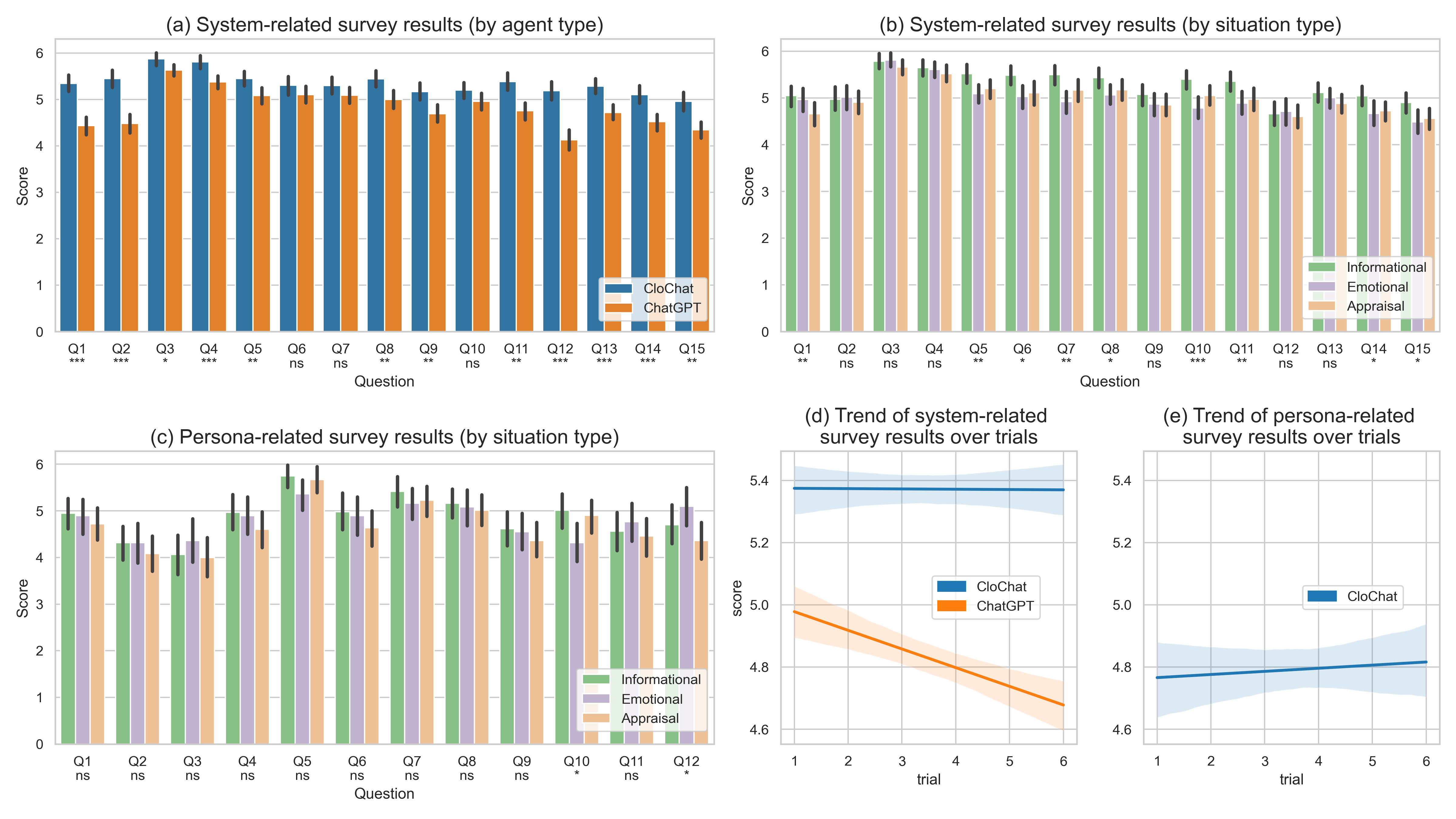} &
        \includegraphics[width=0.18\textwidth]{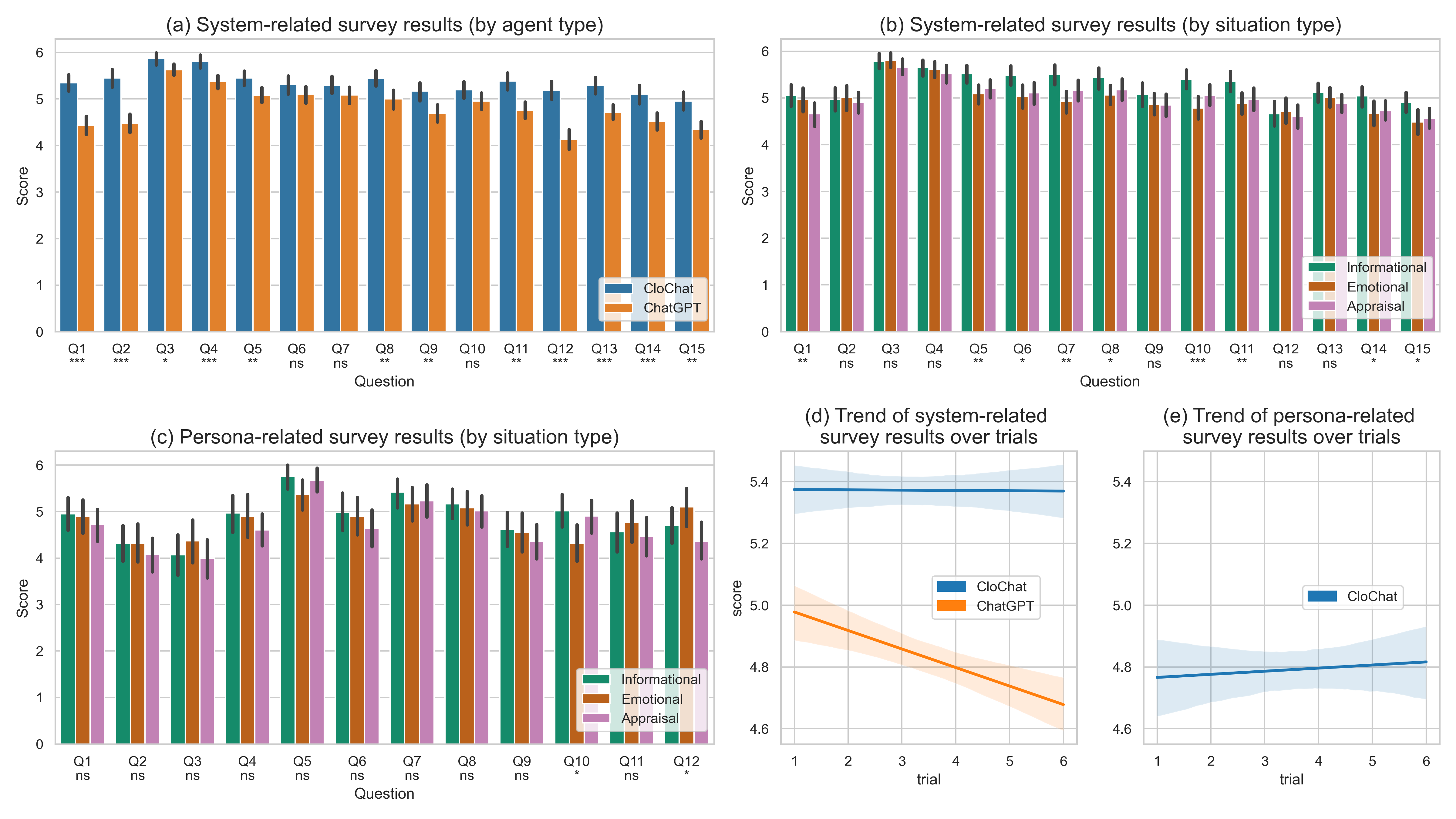} &
        \includegraphics[width=0.18\textwidth]{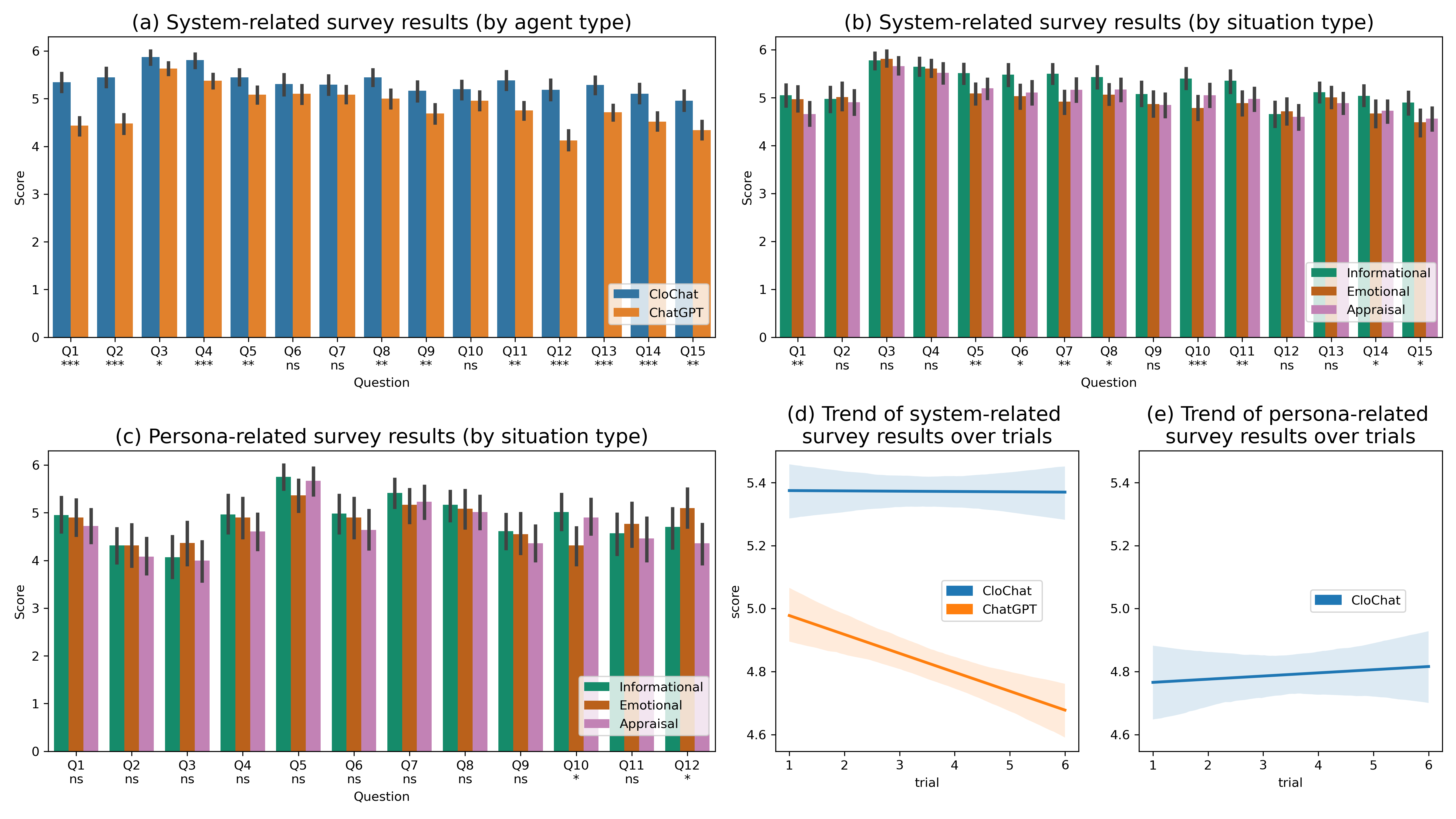} &
        \includegraphics[width=0.18\textwidth]{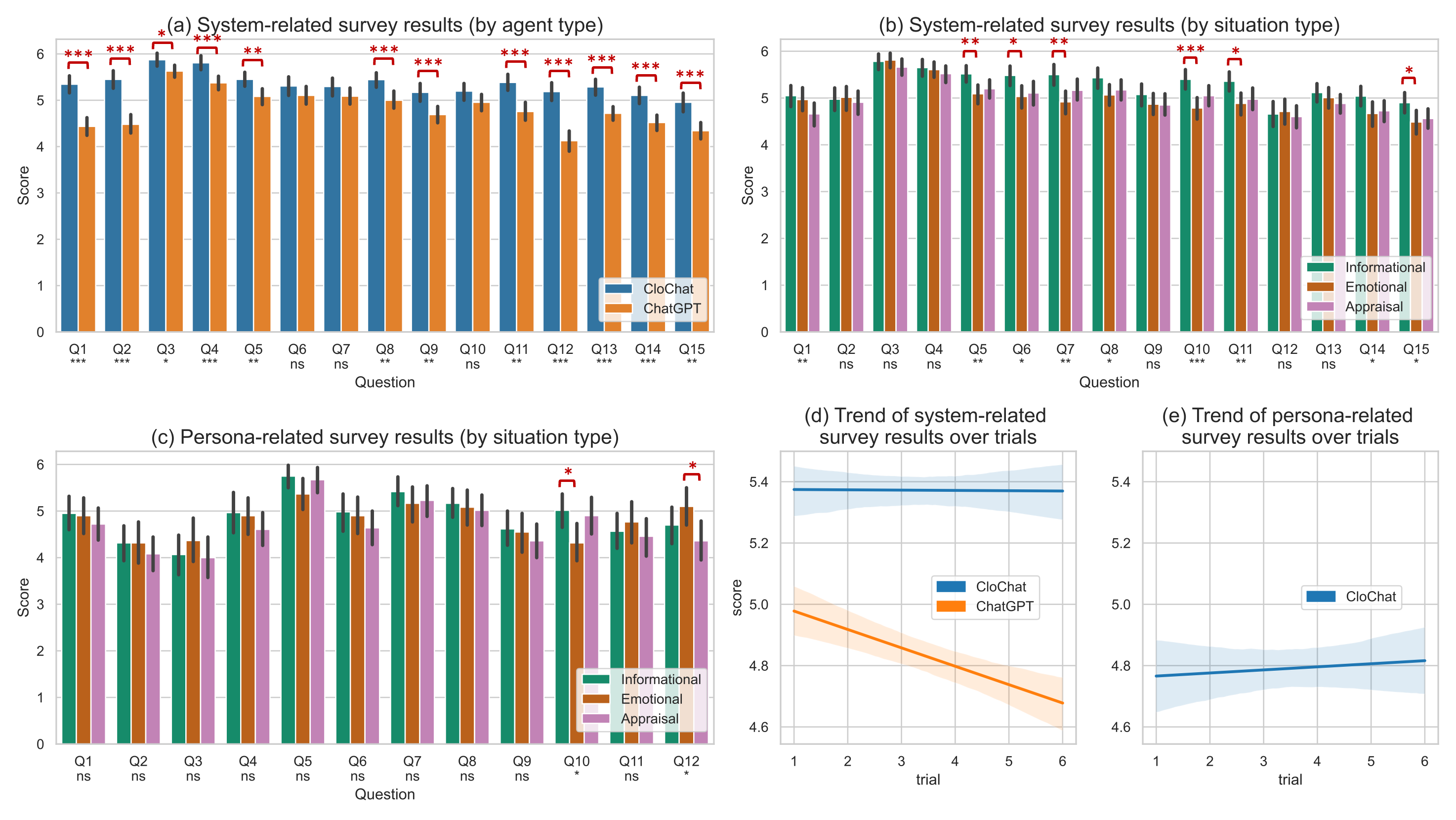}
        \\
        \midrule

        \textbf{N5} & 
        \includegraphics[width=0.18\textwidth]{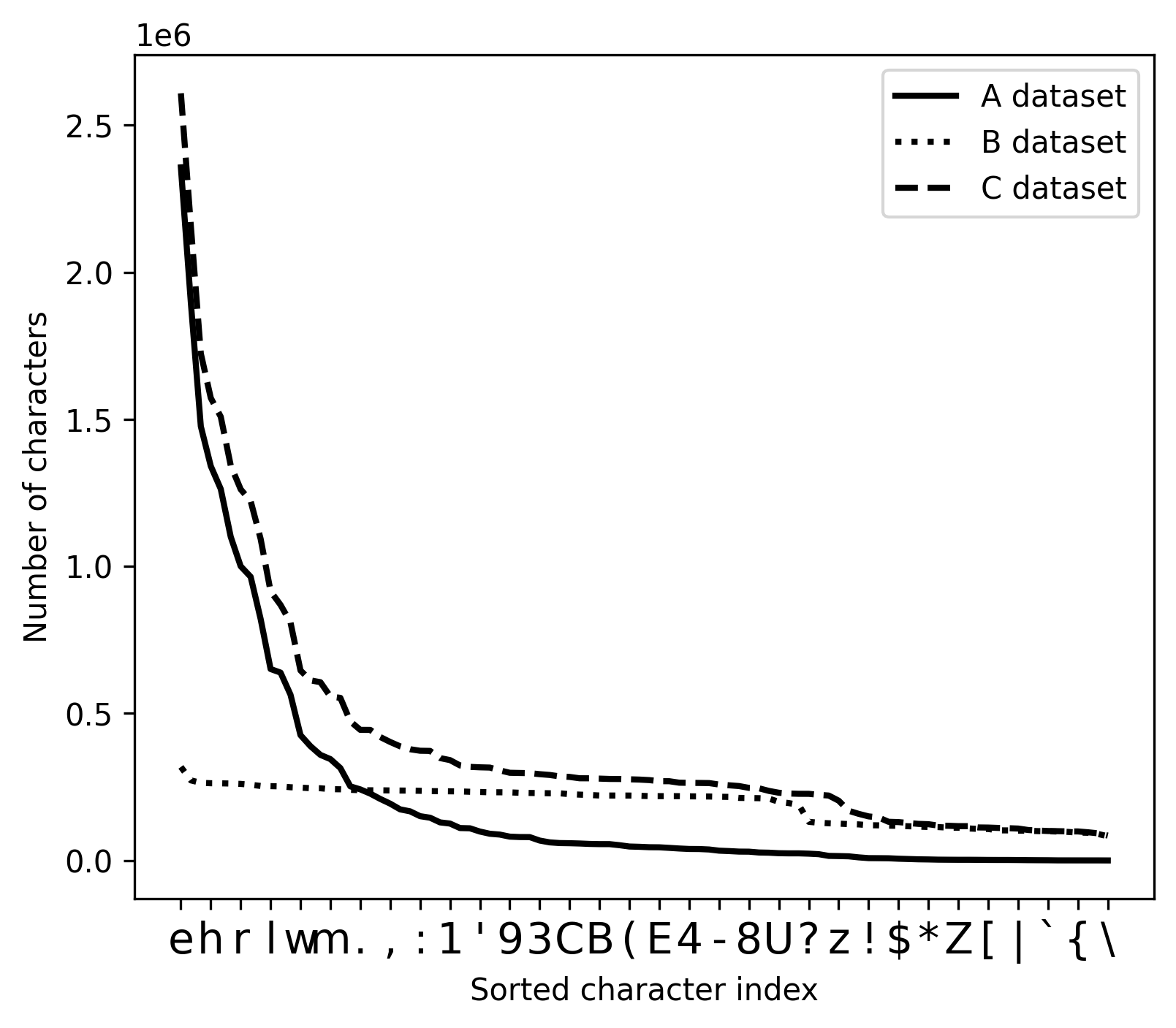} &
        \includegraphics[width=0.18\textwidth]{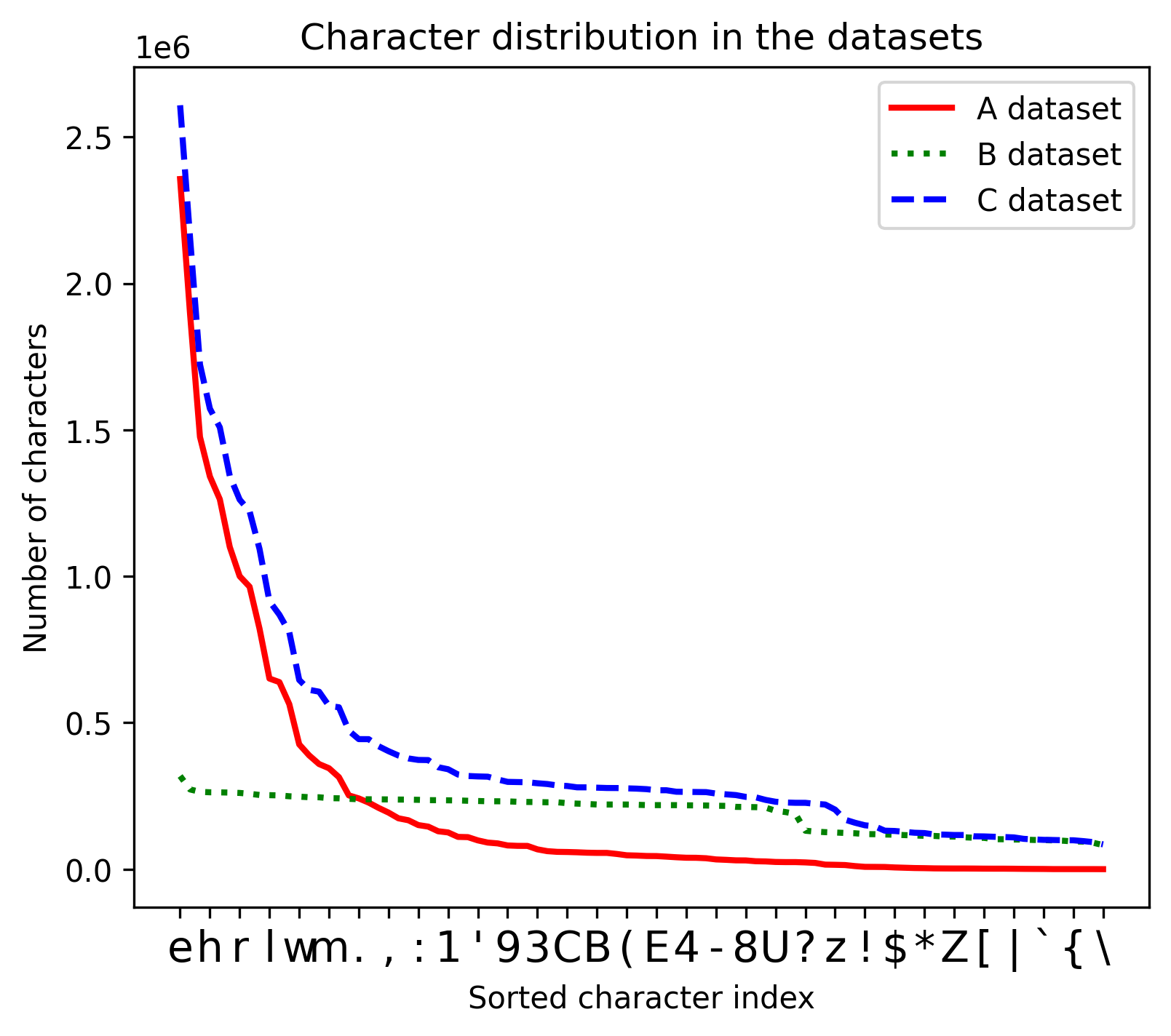} &
        \includegraphics[width=0.18\textwidth]{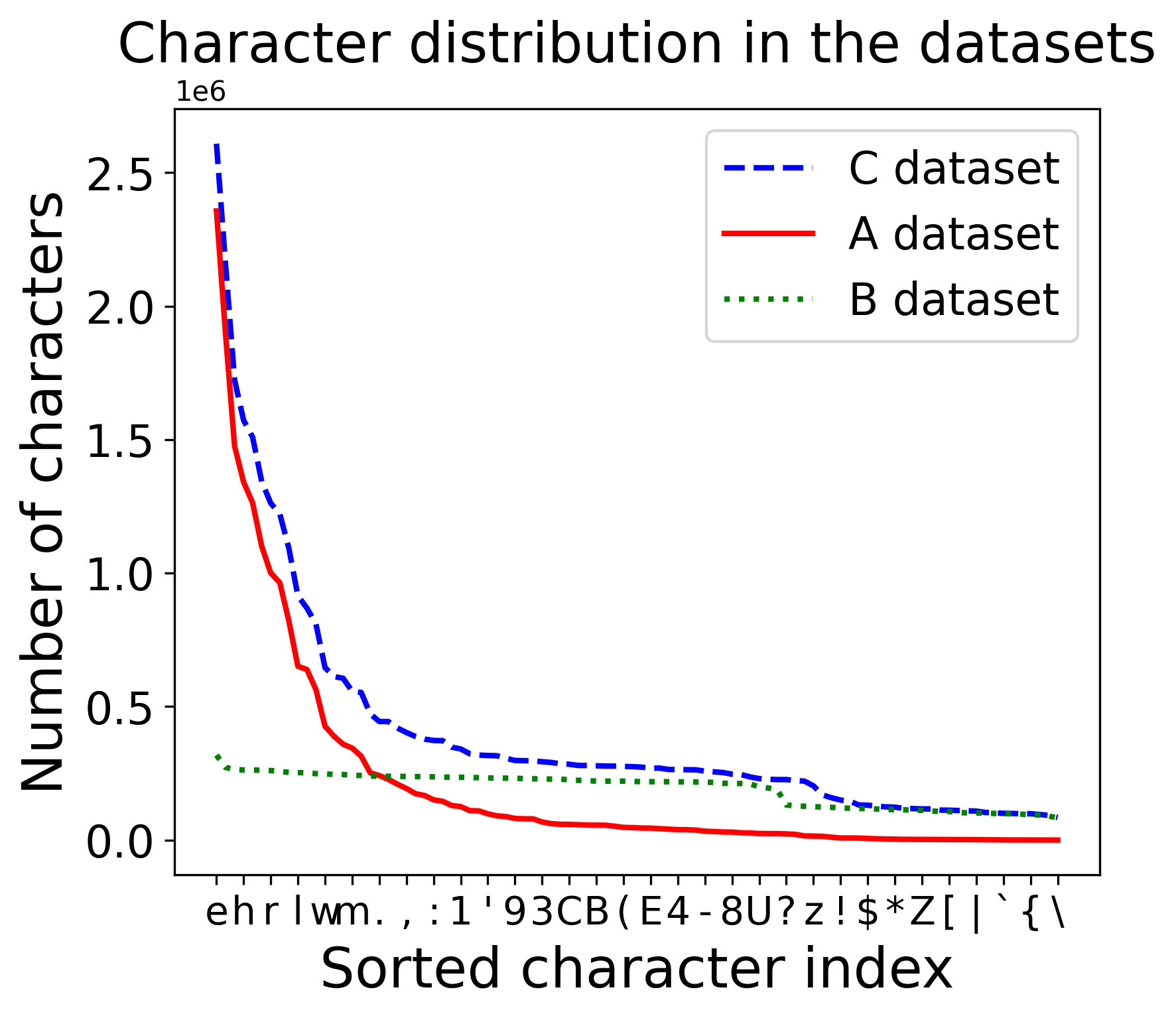} &
        \includegraphics[width=0.18\textwidth]{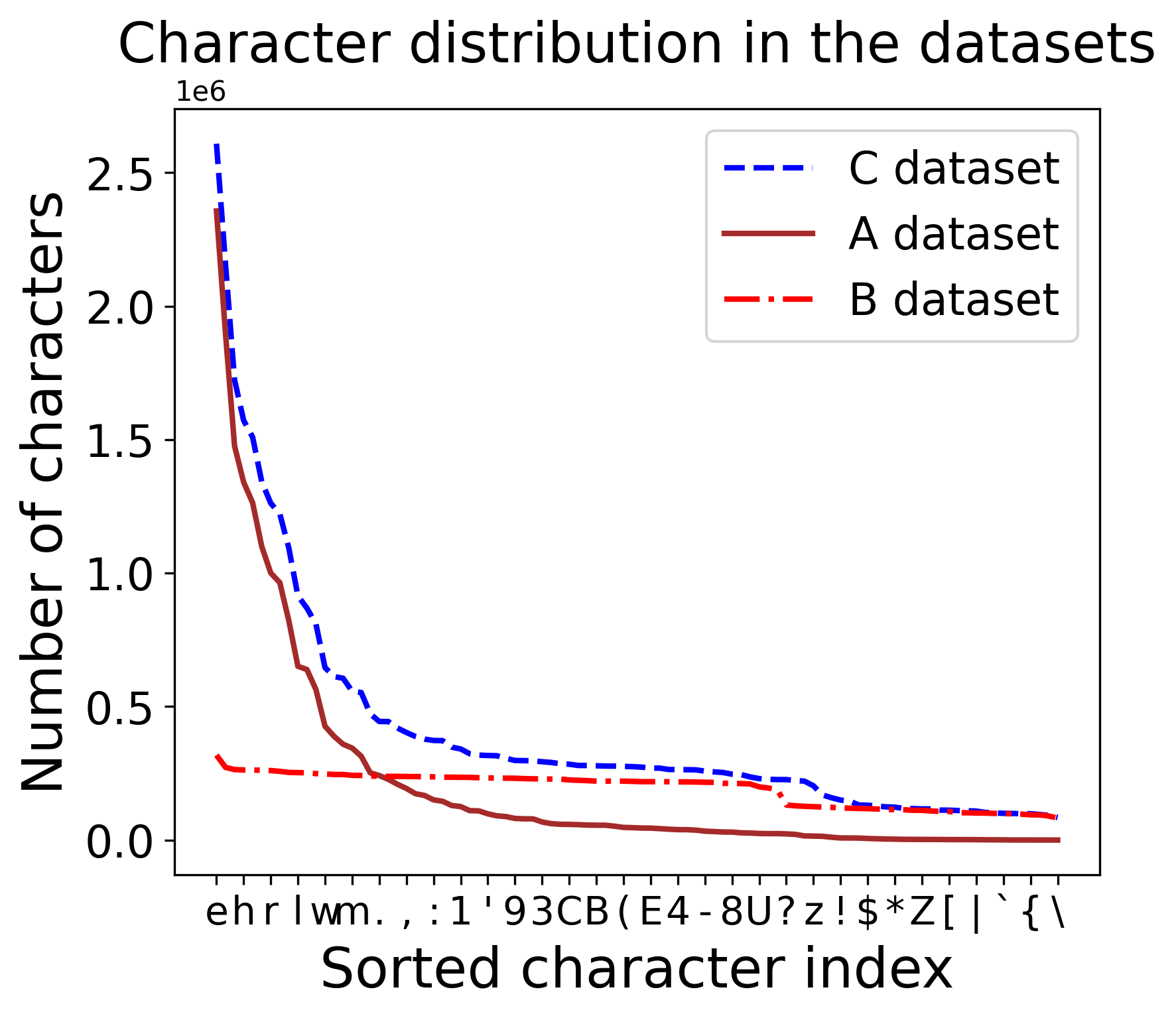} &
        \includegraphics[width=0.18\textwidth]{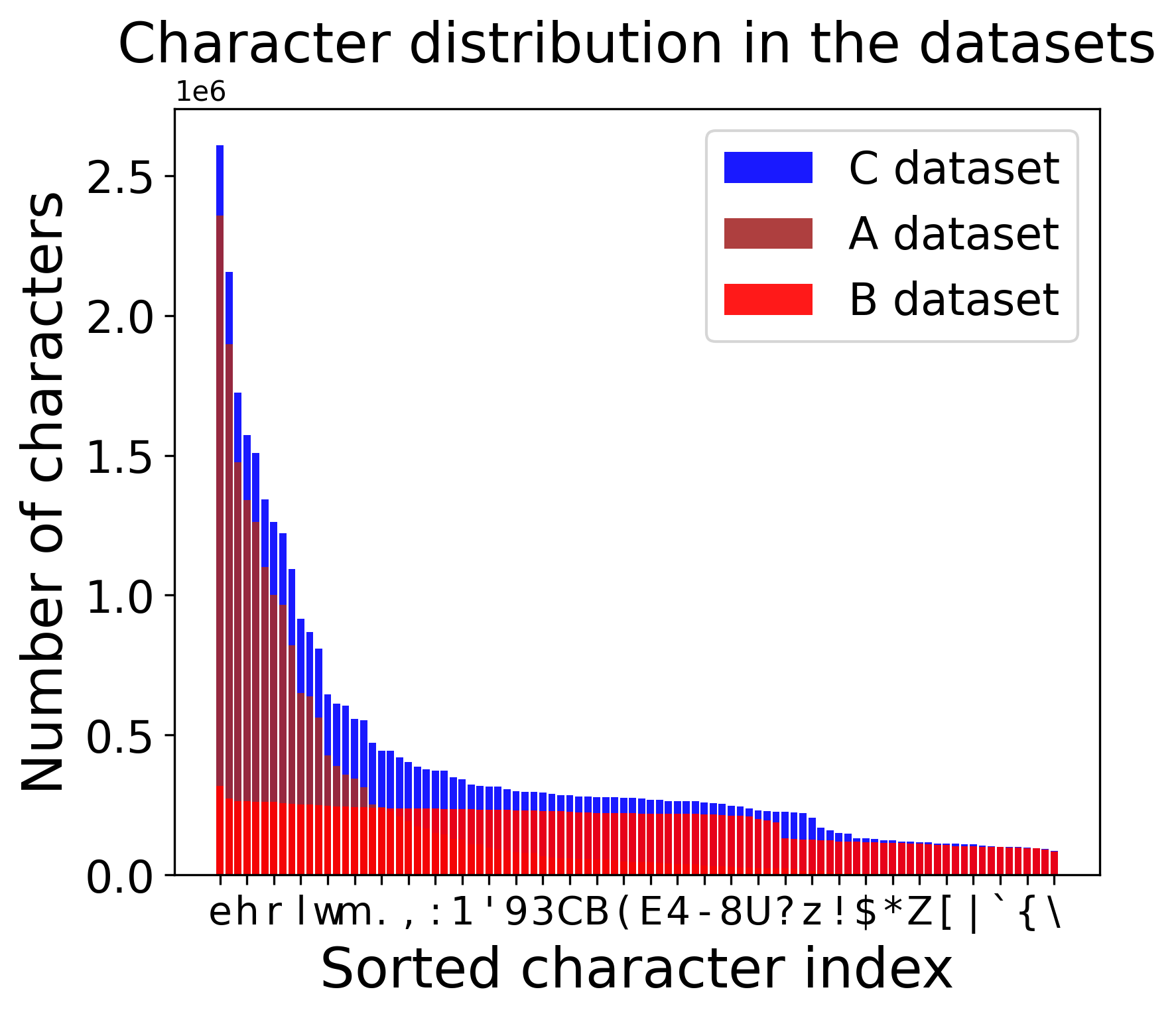}
        \\
        \midrule

        \textbf{N6} & 
        \includegraphics[width=0.18\textwidth]{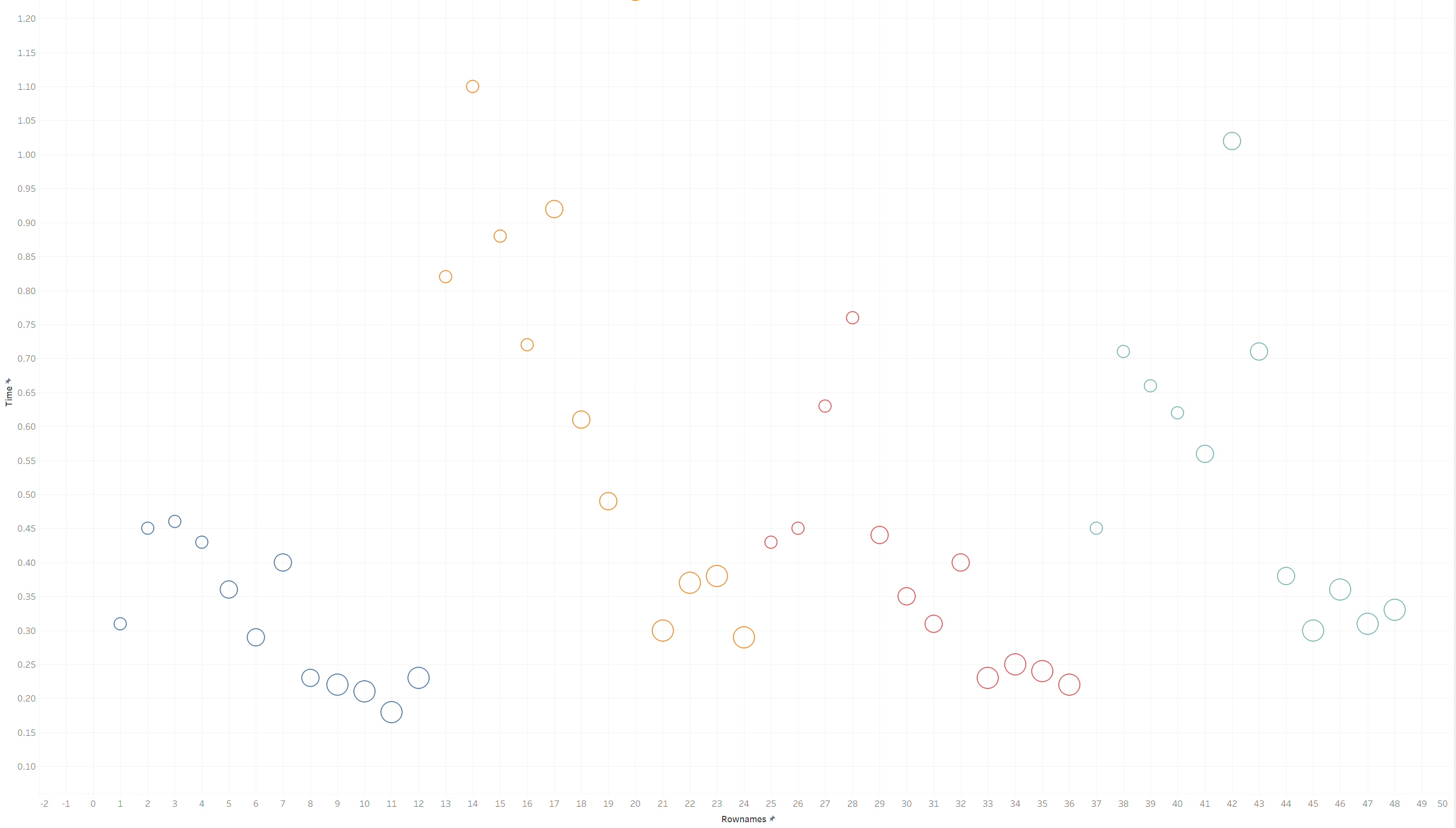} &
        \includegraphics[width=0.18\textwidth]{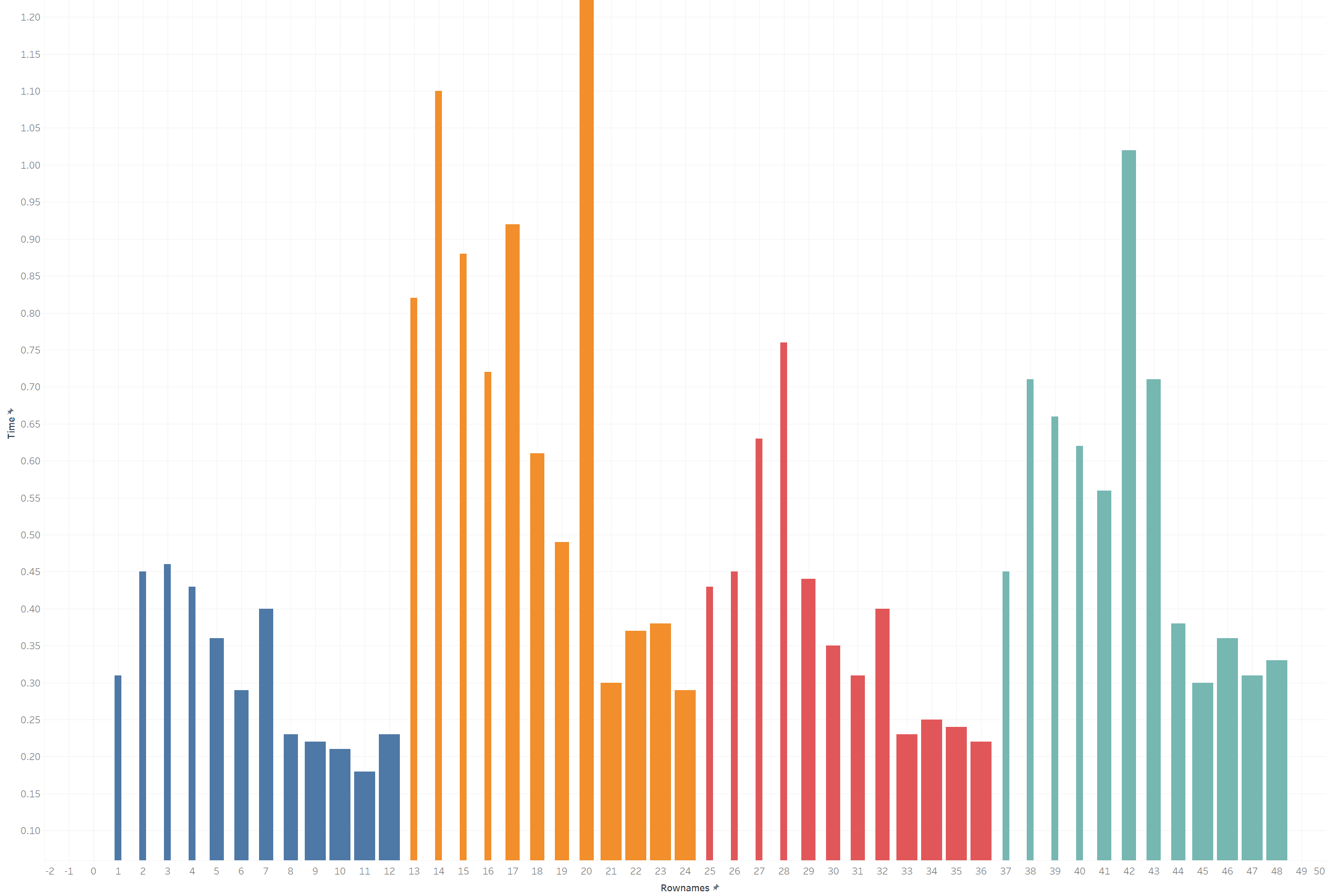} &
        \includegraphics[width=0.18\textwidth]{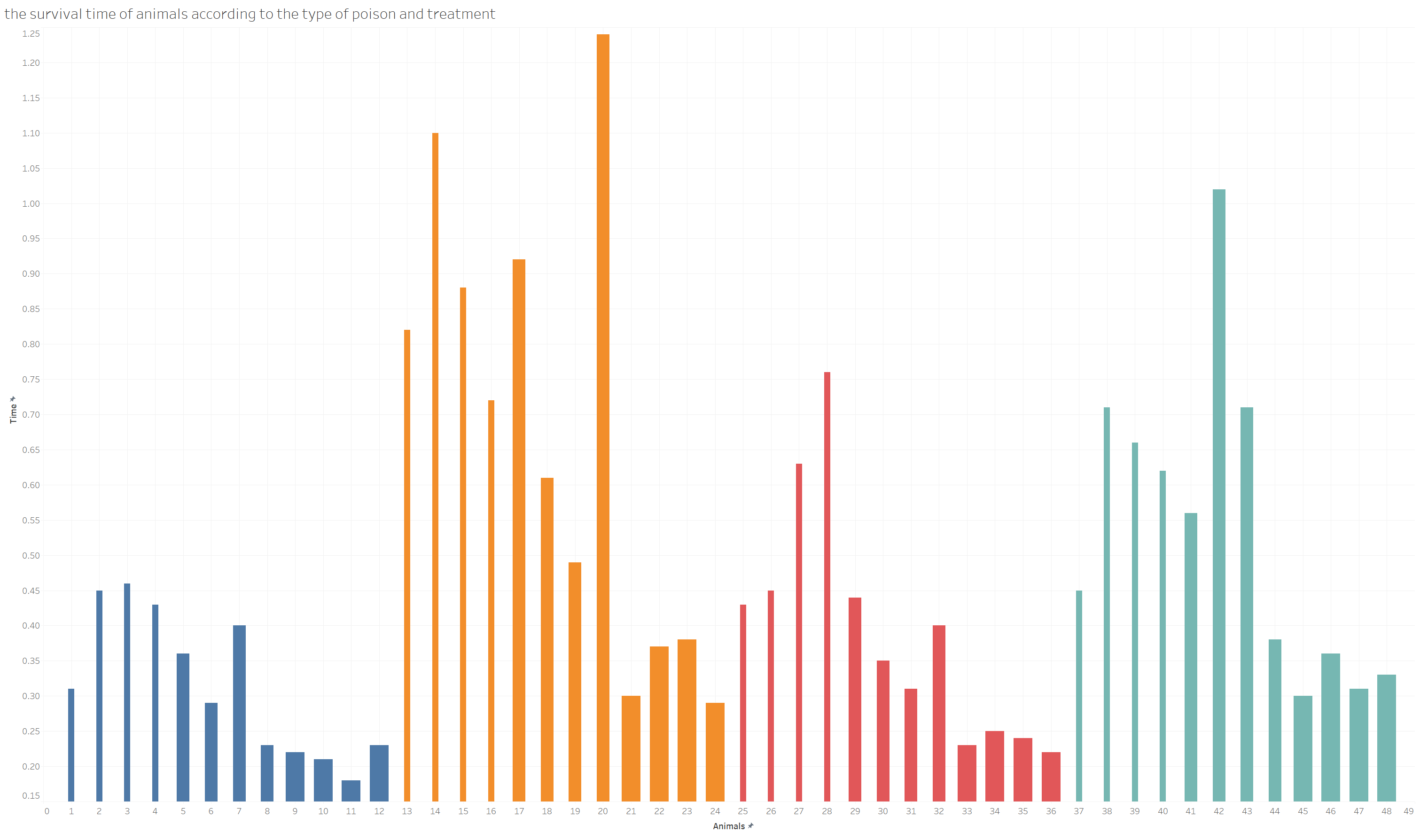} &
        \includegraphics[width=0.18\textwidth]{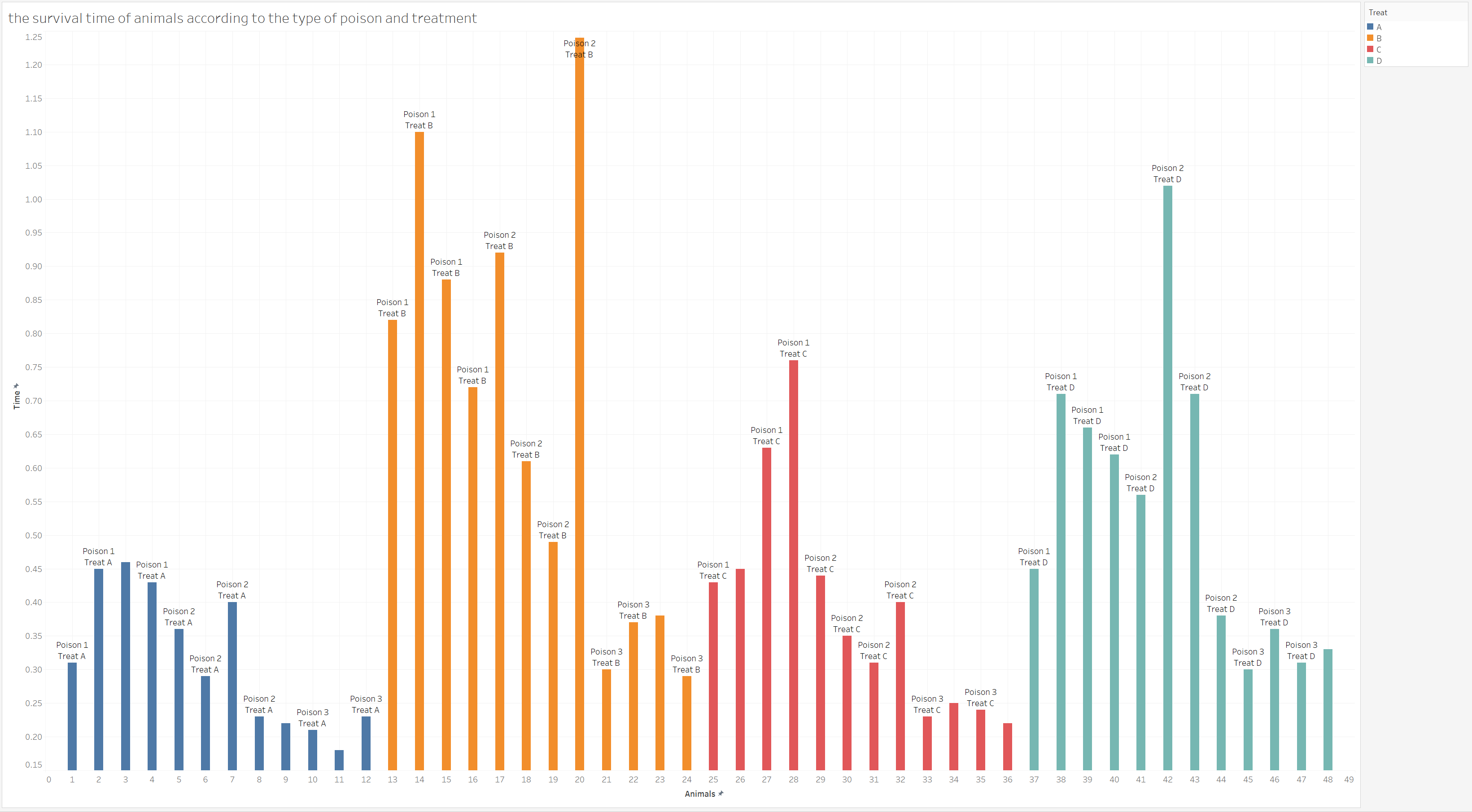} &
        \includegraphics[width=0.18\textwidth]{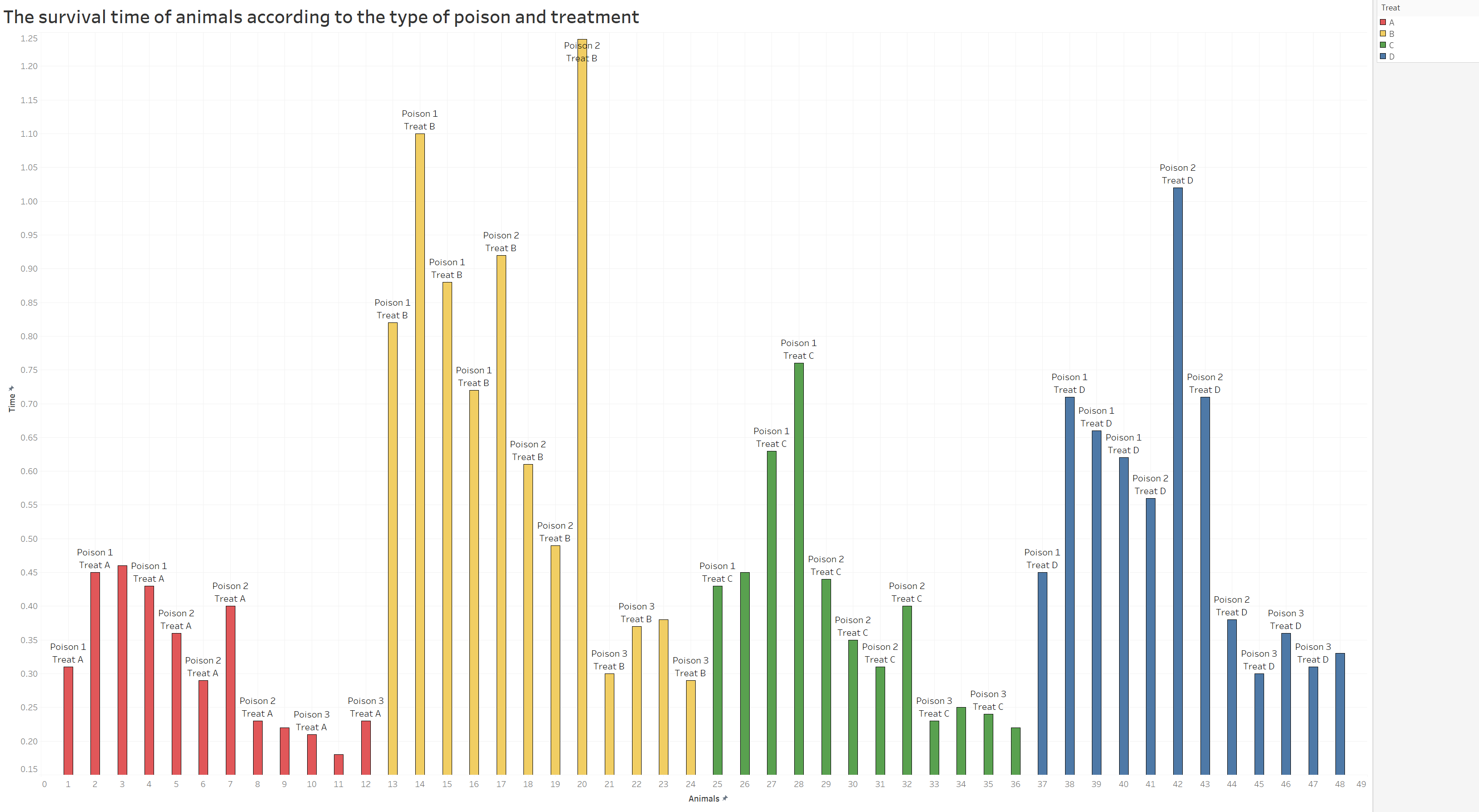}
        \\
        \bottomrule
        
    \end{tabular}

    \caption{\textbf{Examples of visualization design evolution (Novices).}
    These images represent the five most significant versions (as identified by the participants) of novice visualization designers during the design process of creating a new visualization artifact from scratch.
    The explanations of these design iterations can be found in the supplemental material.
    }
    \label{fig:novice-versions}
    \vspace{-0.5cm}
\end{figure*}

\begin{figure*}[ht]
    \setlength{\tabcolsep}{0pt}
    \begin{tabular}{lccccc}
        \hspace{0.4cm} &
        \colorbox{blue!10}{\parbox[t][0.3cm]{0.18\textwidth}{\centering Version 1}} &
        \colorbox{blue!20}{\parbox[t][0.3cm]{0.18\textwidth}{\centering Version 2}} &
        \colorbox{blue!30}{\parbox[t][0.3cm]{0.18\textwidth}{\centering Version 3}} &
        \colorbox{blue!40}{\parbox[t][0.3cm]{0.18\textwidth}{\centering Version 4}} &
        \colorbox{blue!50}{\parbox[t][0.3cm]{0.18\textwidth}{\centering Version 5}}
        \\
        
        \textbf{I1} & 
        \includegraphics[width=0.135\textwidth]{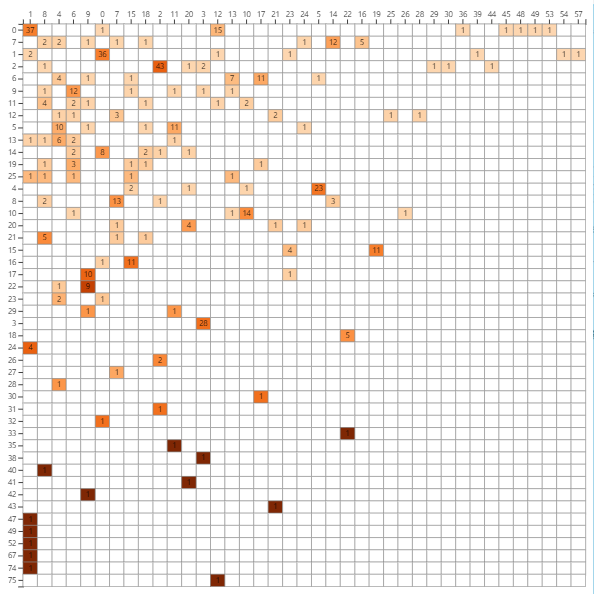} &
        \includegraphics[width=0.135\textwidth]{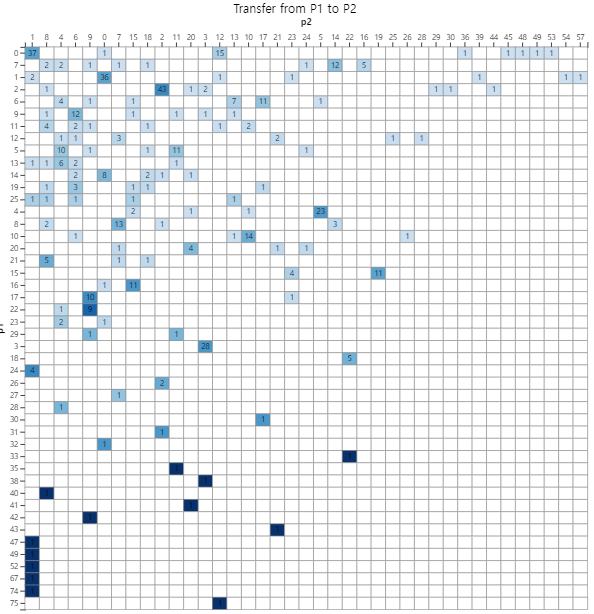} &
        \includegraphics[width=0.135\textwidth]{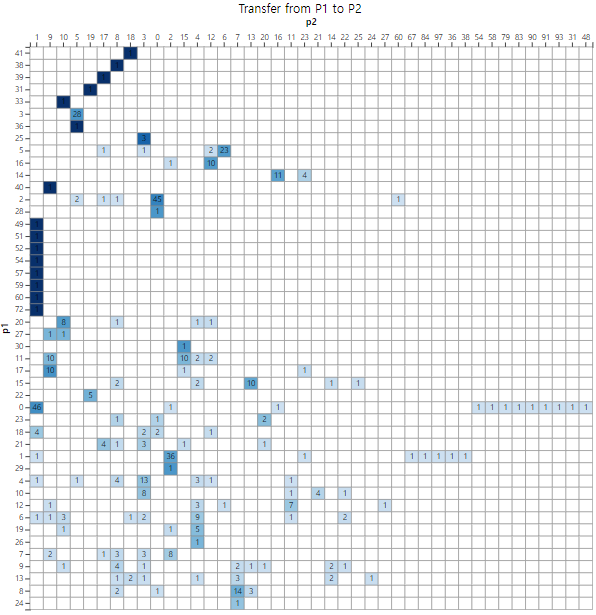} &
        \includegraphics[width=0.135\textwidth]{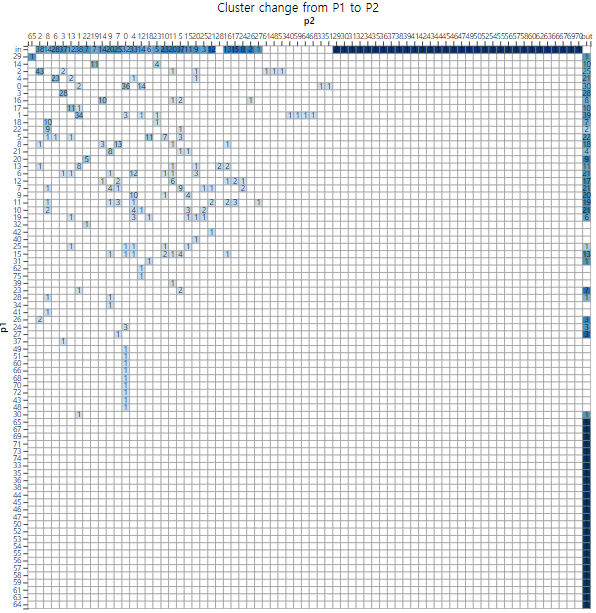} &
        \includegraphics[width=0.135\textwidth]{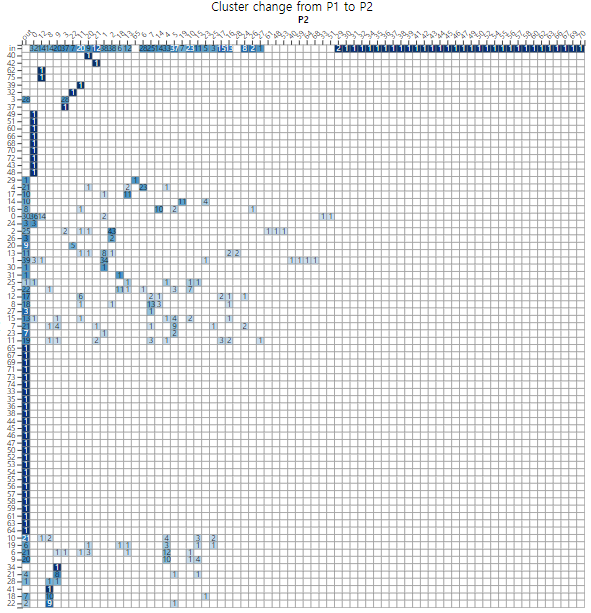}
        %\includegraphics[width=0.17\textwidth]{figures/P1_1} &
        %\resizebox{0.18\textwidth}{!}{\framebox{Vis}} &
        %\includegraphics[width=0.17\textwidth]{figures/P1_2} &
        %\resizebox{0.18\textwidth}{!}{\framebox{Vis}} &
        %\includegraphics[width=0.17\textwidth]{figures/P1_3} &
        %\resizebox{0.18\textwidth}{!}{\framebox{Vis}} &
        %\includegraphics[width=0.17\textwidth]{figures/P1_5} &
        %\resizebox{0.18\textwidth}{!}{\framebox{Vis}} &
        %\includegraphics[width=0.17\textwidth]{figures/P1_6} 
        %\resizebox{0.18\textwidth}{!}{\framebox{Vis}}
        \\
        \midrule

        \textbf{I2} & 
        \includegraphics[width=0.18\textwidth]{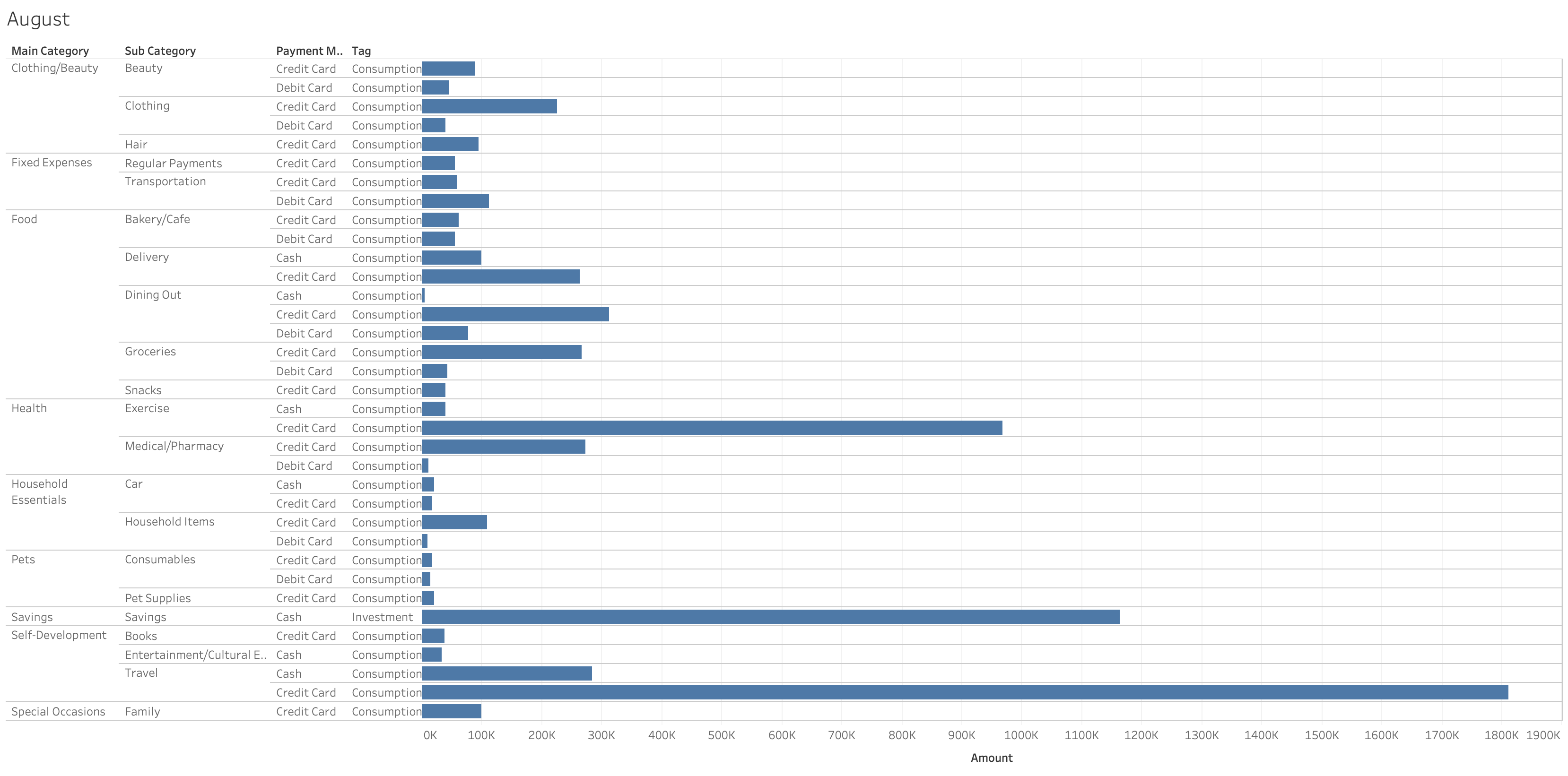} &
        \includegraphics[width=0.18\textwidth]{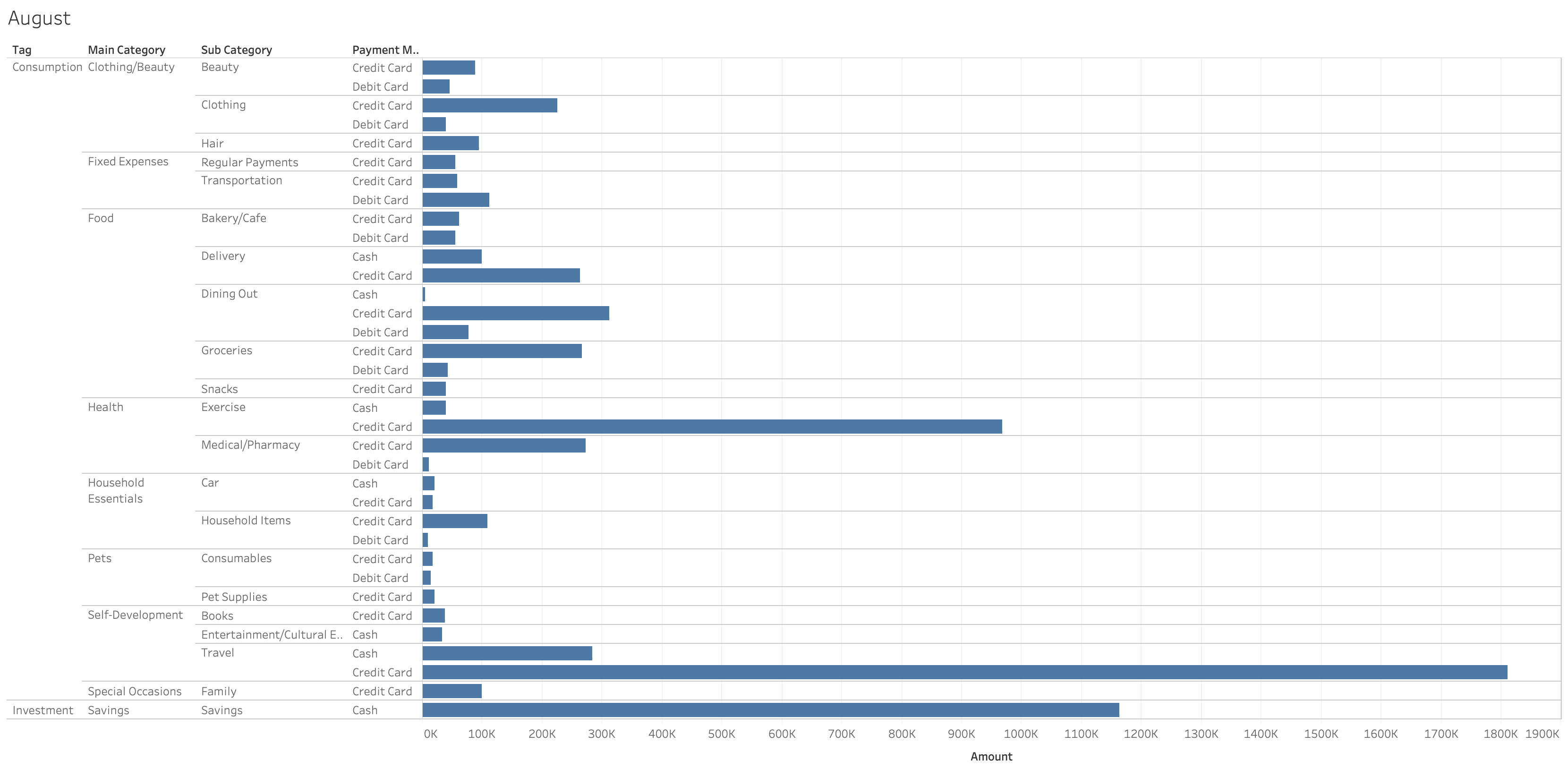} &
        \includegraphics[width=0.18\textwidth]{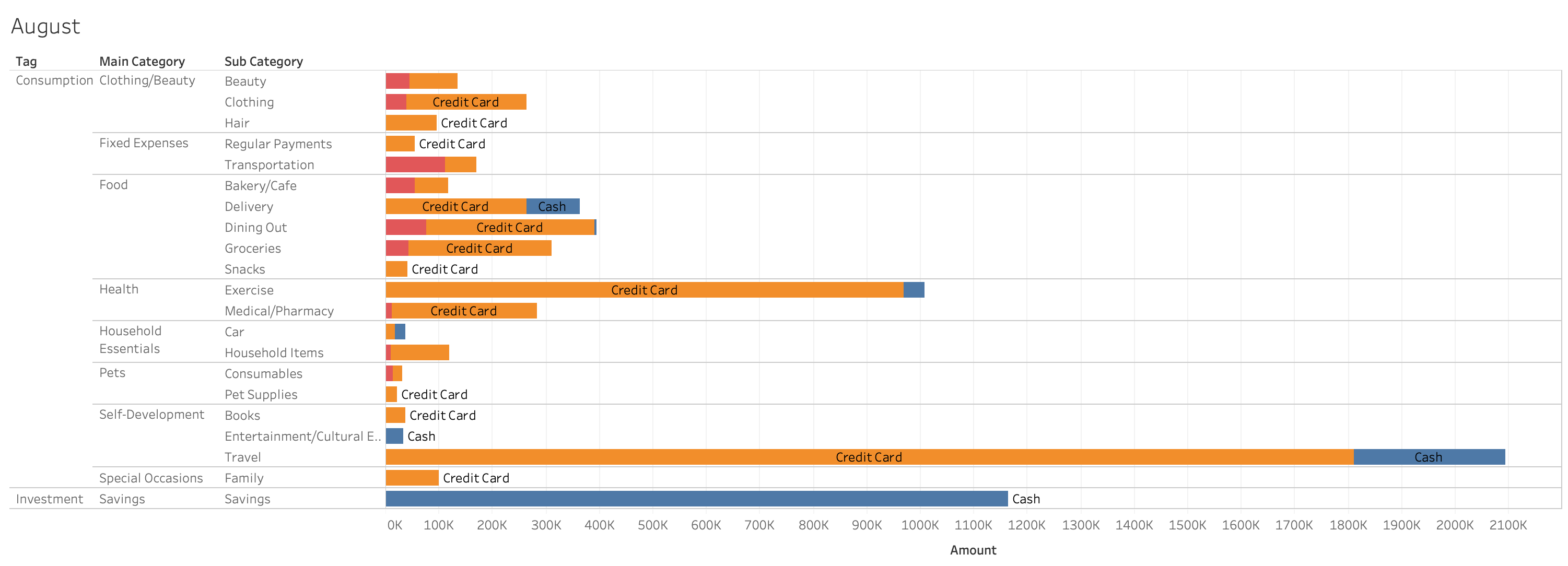} &
        \includegraphics[width=0.18\textwidth]{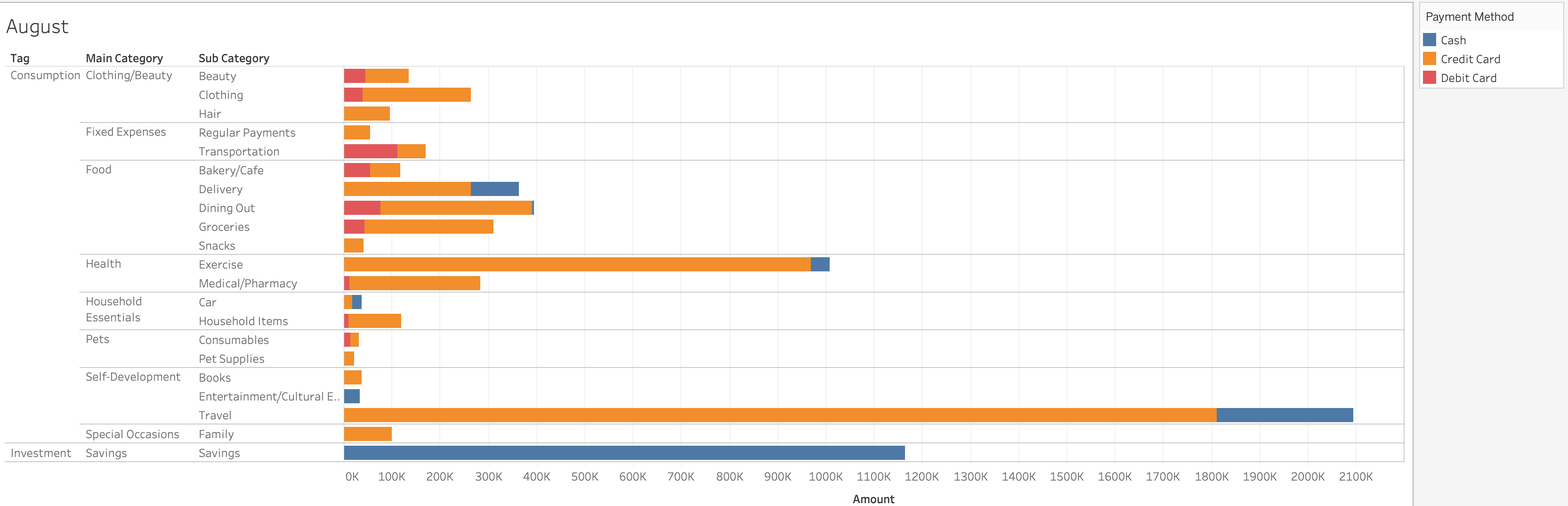} &
        \includegraphics[width=0.18\textwidth]{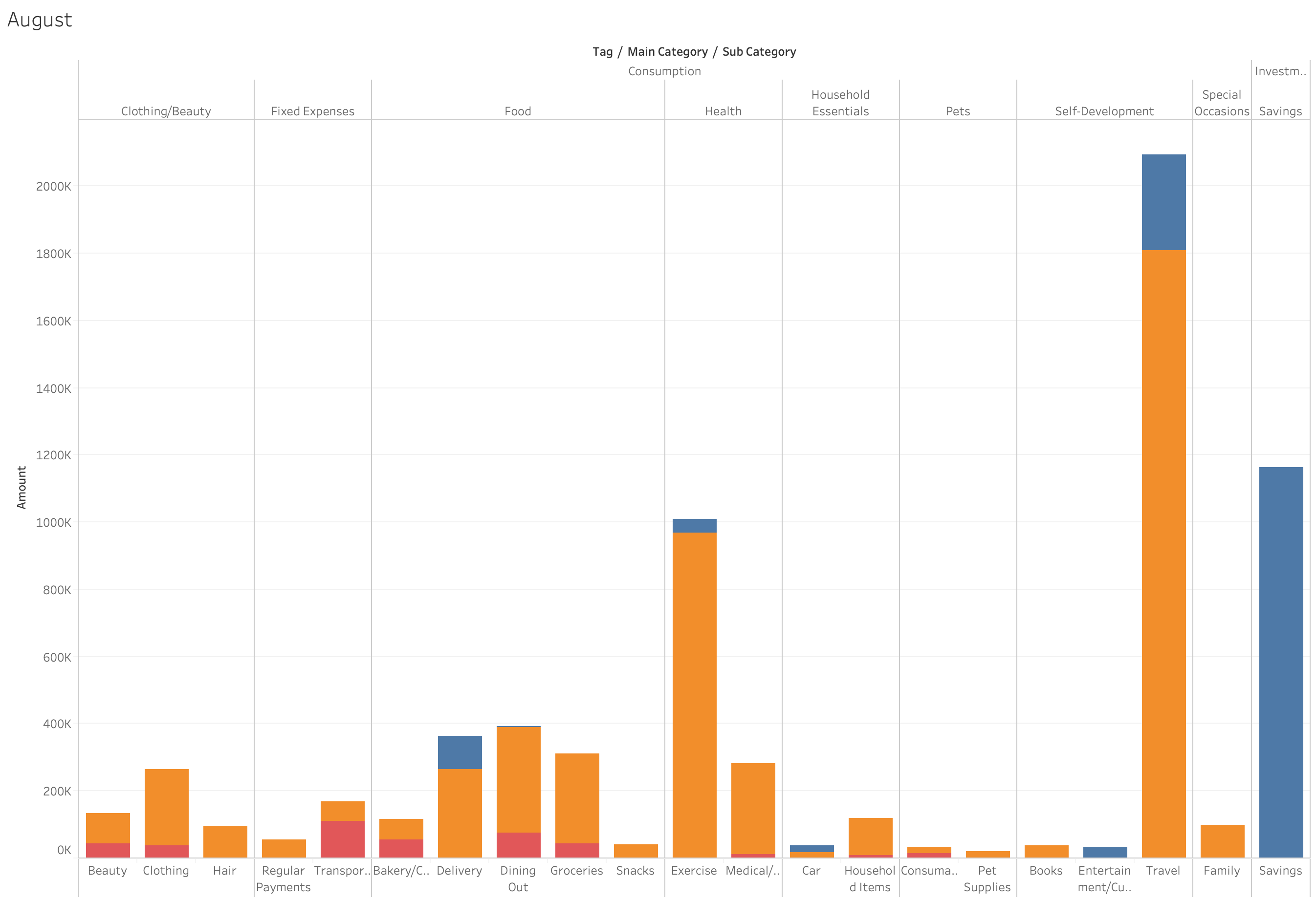}
        \\
        \midrule

        \textbf{I3} & 
        \includegraphics[width=0.18\textwidth]{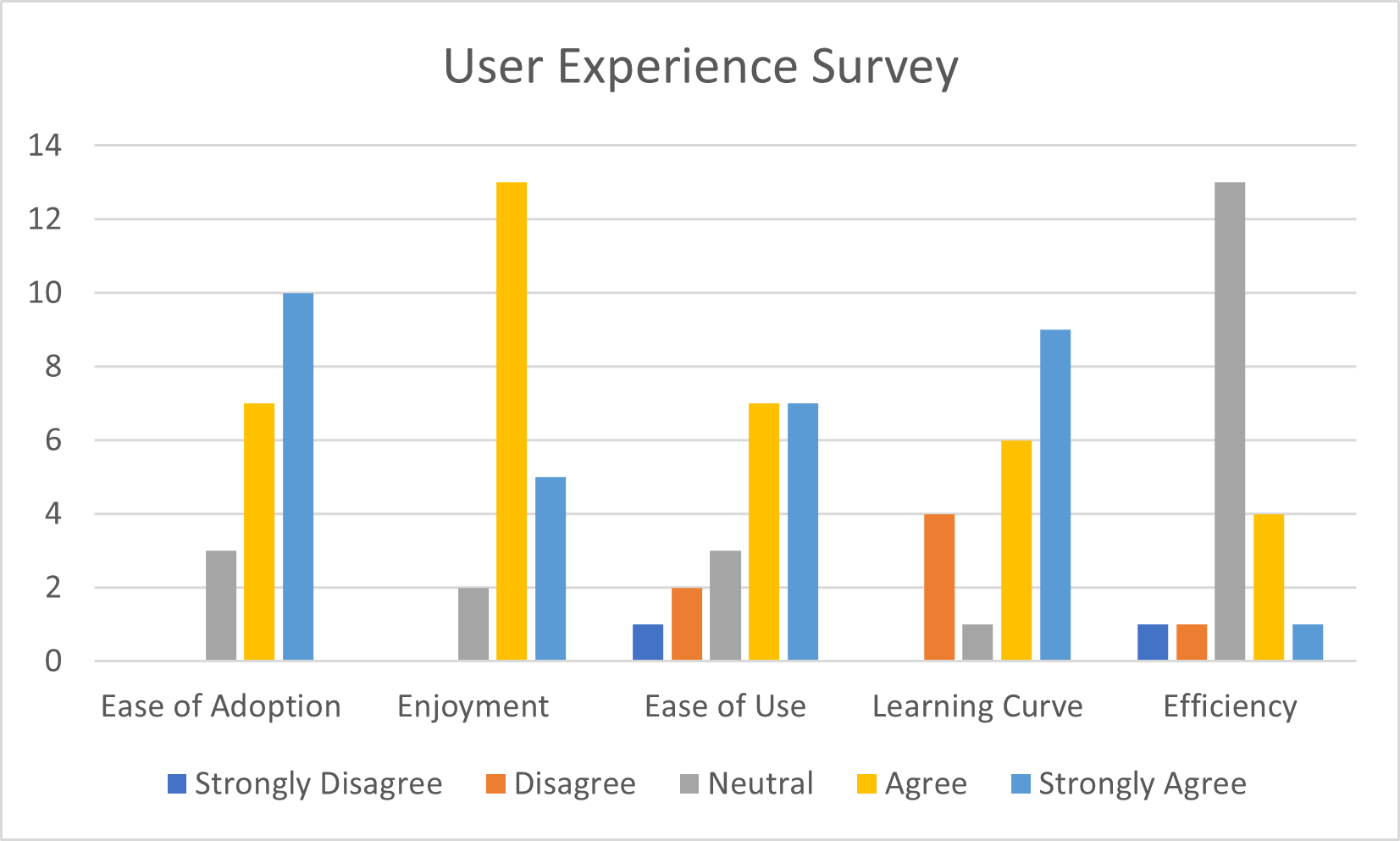} &
        \includegraphics[width=0.18\textwidth]{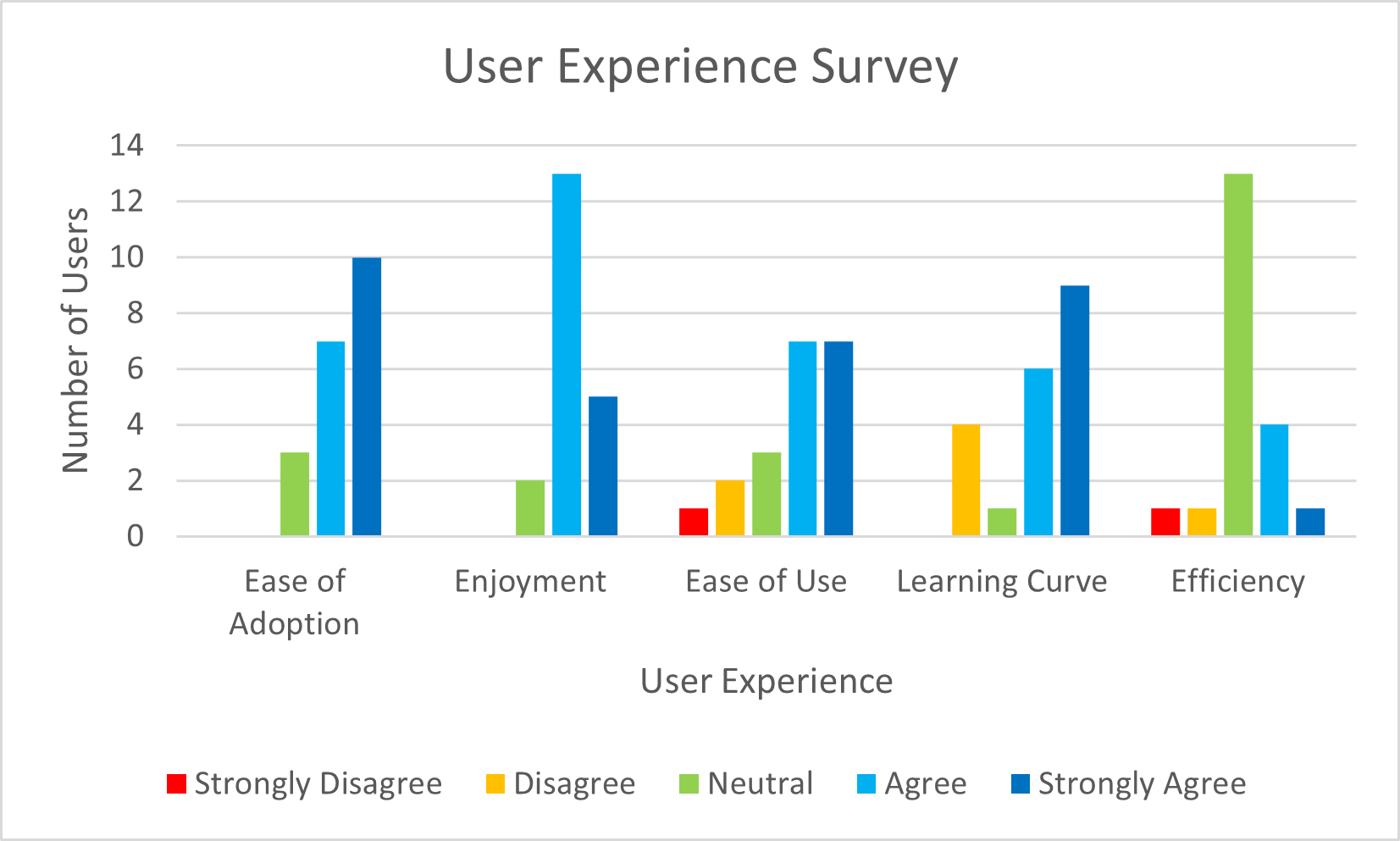} &
        \includegraphics[width=0.18\textwidth]{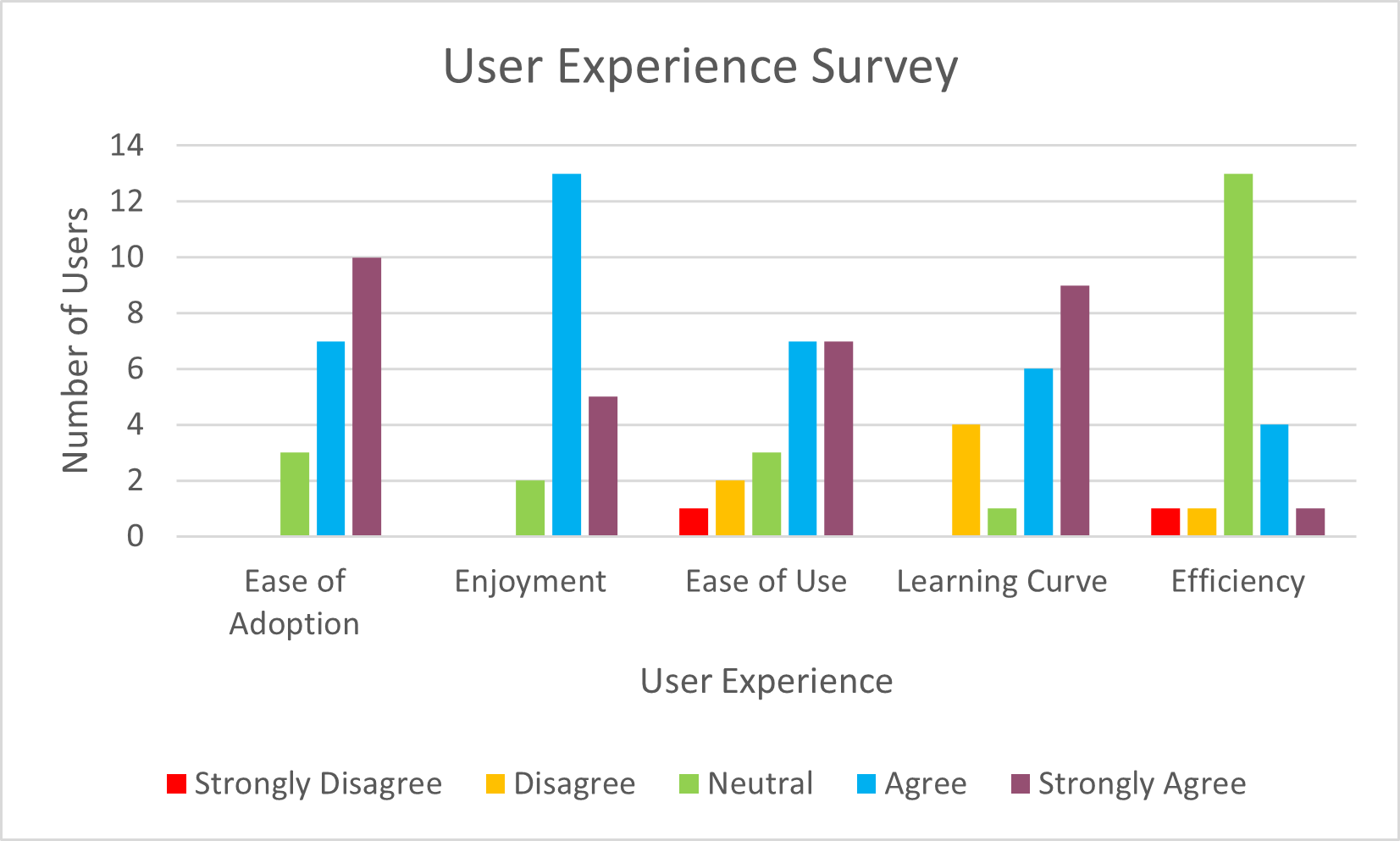} &
        \includegraphics[width=0.18\textwidth]{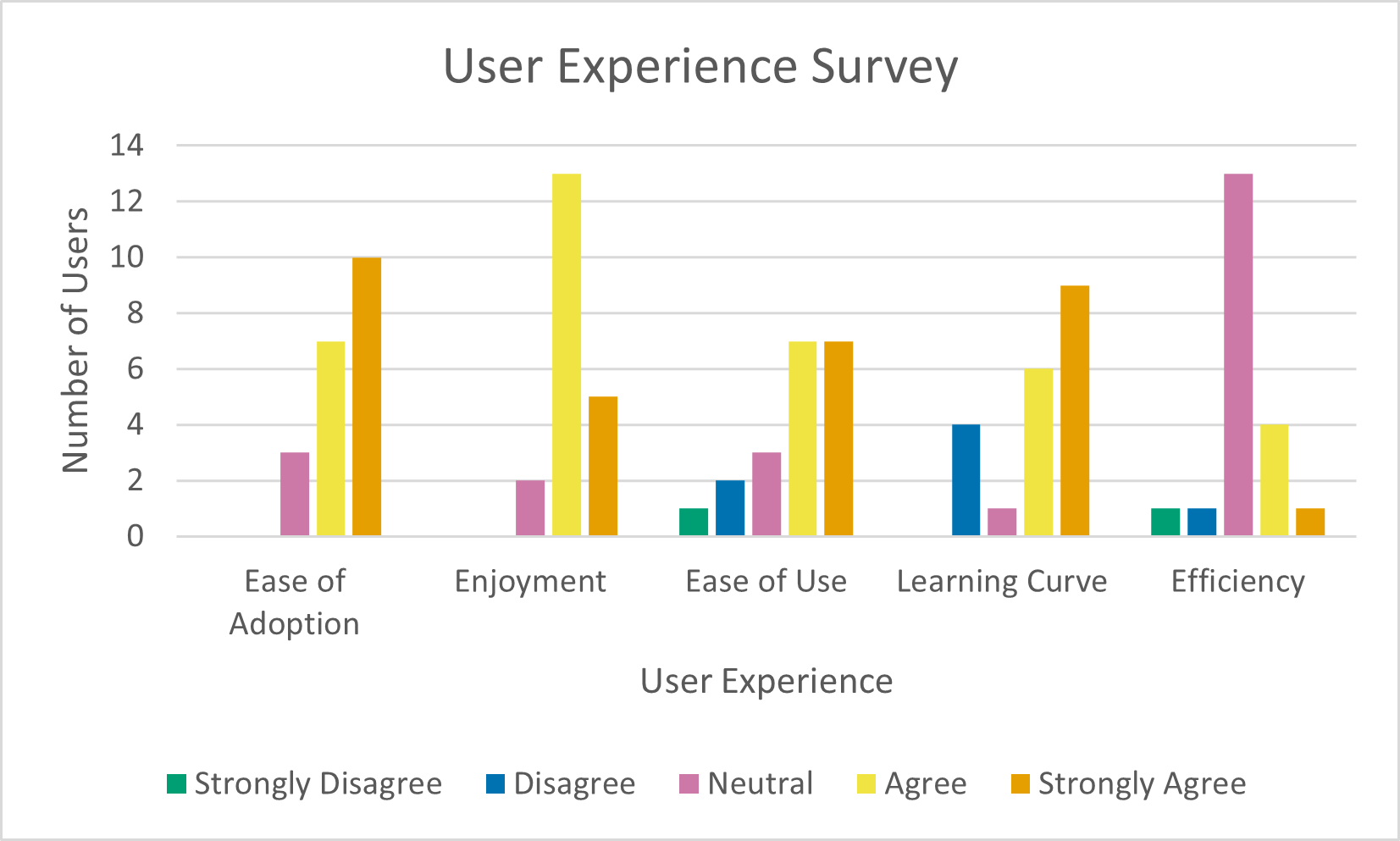} &
        \includegraphics[width=0.18\textwidth]{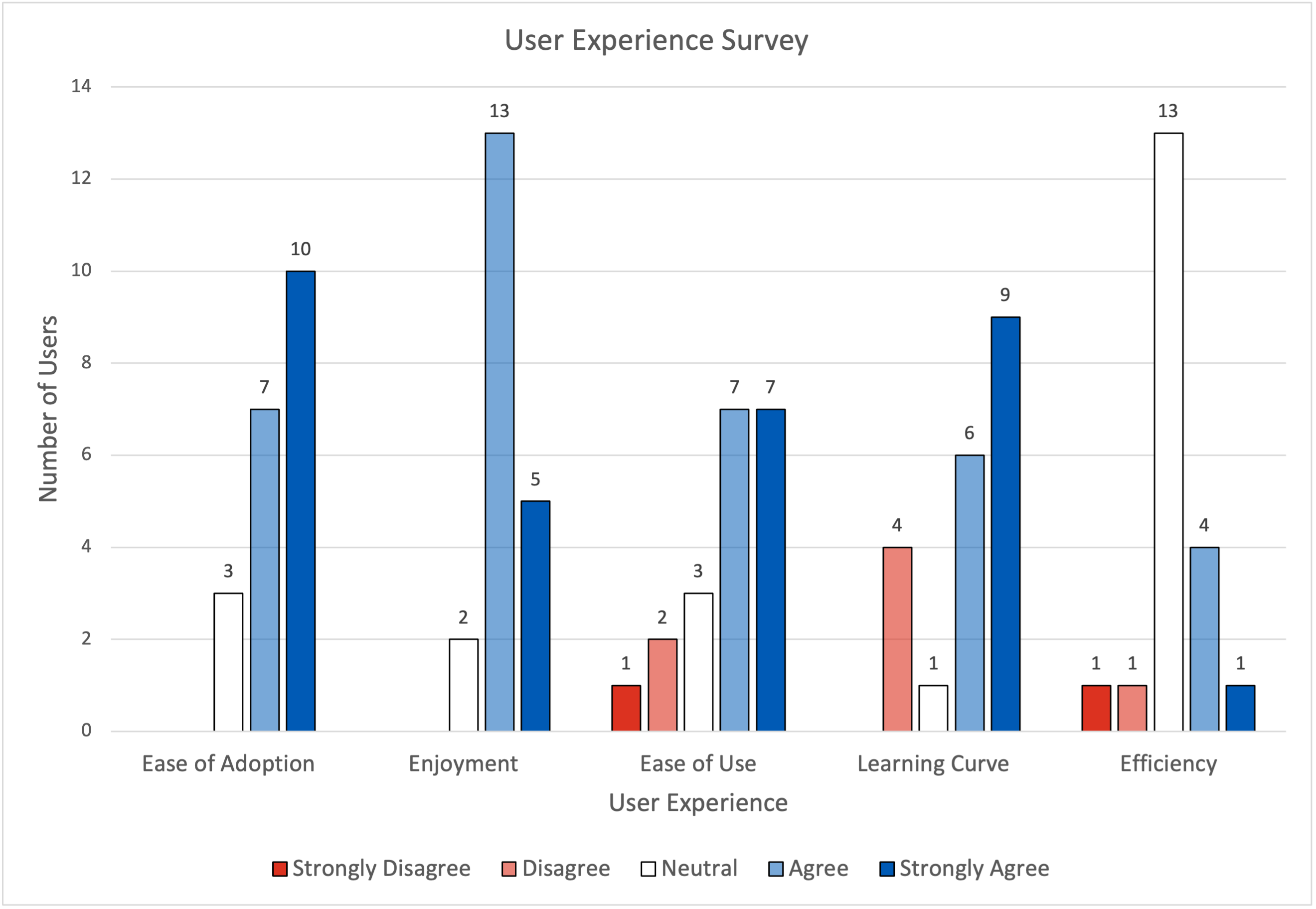}
        \\
        \midrule

        \textbf{I4} & 
        \includegraphics[width=0.18\textwidth]{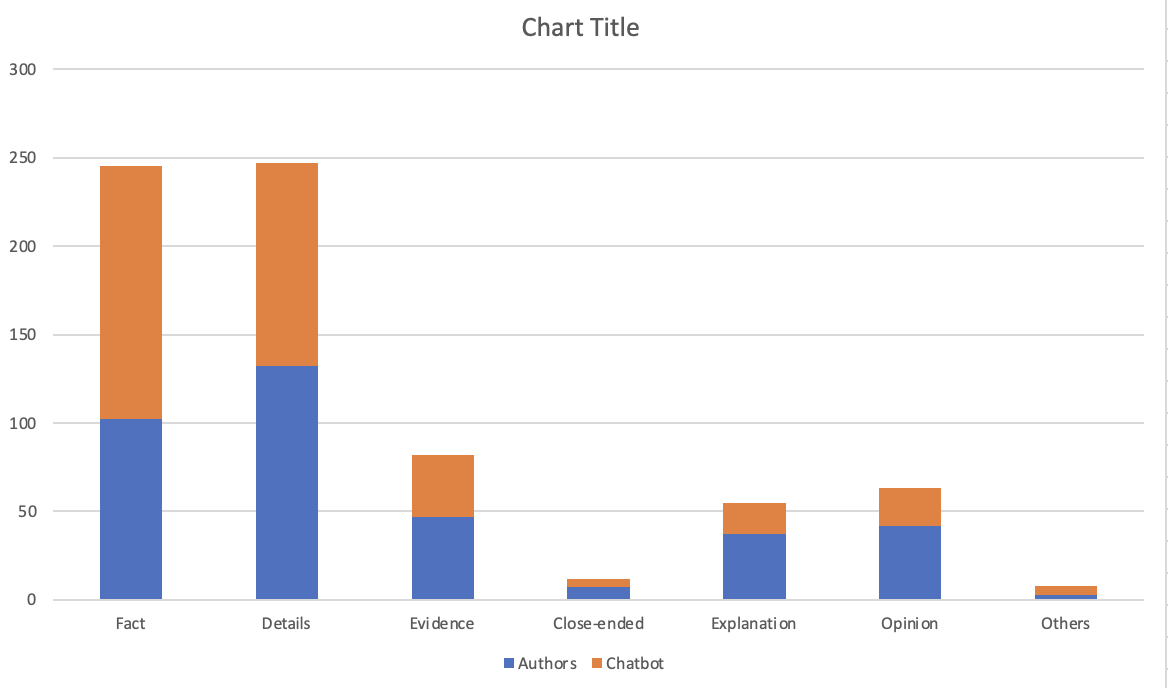} &
        \includegraphics[width=0.18\textwidth]{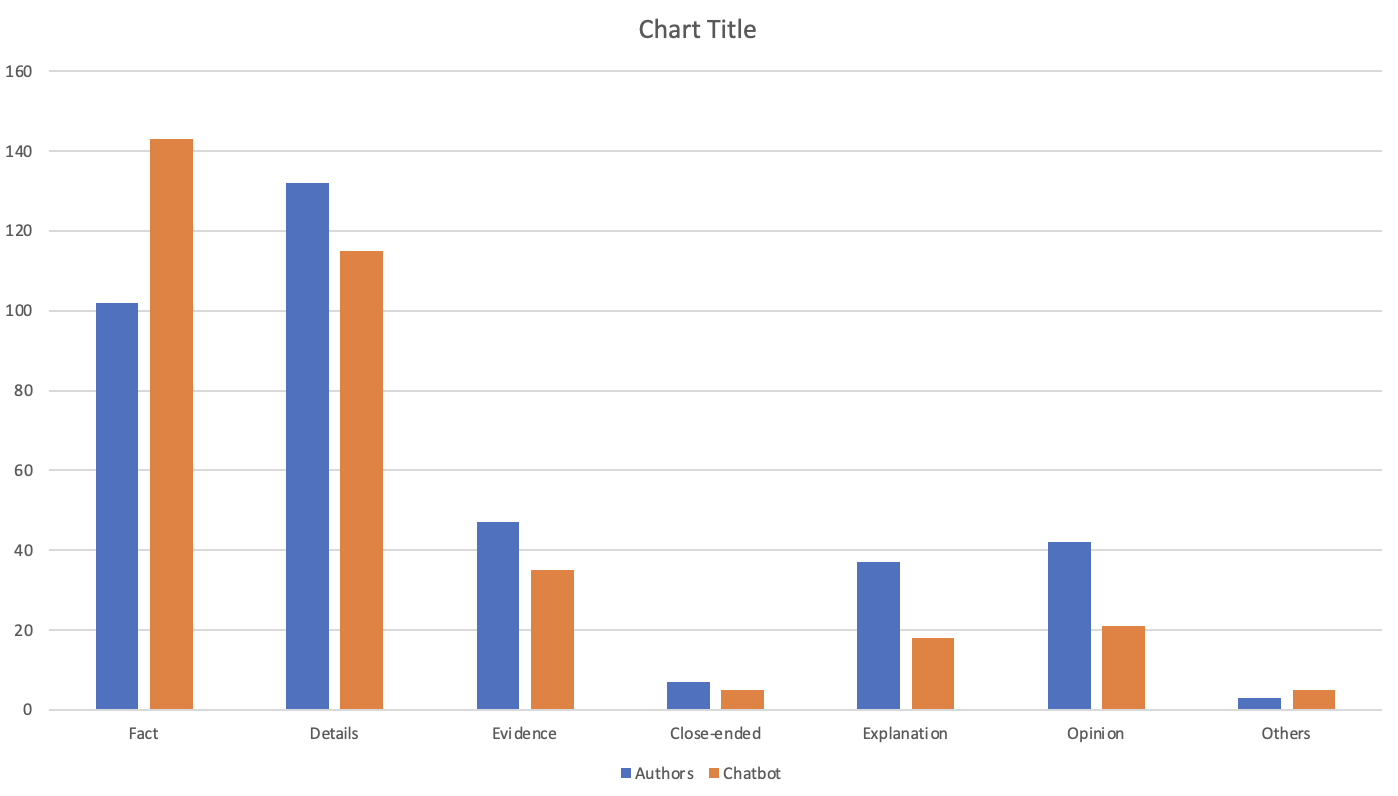} &
        \includegraphics[width=0.18\textwidth]{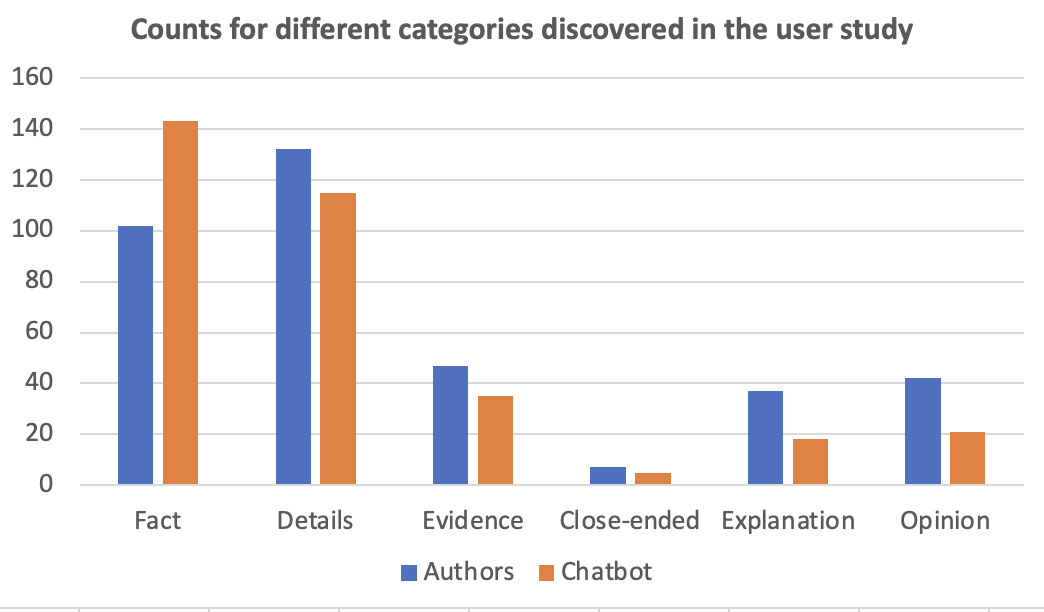} &
        \includegraphics[width=0.18\textwidth]{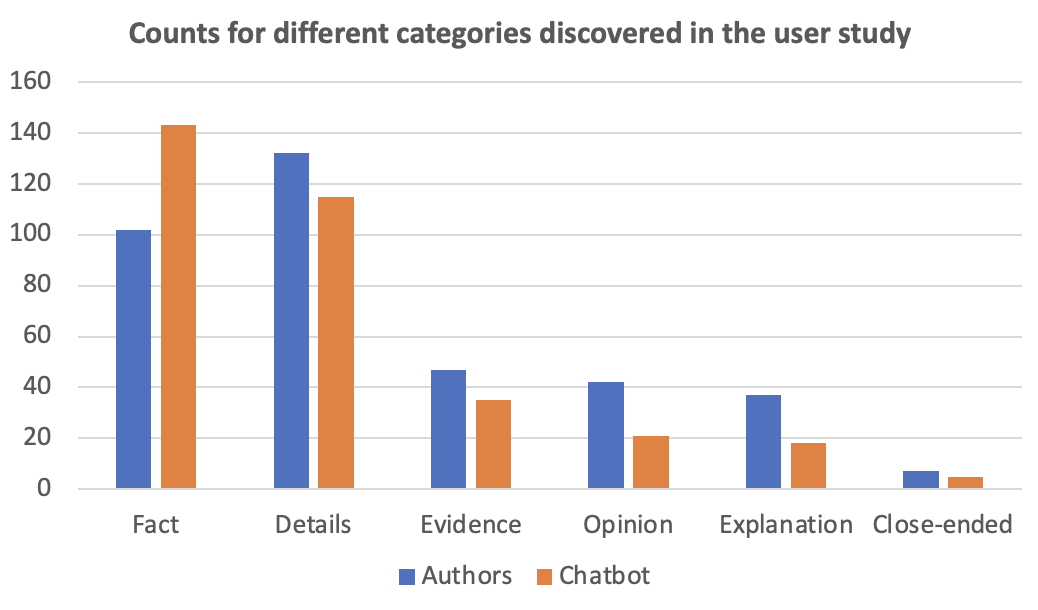} &
        \includegraphics[width=0.18\textwidth]{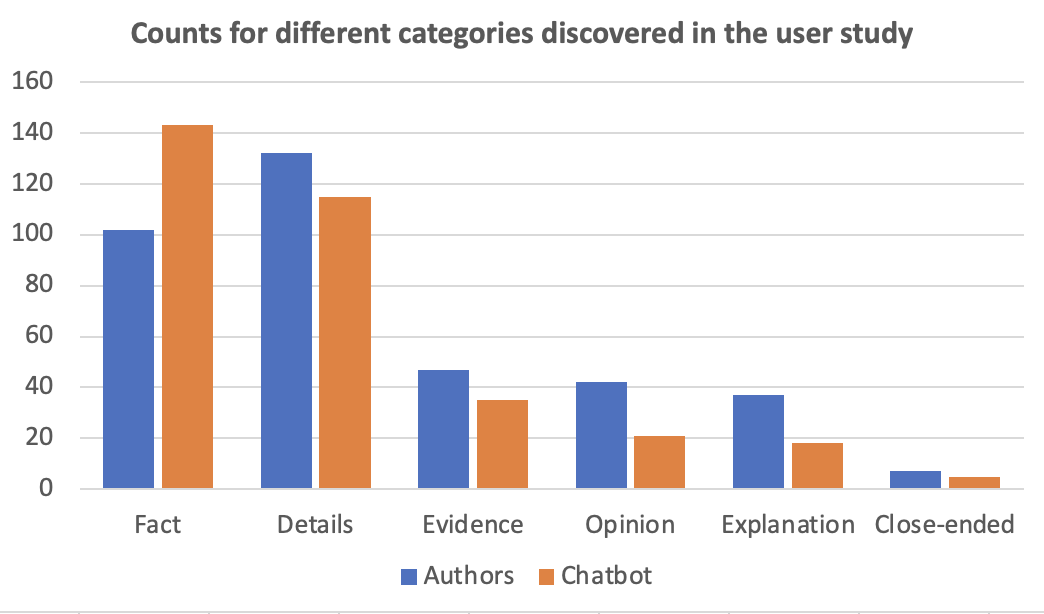}
        \\
        \midrule

        \textbf{E1} & 
        \includegraphics[width=0.18\textwidth]{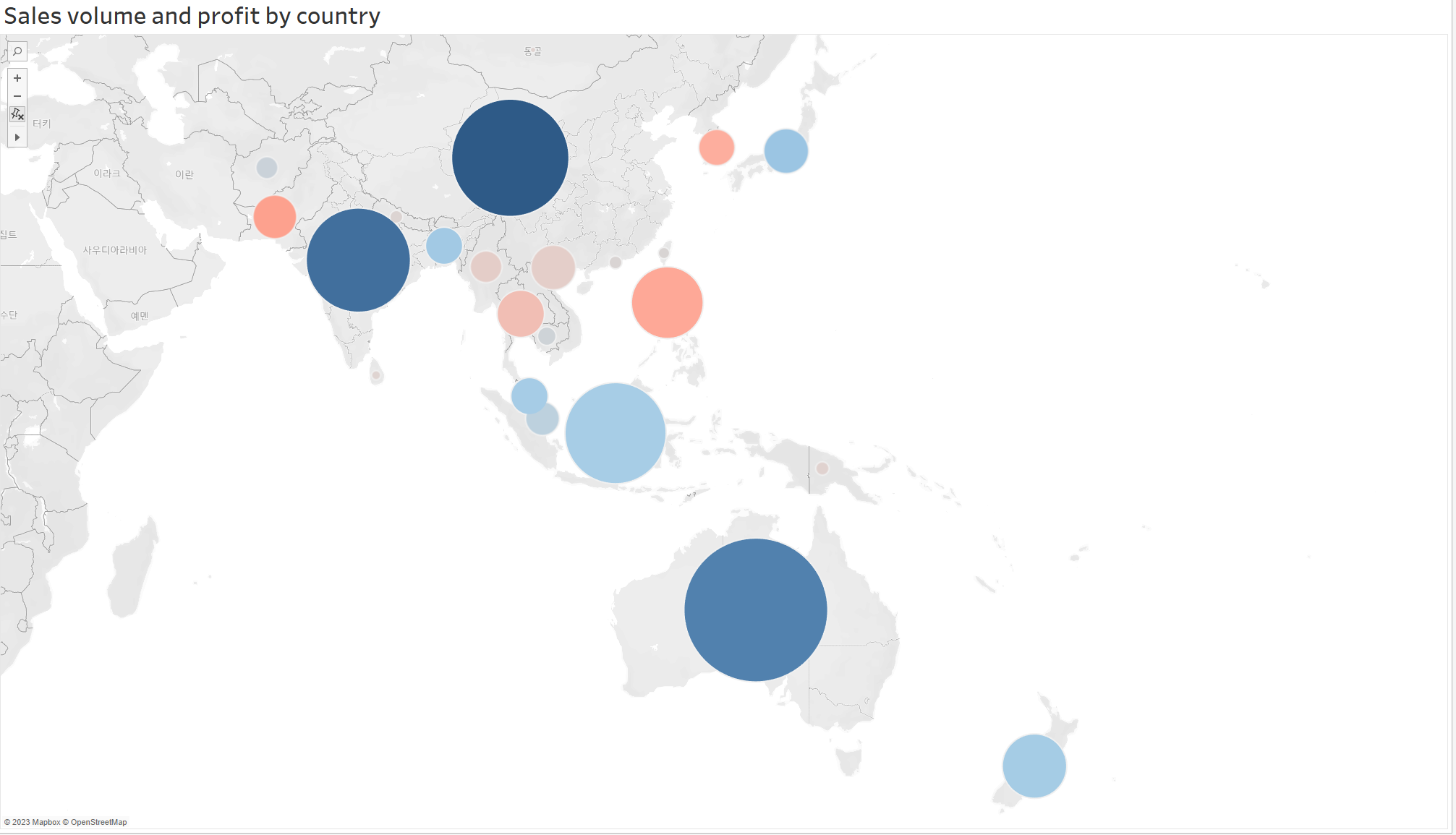} &
        \includegraphics[width=0.18\textwidth]{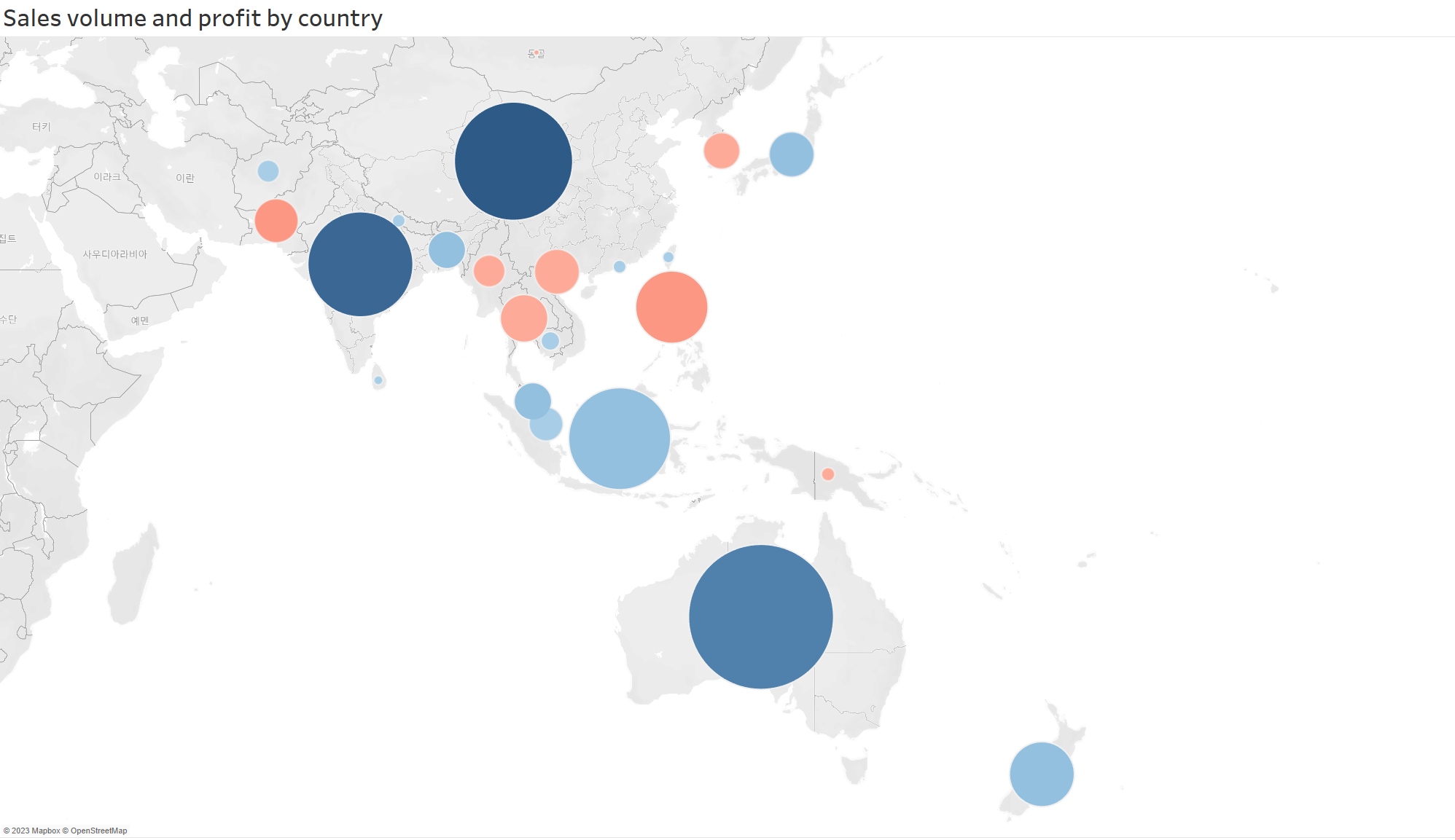} &
        \includegraphics[width=0.18\textwidth]{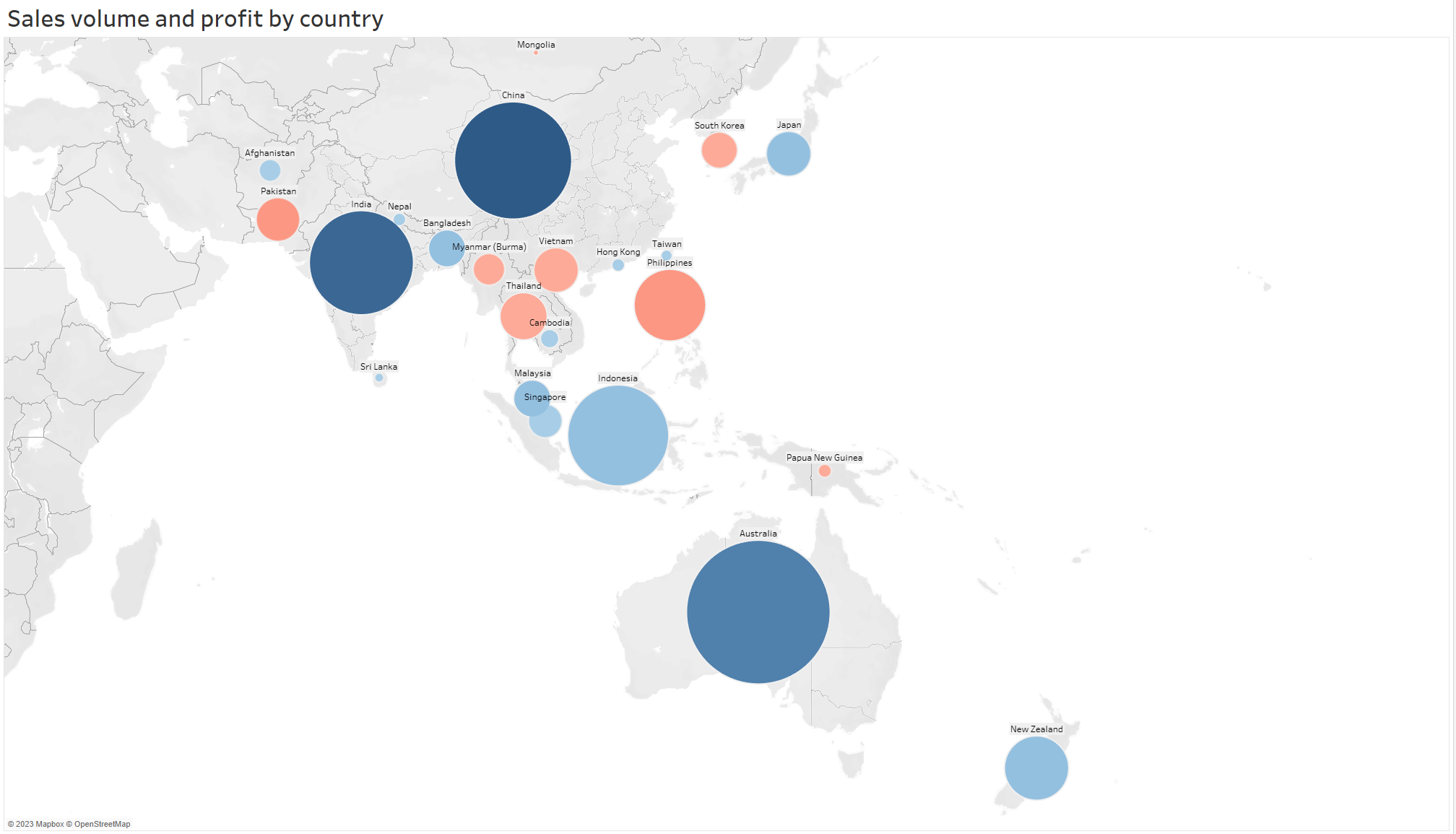} &
        \includegraphics[width=0.18\textwidth]{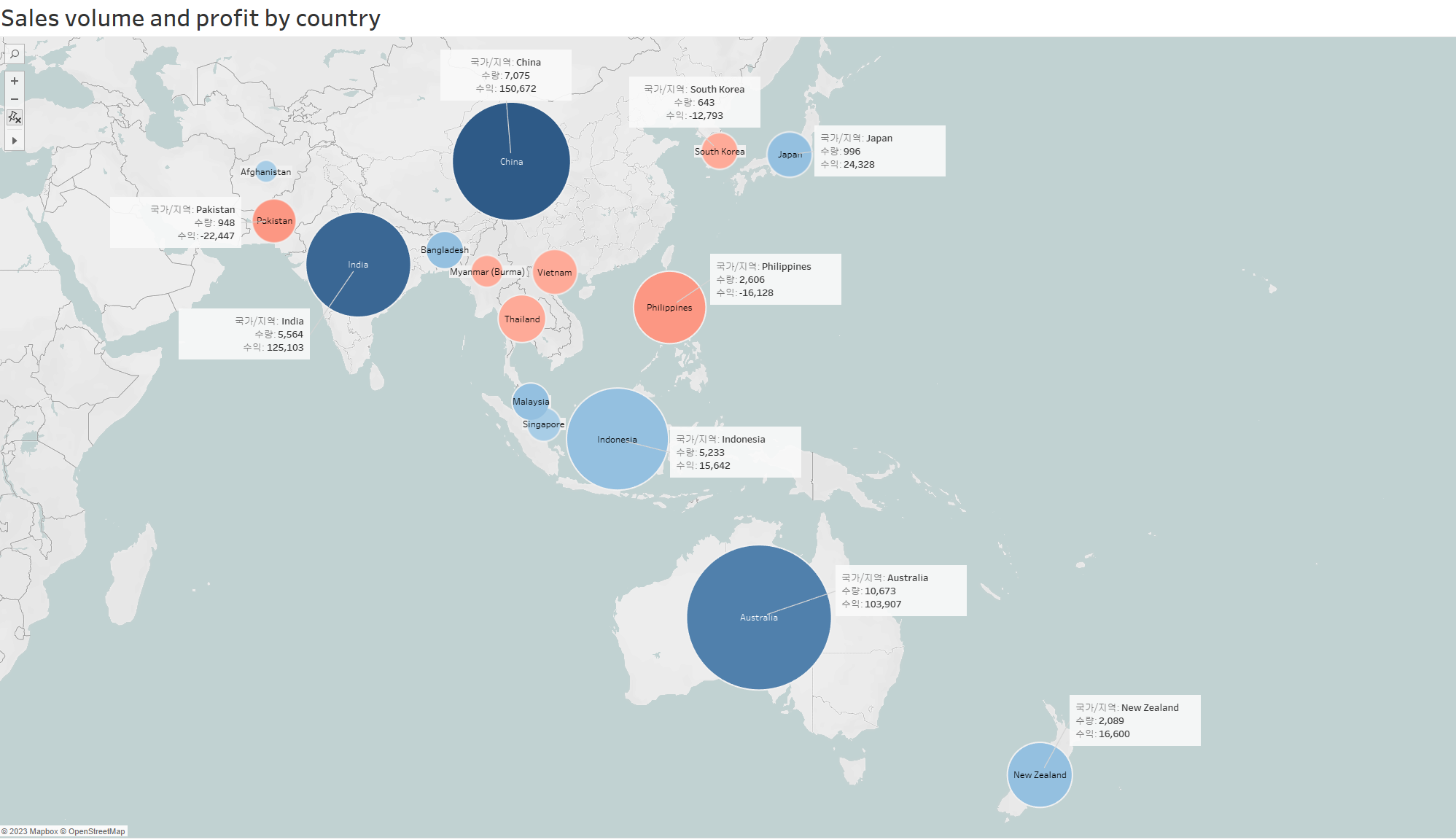} &
        \includegraphics[width=0.18\textwidth]{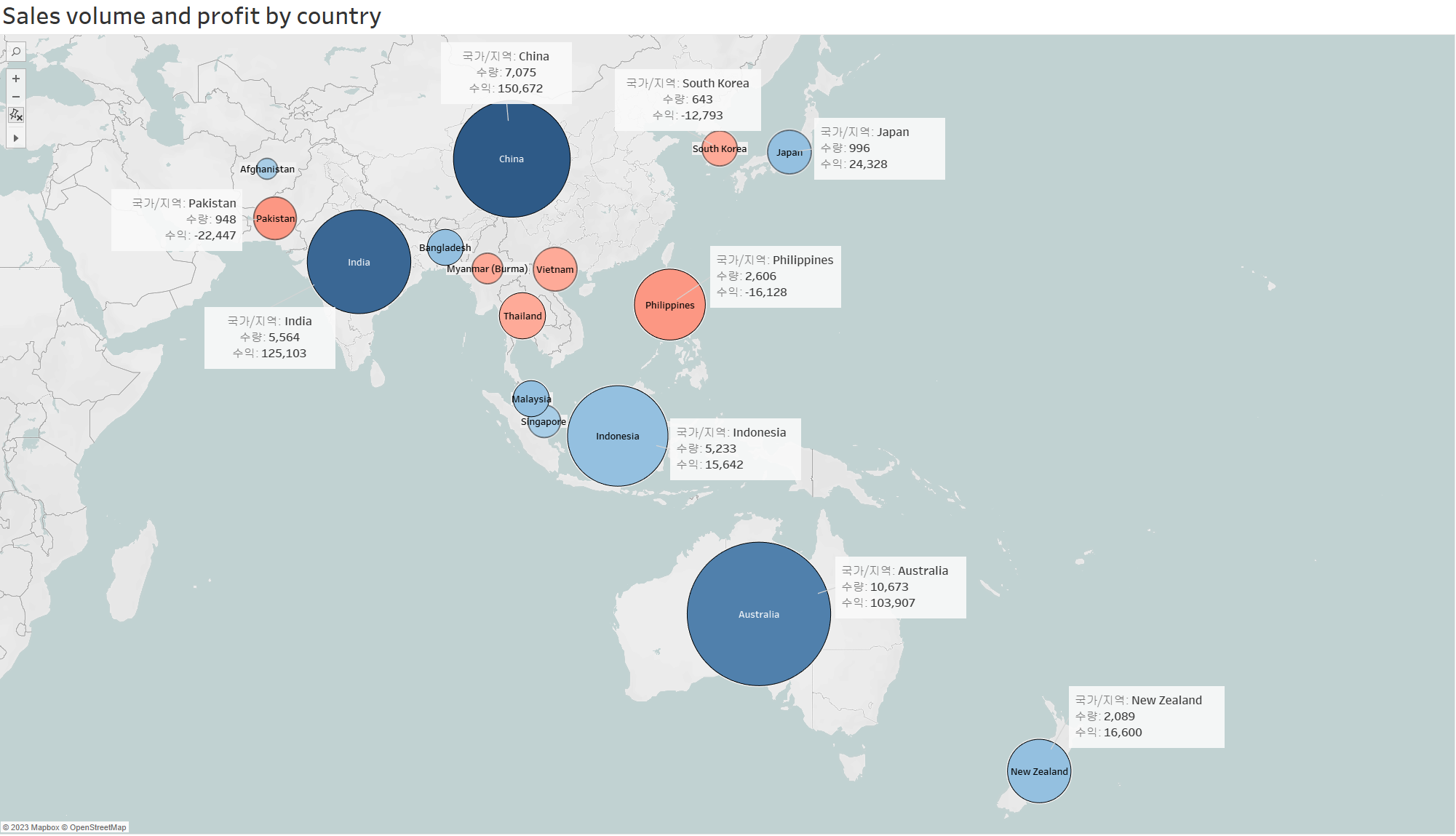}
        \\
        \midrule

        \textbf{E2} & 
        \includegraphics[width=0.18\textwidth]{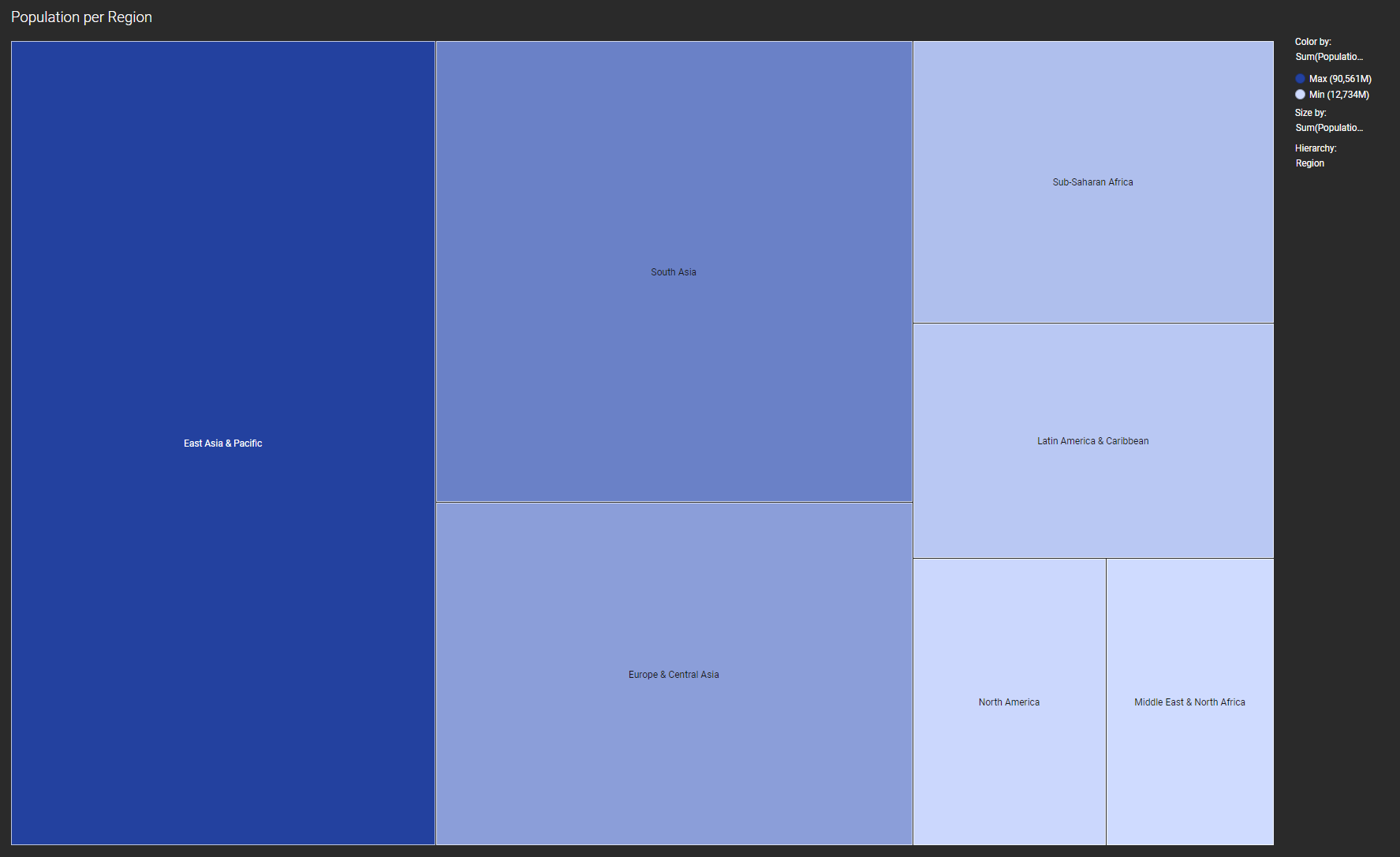} &
        \includegraphics[width=0.18\textwidth]{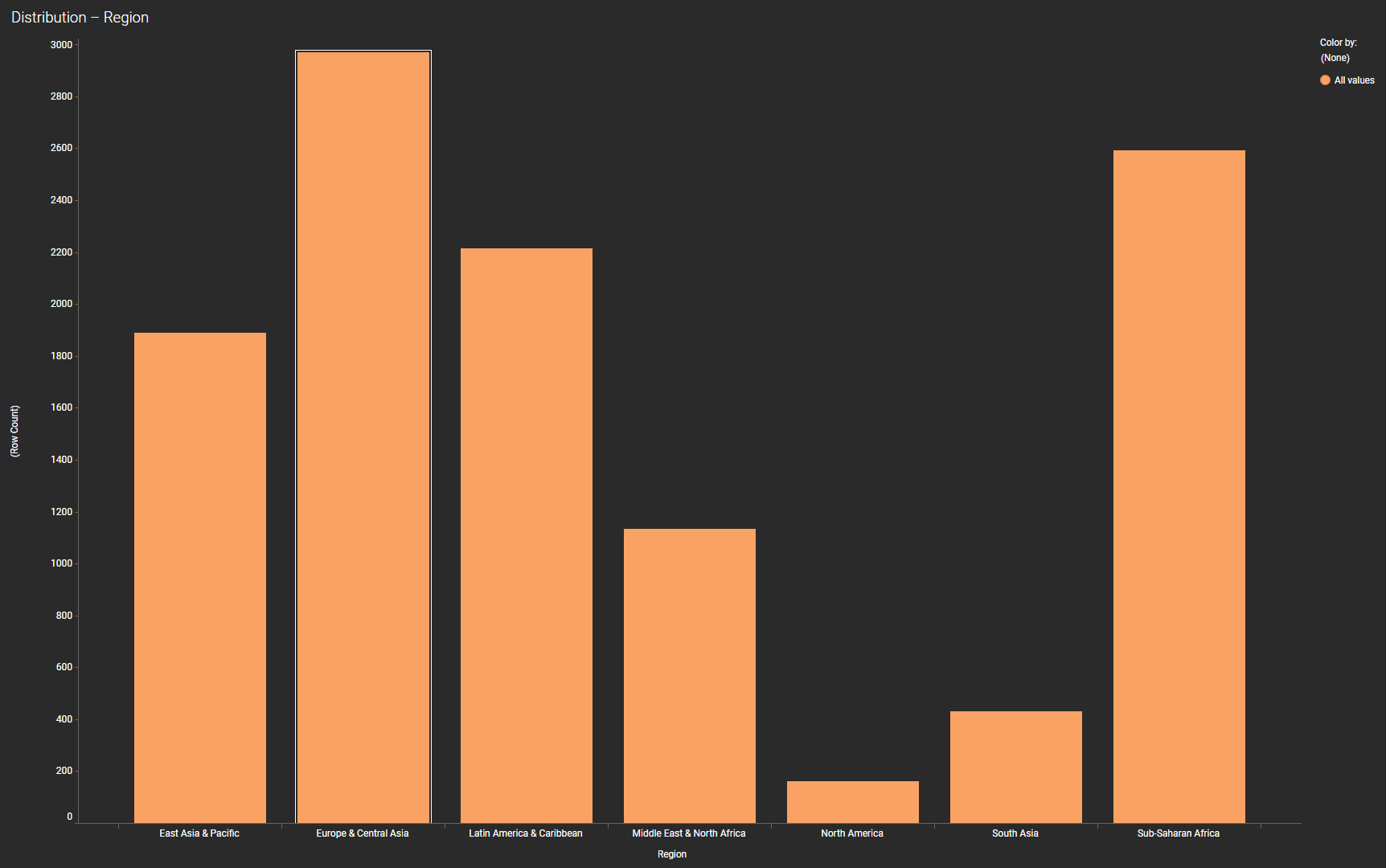} &
        \includegraphics[width=0.18\textwidth]{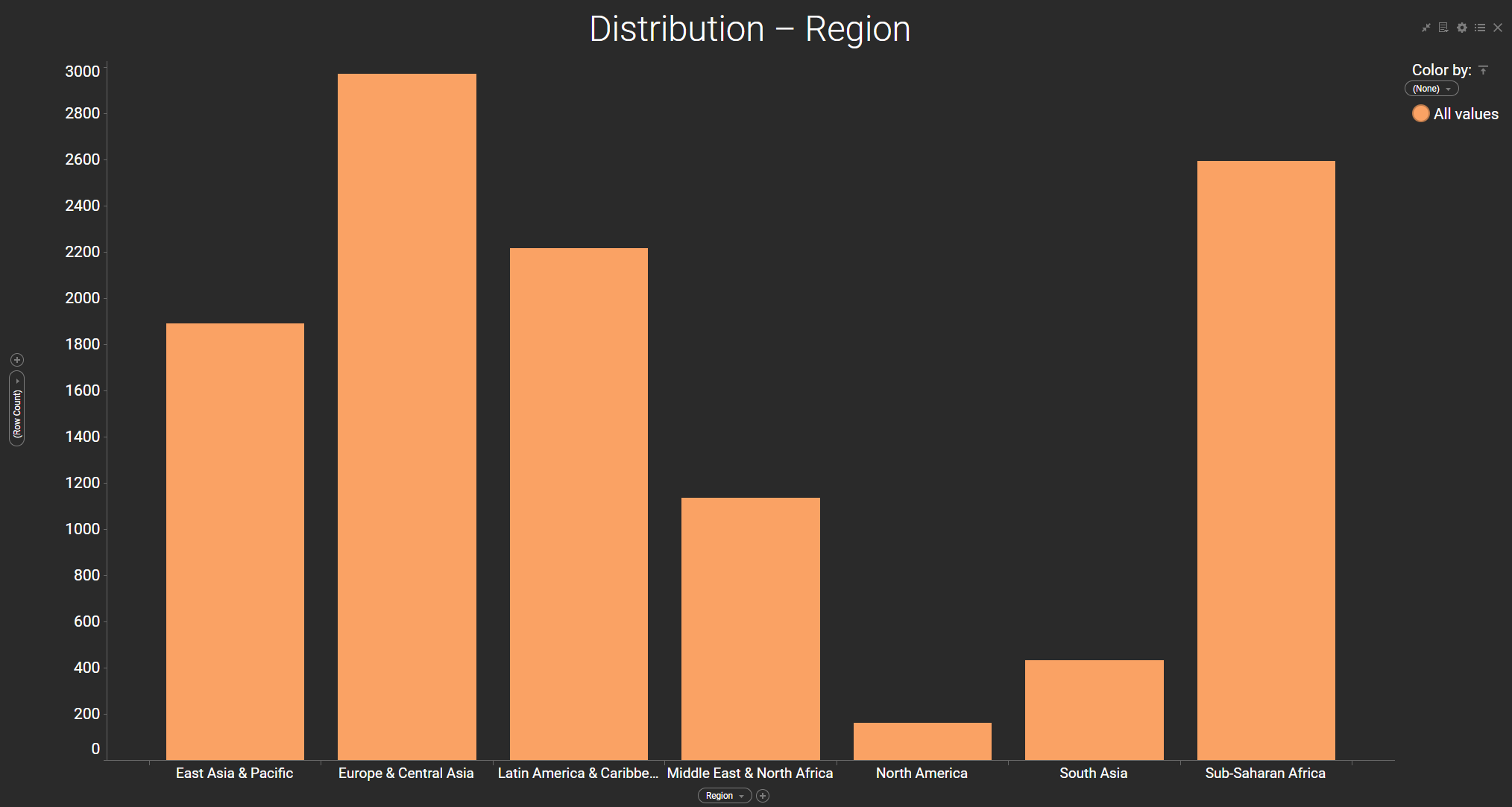} &
        \includegraphics[width=0.18\textwidth]{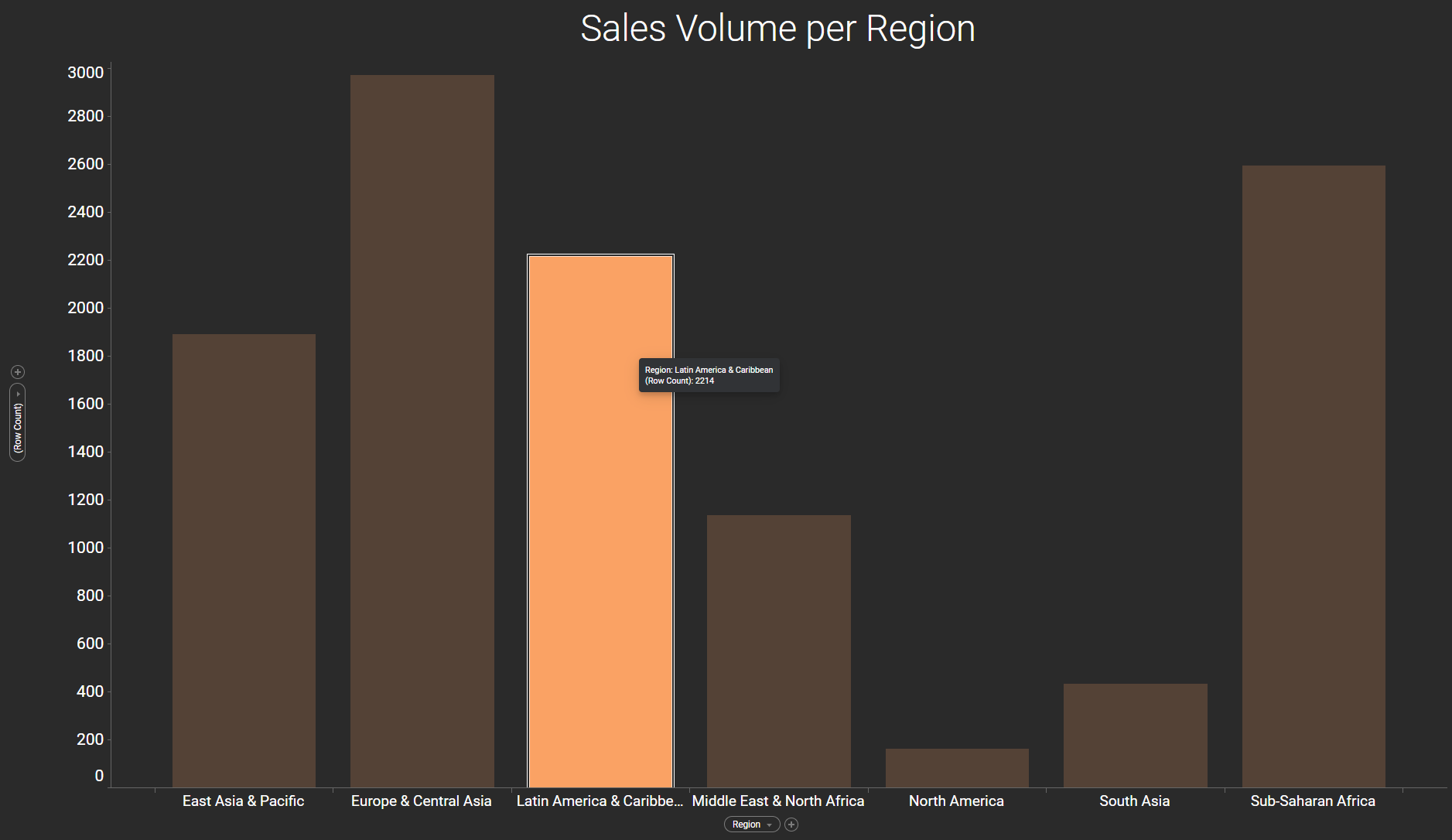} &
        \includegraphics[width=0.18\textwidth]{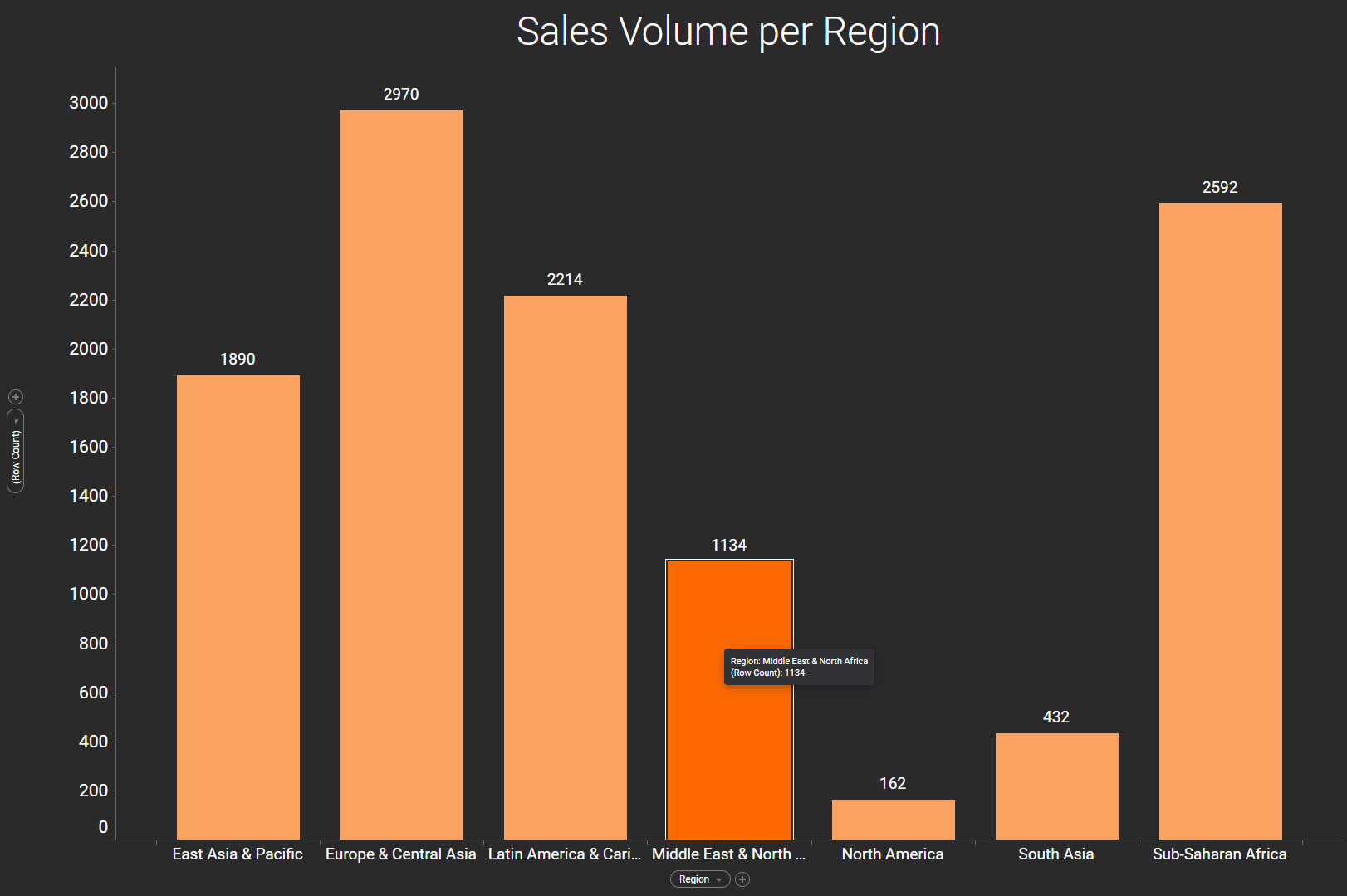}
        \\
        \midrule

        \textbf{E3} & 
        \includegraphics[width=0.18\textwidth]{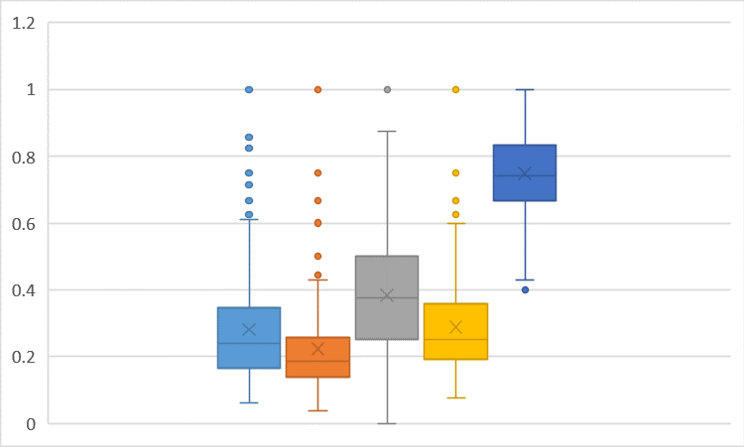} &
        \includegraphics[width=0.18\textwidth]{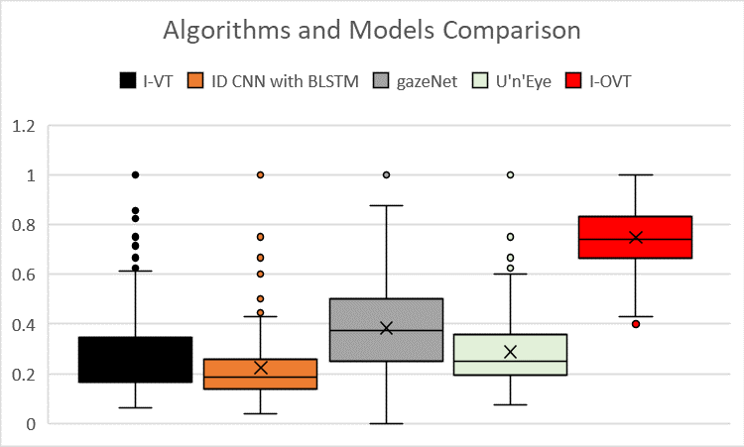} &
        \includegraphics[width=0.18\textwidth]{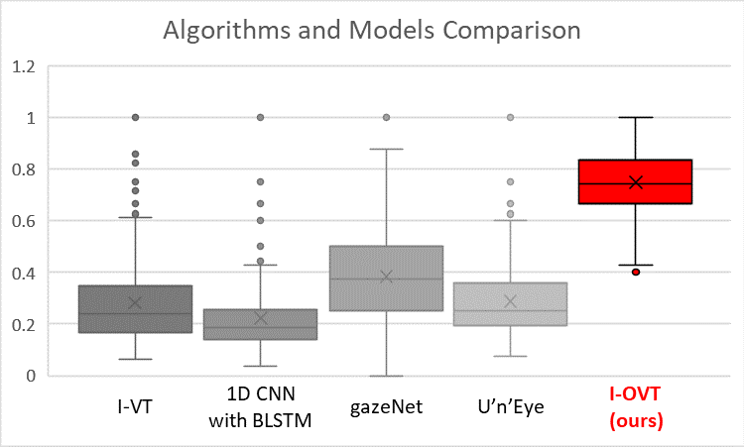} &
        \includegraphics[width=0.18\textwidth]{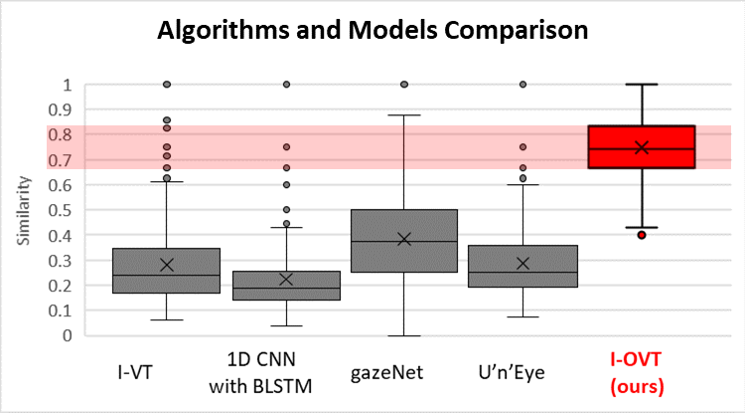} &
        \includegraphics[width=0.18\textwidth]{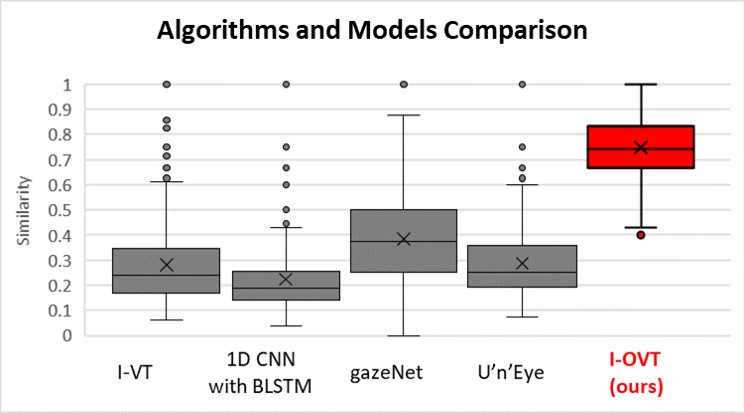}
        \\
        \bottomrule
        
    \end{tabular}

    \caption{\textbf{Examples of visualization design evolution (Intermediate and expert designers).}
    These images represent the five most significant versions (as identified by the participants) of intermediate and expert visualization designers during the design process of creating a new visualization artifact from scratch.
    The explanations of these design iterations can be found in the supplemental material.
   }
    \label{fig:inter-expert-versions}
    \vspace{-0.5cm}
\end{figure*}

% \definecolor{darkestblue}{HTML}{000052} % 0% (start)
% \definecolor{midblue1}{HTML}{54548B}    % 33%
% \definecolor{midblue2}{HTML}{AAAAC6}    % 67%
% \definecolor{lightestblue}{HTML}{FFFFFF}% 100% (end)

\subsection{Apparatus}
\label{subsec:appratus}

The interview and the experiment were conducted using the participants' own computers.
Although we did not enforce strict hardware constraints, we recommended participants use a resolution of at least 1920$\times$1080 and Google Chrome, as the \techname{} interface is optimized for that setting. 
When designing visualizations, the participants had the freedom to choose their favorite authoring tools, as long as they could upload a screenshot of their visualization image in \verb|.jpeg| or \verb|.png| format.

%% -------------------------------------------------------------
%% RESULTS
%% -------------------------------------------------------------
% ----------------------------------------------------------------
% Results
% ----------------------------------------------------------------
\section{Results}
\label{sec:results}

We now present the results from our longitudinal study, including design iterations, subjective comments, and observations.

\subsection{Design Evolution}
\label{subsec:qual-evaluation}

\begin{table*}[ht]
    \centering
    \caption{\textbf{Main feedback taken by participants. }
    Here we summarize the list of feedback mainly taken by participants. We detail the type, and detail the categories (\Chart, \Sal, \Col, \Text, and \CVD) of, and filters (--\textsc{rec} (recommendation), --\textsc{center/text/poi mismatch} (focus on center/focus on text/mismatch of salience patterns), \textsc{title/text} (no text/no title), and \textsc{sim/var} (similarity, variability of colors) themselves used per iteration.
    As a case in point, Iteration 1 refers to the feedback taken by the participant to create their second version.}
    \scalebox{0.86}{
    \begin{tabular}{lllll}
        % \rowcolor{SteelBlue}
        \toprule
        \textbf{ID} &
        \textbf{Iteration 1} &
        \textbf{Iteration 2} &
        \textbf{Iteration 3} &
        \textbf{Iteration 4} \\
        \midrule 
        N1 & \Chart--\textsc{rec} & \Sal--\textsc{center} & \Sal--\textsc{center} & \Sal--\textsc{poi mismatch}\\
        % \rowcolor{gray!20}
        N2 & \Sal--\textsc{center} & \CVD, \Sal--\textsc{center} & \CVD, \Sal--\textsc{center} & \Sal--\textsc{poi mismatch} \\
        
        N3 & \Sal--\textsc{center}, \Col--\textsc{var} & \Sal--\textsc{center}, \CVD & \Sal--\textsc{poi mismatch} & \Sal--\textsc{text} \\
        N4 & \Col--\textsc{sim}, \Text--\textsc{title} & \CVD & \Sal--\textsc{poi mismatch} & \Sal--\textsc{poi mismatch} \\
        % \rowcolor{LightSteelBlue!30}
        N5 & \Col-\textsc{var} & \Text--\textsc{title} & \Sal--\textsc{poi mismatch} & \CVD \\
        N6 & \Chart--\textsc{rec} & \Sal--\textsc{poi mismatch}, \Text--\textsc{title} & \Text--\textsc{text} & \Col--\textsc{var}, \Sal--\textsc{poi mismatch} \\
        \midrule
        % \rowcolor{LightSteelBlue!30}
        I1 & \Text--\textsc{title}, \Sal--\textsc{center} & \Sal--\textsc{poi mismatch}, \Chart--\textsc{rec} & \Sal--\textsc{poi mismatch} & \Sal--\textsc{poi mismatch} \\
        I2 & \Sal--\textsc{text} & \Sal--\textsc{text} & \Sal--\textsc{text} & \Sal--\textsc{text} \\
        % \rowcolor{LightSteelBlue!30}
        I3 & \Col--\textsc{sim}, \Text--\textsc{text} & \Col--\textsc{sim} & \CVD & \CVD, \Sal--\textsc{center} \\
        I4 & \Chart--\textsc{rec} & \Text--\textsc{title}, \Sal--\textsc{poi mismatch} & \Sal--\textsc{center}, \Col-\textsc{var} & \Sal--\textsc{poi mismatch} \\
        \midrule
        % \rowcolor{LightSteelBlue!30}
        E1 & \CVD & \Sal--\textsc{poi mismatch} & \Text--\textsc{text} & \Sal--\textsc{poi mismatch} \\
        E2 & \Chart--\textsc{rec} & \Text--\textsc{title} & \Sal--\textsc{poi mismatch}, \Text--\textsc{title} & \Text-\textsc{text}, \Sal--\textsc{poi mismatch} \\
        % \rowcolor{LightSteelBlue!30}
        E3 & \Text--\textsc{text}, \Text--\textsc{title}, \Sal--\textsc{center} & \Col--\textsc{var} & \Col--\textsc{var} & \Sal--\textsc{poi mismatch} \\
        \bottomrule
    \end{tabular}
    }
    \label{tab:filters-used}
\end{table*}

Here, we describe how the participants’ designs evolved. 
First, we provide an overview of their experimental setup. 
Table \ref{tab:participant-setup} details the tools used, dataset themes, chart types, and the time spent iterating each design. 
We found that a variety of tools (6 in total) are employed regardless of expertise level, and the datasets covered diverse topics. 
This justifies our decision to upload visualizations as image files, since this approach keeps the feedback independent of the tool.
Bar charts were the most common chart type, adopted by eight participants, though nine different visualization types appeared in total.
The time spent is calculated as the interval between the first and last upload. 
While we cannot determine exactly how much effort was allocated to each design within this period, a few patterns emerged. 
Participants worked in one long session, and it lasted approximately 90 to 150 minutes.

Second, we present information about participants' design evolution processes. 
Figs. \ref{fig:novice-versions} and \ref{fig:inter-expert-versions} show visualization snapshots from all 13 participants, categorized into novice, intermediate, and expert groups. 
Larger snapshots and full descriptions appear in the supplementary material. 
During interviews, we asked participants to explain the changes they made in each iterative step and to discuss their intentions.
Below we discuss the design evolution for three selected participants, N1, I2, and E1 as follows: the visualization, authoring tool, and dataset used, a summary of each version, including the feedback from \techname{}, the changes made, and---if given---their rationale. 
Results for participants not covered here can be found in the supplementary material.

\smallskip
\noindent\textbf{Participant N1 (novice).}
N1 created a stacked area chart (on Microsoft Excel's recommendation) to visualize daily activities.

\begin{itemize}%[noitemsep, topsep=0.em, leftmargin=1.2em]

    \item\textbf{Version 1$\rightarrow$2:}
    \textit{Feedback:} \techname{} recommended bar charts to enhance value comparisons. 
    \textit{Revision:} N1 changed from stacked line chart to a stacked bar chart. 

    \item\textbf{Version 2$\rightarrow$3:} 
    \textit{Feedback:} \techname{} noted that there is too much focus of salience towards the center.
    \textit{Revision:} \rev{N1} adjusted the chart's size ratio and incorporated a new line chart, which displayed the cumulative average hours spent on a specific activity (``work'').
    \rev{N1} \rev{introduced} a vertical axis to represent the line value for examining influences on work.
    
    \item\textbf{Version 3$\rightarrow$4:} 
    \textit{Feedback:} Again, \techname{} observed excessive concentration of salience in the central area.
    \textit{Revision:} \rev{N1} re-adjusted chart dimensions and highlighted labels. 
    
    \item\textbf{Version 4$\rightarrow$5:} 
    \textit{Feedback:} Saliency is focused in unimportant areas. 
    \textit{Revision:} N1 increased title and legend font sizes.

\end{itemize}

\noindent\textbf{Participant I2 (intermediate).}
I2 wanted to analyze their monthly spending based on two factors: (1) category, and (2) payment method. 
They used Tableau to author the visualization. 

\begin{itemize}%[noitemsep, topsep=0.em, leftmargin=1.2em]
    \item\textbf{Version 1$\rightarrow$2:}
    \textit{Feedback:} \techname{} asserted that there is too much focus on text.
    \textit{Revision:} I2 changed the order of the hierarchy to remove repetitive labels.
    
    \item\textbf{Version 2$\rightarrow$3:} 
    \textit{Feedback:} Again, \techname{} stated that there is too much focus on text.
    \textit{Revision:} I2 removed the ``payment method'' category, consolidated its values into a stacked bar chart, and labeled each bar segment with its corresponding payment method.
    
    \item\textbf{Version 3$\rightarrow$4:}
    \textit{Feedback:} And once again, \techname{} said that there is still too much focus on text.
    \textit{Revision:} I2 eliminated the text in the bar chart and introduced a legend.
    
    \item\textbf{Version 4$\rightarrow$5:}
    \textit{Feedback:} And again, \techname{} maintained that there is still too much text.
    \textit{Revision:} I2 rotated the bars vertically and separated the category labels.
    
\end{itemize}

\textbf{Participant E1 (expert).}
Using Tableau, E1 chose to display a multinational company's profits in Asia on a geographic map.

\begin{itemize}%[noitemsep, topsep=0.em, leftmargin=1.2em]
    \item\textbf{Version 1$\rightarrow$2:}
    \textit{Feedback:} From the protanopia-simulated image in the tool, several circle cues were hard to detect.
    \textit{Revision:} The colors in each circle were made darker.
    
    \item\textbf{Version 2$\rightarrow$3:}
    \textit{Feedback:} The salience map in \techname{} directed attention to country names, but several were unrelated to the data. 
    \textit{Revision:} Country names were added to each circle, and irrelevant country names were removed.
    
    \item\textbf{Version 3$\rightarrow$4:}
    \textit{Feedback:} \techname{} noted that textual content would make the chart more readable. 
    \textit{Revision:} E1 relocated country names to circle centers, added explanatory boxes, and differentiated ocean colors for clarity.
    
    \item\textbf{Version 4$\rightarrow$5:}
    \textit{Feedback:} Scanner Deeply did not detect some circles.
    \textit{Revision:} Strokes were added to circles.
\end{itemize}

\subsection{Feedback Usage Patterns}
\label{subsec:feedback-patterns}

Below we present a summary of feedback usage patterns. 
Table \ref{tab:filters-used} outlines the filters each participant used during the design process.
To begin with, we describe iteration patterns, and then we also analyze cases when participants do not take the feedback.

\noindent\textbf{Iteration patterns. } 
Across the 13 participants, each design iteration typically incorporated one or two feedback changes, occasionally reaching three in a single iteration. 
Of the 68 total feedback instances, 33 were related to ``Scanner Deeply,'' with 18 of these examining whether salience aligned with the designer’s intended focal points. 
Notably, salience-related feedback often required multiple iterations to complete, possibly because shifting salience in a single step is not straightforward. 
Chart recommendations appeared only five times—primarily in the first iteration.
This seems to reflect the tendency to make major design choices (e.g., chart type) prior to making minor design choices~\cite{Sedlmair2012}. 
The chartjunk filter was never triggered, as no chartjunk elements were introduced in the participants' designs. 
Although usage patterns were relatively consistent across skill levels, a slight decrease in implemented feedback was observed from early to late iterations (an average of 1.46 changes in the first iteration versus 1.23 in the last), which may indicate that designs converged over time, reducing the need for additional modifications.

\noindent\textbf{When participants do not listen to the feedback. } We observe several instances where participants noted the system’s feedback but ultimately chose not to implement it. 
The most frequent reason was that the model did not fully align with the user’s intent. 
When I4 was advised to use additional colors for separate categories, they declined because only two colors were necessary for their comparative task.
In other cases, users felt the system’s suggestions conflicted with their design objectives. 
For instance, N4 wanted to include a grid for comparison purposes, despite realizing it might reduce salience in key areas. 
Some participants also proceeded with their own experiments unrelated to any feedback, such as N1 adding a line chart to meet personal objectives.
Finally, another example is when the system presents incorrect recommendations. 
For instance, participant I1 was advised to replace a heatmap with a bar chart but did not know how to create this chart for their data.

Overall, these observations indicate that the system is not as effective when users have their own objectives, as it currently lacks the capacity to capture their intentions. 
They also reveal that some filters within the system are not optimal.

\definecolor{myNavy}{RGB}{0,0,128}

\newcommand{\minval}{3}
\newcommand{\maxval}{5}

\newcommand{\heatmap}[1]{%
  \pgfmathsetmacro{\midval}{3} % White midpoint
  \pgfmathsetmacro{\minval}{0} % Blue end
  \pgfmathsetmacro{\maxval}{5} % Red end
  \ifdim #1 pt=\midval pt
    \cellcolor{white}{#1}%
  \else
    \ifdim #1 pt>\midval pt
      \pgfmathparse{70*(#1-\midval)/(\maxval-\midval)} % scale from white to red
      \xdef\colval{\pgfmathresult}%
      \cellcolor{red!\colval!white}{#1}%
    \else
      \pgfmathparse{70*(\midval-#1)/(\midval-\minval)} % scale from white to blue
      \xdef\colval{\pgfmathresult}%
      \cellcolor{blue!\colval!white}{#1}%
    \fi
  \fi
}

\renewcommand{\hmCell}[1]{%
  \pgfmathsetmacro{\midval}{3} % middle point (white)
  \pgfmathsetmacro{\maxval}{5} % upper bound (fully red)
  \pgfmathsetmacro{\minval}{0} % lower bound (fully blue)
  \ifdim #1 pt=\midval pt
    \cellcolor{white}{#1}%
  \else
    \ifdim #1 pt>\midval pt % positive side (red)
      \pgfmathparse{70*(#1-\midval)/(\maxval-\midval)} % 0 (white) to 100 (red)
      \xdef\colval{\pgfmathresult}%
      \cellcolor{applerednormal!\colval!white}{#1}%
    \else % negative side (blue)
      \pgfmathparse{70*(\midval-#1)/(\midval-\minval)} % 0 (white) to 100 (blue)
      \xdef\colval{\pgfmathresult}%
      \cellcolor{blue!\colval!white}{#1}%
    \fi
  \fi
}

\begin{table*}[ht]
    \centering
    \caption{\textbf{Expert evaluation of visualization designs.}
    We deployed 3 seasoned visualization experts to evaluate visualization designs authored by 13 participants.
    The scores from second to fifth columns show the evaluation of visualization designer per iteration.
    We present the trends of scores per participant in the sixth column.
    We also asked the experts to pick the best design from each designer's iterations.
    This is shown in the seventh column.
    Finally, the last column shows the overall improvement scores of visualization designs.
    \rev{The three numbers in brackets in the third and fifth column represent the scores obtained by the experts, sorted per different expert evaluator. 
    The 1--5 Likert scale is defined as follows: 1 = significant decline in quality, 3 = neutral, and 5 = significant improvement in quality.}}
    \begin{tabular}{lllllccc}

        \toprule
        \multirow{2}{*}{\textbf{\textcolor{black}{ID}}} &
        \multicolumn{4}{c}{ \textbf{\textcolor{black}{Expert evaluation scores per iteration}}}&\multirow{2}{*}{\textbf{\textcolor{black}{Trends}}} &
        \multirow{2}{*}{ \textbf{\textcolor{black}{
        \makecell{Votes for the\\best version}
        }}}&
        \multirow{2}{*}{\textbf{\textcolor{black}{\makecell{Overall score\\(V1 - V5)}}}}\\
         &
        \textbf{\textcolor{black}{V1 $\rightarrow$ V2}}& 
        \textbf{\textcolor{black}{V2 $\rightarrow$ V3}}& 
        \textbf{\textcolor{black}{V3 $\rightarrow$ V4}}& 
        \textbf{\textcolor{black}{V4 $\rightarrow$ V5}}& & & \\
        \midrule
                N1  & \hmCell{3.00} (3/4/2) & \hmCell{3.00} (5/2/4) & \hmCell{3.67} (3/4/4) & \hmCell{3.33} (3/3/4) & \resizebox{0.8cm}{0.2cm}{\begin{sparkline}{2.5}
\sparkspike .00 .3
\sparkspike .33 .30
\sparkspike .67 .5
\sparkspike 1.0 .4
\end{sparkline}}
&{\scriptsize % Slightly smaller text so squares remain small
\begin{tabular}{%
  @{}c% no left gap
  @{\hskip 2pt}c%
  @{\hskip 2pt}c%
  @{\hskip 2pt}c%
  @{\hskip 2pt}c@{}}
  \ZeroCell{~} &
  \SquareCell{midblue2}{1} &
  \SquareCell{midblue2}{1} &
  \SquareCell{midblue2}{1} &
  \ZeroCell{~} \\
\end{tabular}}&

\hmCell{3.67} (4/3/4) \\
        % \rowcolor{LightSteelBlue!30}
        N2 & \hmCell{2.33} (4/1/2) & \hmCell{2.67} (3/2/3) & \hmCell{3.00} (4/3/2) & \hmCell{3.00} (4/3/2) & \resizebox{0.8cm}{0.2cm}{\begin{sparkline}{2.5}
\sparkspike .00 .2
\sparkspike .33 .25
\sparkspike .67 .3
\sparkspike 1.0 .3
\end{sparkline}}
&{\scriptsize % Slightly smaller text so squares remain small
\begin{tabular}{%
  @{}c% no left gap
  @{\hskip 2pt}c%
  @{\hskip 2pt}c%
  @{\hskip 2pt}c%
  @{\hskip 2pt}c@{}}
  \SquareCell{midblue1}{2} &
  \ZeroCell{~} &
  \ZeroCell{~} &
  \ZeroCell{~} &
  \SquareCell{midblue2}{1} \\
\end{tabular}
}& 
\hmCell{2.67} (3/3/2) \\
        N3  & \hmCell{3.33} (3/3/4) & \hmCell{1.67} (1/2/2) & \hmCell{4.67} (4/5/5) & \hmCell{3.33} (4/2/4) & %5/4/5
        \resizebox{0.8cm}{0.2cm}{\begin{sparkline}{2.5}
\sparkspike .00 .4
\sparkspike .33 .1
\sparkspike .67 1.0
\sparkspike 1.0 .4
\end{sparkline}}
&{\scriptsize % Slightly smaller text so squares remain small
\begin{tabular}{%
  @{}c% no left gap
  @{\hskip 2pt}c%
  @{\hskip 2pt}c%
  @{\hskip 2pt}c%
  @{\hskip 2pt}c@{}}
  \ZeroCell{~} &
  \ZeroCell{~} &
  \ZeroCell{~} &
  \SquareCell{midblue2}{1} &
  \SquareCell{midblue1}{2} \\
\end{tabular}
}& 
\hmCell{4.00} (4/4/4) \\
        
    % \rowcolor{LightSteelBlue!30}
    N4 & \hmCell{3.00} (3/4/2) & \hmCell{2.67} (2/3/3) & \hmCell{2.67} (3/3/2) & \hmCell{3.00} (4/3/2) & \resizebox{0.8cm}{0.2cm}{\begin{sparkline}{2.5}
\sparkspike .00 .3
\sparkspike .33 .2
\sparkspike .67 0.2
\sparkspike 1.0 .3
\end{sparkline}}
&{\scriptsize % Slightly smaller text so squares remain small
\begin{tabular}{%
  @{}c% no left gap
  @{\hskip 2pt}c%
  @{\hskip 2pt}c%
  @{\hskip 2pt}c%
  @{\hskip 2pt}c@{}}
  \SquareCell{midblue2}{1} &
  \SquareCell{midblue2}{1} &
  \ZeroCell{~} &
  \ZeroCell{~} &
  \SquareCell{midblue2}{1} \\
\end{tabular}
}& 
\hmCell{2.67} (3/3/2) \\
N5  & \hmCell{4.33} (4/4/5) & \hmCell{2.33} (3/2/2) & \hmCell{3.00} (4/1/4) & \hmCell{3.00} (2/3/4) & \resizebox{0.8cm}{0.2cm}{\begin{sparkline}{2.5}
\sparkspike .00 .9
\sparkspike .33 .1
\sparkspike .67 0.3
\sparkspike 1.0 .3
\end{sparkline}}
&%4/4/2
        {\scriptsize % Slightly smaller text so squares remain small
\begin{tabular}{%
  @{}c% no left gap
  @{\hskip 2pt}c%
  @{\hskip 2pt}c%
  @{\hskip 2pt}c%
  @{\hskip 2pt}c@{}}
  \ZeroCell{~} &
  \SquareCell{midblue2}{1} &
  \ZeroCell{~} &
  \SquareCell{midblue1}{2} &
  \ZeroCell{~} \\
\end{tabular}
}& \hmCell{3.00} (3/3/3) \\
% \rowcolor{LightSteelBlue!30}
N6  & \hmCell{4.33} (4/4/5) & \hmCell{4.33} (4/5/4) & \hmCell{4.00} (3/5/4) & \hmCell{4.00} (4/4/4) & \resizebox{0.8cm}{0.2cm}{\begin{sparkline}{2.5}
\sparkspike .00 .9
\sparkspike .33 .9
\sparkspike .67 0.7
\sparkspike 1.0 .7
\end{sparkline}}
&{\scriptsize % Slightly smaller text so squares remain small
\begin{tabular}{%
  @{}c% no left gap
  @{\hskip 2pt}c%
  @{\hskip 2pt}c%
  @{\hskip 2pt}c%
  @{\hskip 2pt}c@{}}
  \ZeroCell{~} &
  \ZeroCell{~}  &
  \ZeroCell{~}  &
  \ZeroCell{~}  &
  \SquareCell{darkestblue}{3} \\
\end{tabular}
}& 
\hmCell{4.00} (4/4/4) \\
\midrule
I1  & \hmCell{3.67} (3/4/4) & \hmCell{3.00} (4/3/2) & \hmCell{2.67} (2/3/3) & \hmCell{3.33} (3/4/3) & %2/3/5
       \resizebox{0.8cm}{0.2cm}{\begin{sparkline}{2.5}
\sparkspike .00 .5
\sparkspike .33 .3
\sparkspike .67 0.2
\sparkspike 1.0 .4
\end{sparkline}}
& {\scriptsize % Slightly smaller text so squares remain small
\begin{tabular}{%
  @{}c% no left gap
  @{\hskip 2pt}c%
  @{\hskip 2pt}c%
  @{\hskip 2pt}c%
  @{\hskip 2pt}c@{}}
  \ZeroCell{~} &
  \SquareCell{midblue2}{1} &
  \SquareCell{midblue2}{1} &
  \ZeroCell{~} &
  \SquareCell{midblue2}{1} \\
\end{tabular}
}& \hmCell{2.67} (3/3/2) \\
% \rowcolor{LightSteelBlue!30}
        I2 & \hmCell{4.33} (4/4/5) & \hmCell{4.00} (5/3/4) & \hmCell{3.67} (2/4/5) & \hmCell{2.33} (2/3/2) & \resizebox{0.8cm}{0.2cm}{\begin{sparkline}{2.5}
\sparkspike .00 .9
\sparkspike .33 .7
\sparkspike .67 .4
\sparkspike 1.0 .1
\end{sparkline}}
&{\scriptsize % Slightly smaller text so squares remain small
\begin{tabular}{%
  @{}c% no left gap
  @{\hskip 2pt}c%
  @{\hskip 2pt}c%
  @{\hskip 2pt}c%
  @{\hskip 2pt}c@{}}
  \ZeroCell{~} &
  \ZeroCell{~} &
  \ZeroCell{~} &
  \SquareCell{darkestblue}{3} &
  \ZeroCell{~} \\
\end{tabular}
}& 
\hmCell{3.67} (4/4/3) \\
        
        I3  & \hmCell{3.33} (4/2/4) & \hmCell{3.33} (4/3/3) & \hmCell{2.33} (1/3/3) & \hmCell{5.00} (5/5/5) & \resizebox{0.8cm}{0.2cm}{\begin{sparkline}{2.5}
\sparkspike .00 .4
\sparkspike .33 .4
\sparkspike .67 .1
\sparkspike 1.0 1.0
\end{sparkline}}
&{\scriptsize % Slightly smaller text so squares remain small
\begin{tabular}{%
  @{}c% no left gap
  @{\hskip 2pt}c%
  @{\hskip 2pt}c%
  @{\hskip 2pt}c%
  @{\hskip 2pt}c@{}}
  \ZeroCell{~} &
  \ZeroCell{~} &
  \ZeroCell{~} &
  \ZeroCell{~} &
  \SquareCell{darkestblue}{3} \\
\end{tabular}
}& \hmCell{5.00} (5/5/5) \\
% \rowcolor{LightSteelBlue!30}
        I4 & \hmCell{4.00} (4/5/3) & \hmCell{4.00} (4/4/4) & \hmCell{4.33} (5/4/4) & \hmCell{4.33} (3/3/4) & \resizebox{0.8cm}{0.2cm}{\begin{sparkline}{2.5}
\sparkspike .00 .7
\sparkspike .33 .7
\sparkspike .67 .8
\sparkspike 1.0 .8
\end{sparkline}}
&{\scriptsize % Slightly smaller text so squares remain small
\begin{tabular}{%
  @{}c% no left gap
  @{\hskip 2pt}c%
  @{\hskip 2pt}c%
  @{\hskip 2pt}c%
  @{\hskip 2pt}c@{}}
  \ZeroCell{~} &
  \ZeroCell{~} &
  \ZeroCell{~} &
  \SquareCell{midblue1}{2} &
  \SquareCell{midblue2}{1} \\
\end{tabular}
}& 
\hmCell{4.33} (5/4/4) \\
        \midrule

                E1 & \hmCell{3.00} (2/3/4) & \hmCell{4.33} (4/5/4) & \hmCell{4.00} (5/3/4) & \hmCell{3.67} (3/4/4) & \resizebox{0.8cm}{0.2cm}{\begin{sparkline}{2.5}
\sparkspike .00 .3
\sparkspike .33 .8
\sparkspike .67 .7
\sparkspike 1.0 .6
\end{sparkline}}
&{\scriptsize % Slightly smaller text so squares remain small
\begin{tabular}{%
  @{}c% no left gap
  @{\hskip 2pt}c%
  @{\hskip 2pt}c%
  @{\hskip 2pt}c%
  @{\hskip 2pt}c@{}}
  \ZeroCell{~} &
  \ZeroCell{~} &
  \ZeroCell{~} &
  \ZeroCell{~} &
  \SquareCell{darkestblue}{3} \\
\end{tabular}
}& \hmCell{4.67} (4/5/5) \\
% \rowcolor{LightSteelBlue!30}
        E2 & \hmCell{3.00} (2/5/2) & \hmCell{3.33} (3/3/4) & \hmCell{4.33} (5/4/4) & \hmCell{4.33} (4/4/5) & \resizebox{0.8cm}{0.2cm}{\begin{sparkline}{2.5}
\sparkspike .00 .3
\sparkspike .33 .3
\sparkspike .67 .8
\sparkspike 1.0 .8
\end{sparkline}}
&{\scriptsize % Slightly smaller text so squares remain small
\begin{tabular}{%
  @{}c% no left gap
  @{\hskip 2pt}c%
  @{\hskip 2pt}c%
  @{\hskip 2pt}c%
  @{\hskip 2pt}c@{}}
  \ZeroCell{~} &
  \ZeroCell{~} &
  \ZeroCell{~} &
  \ZeroCell{~} &
  \SquareCell{darkestblue}{3} \\
\end{tabular}
}&
\hmCell{3.00} (1/4/4) \\

E3 & \hmCell{4.33} {(4/4/5)} & \hmCell{3.67} (3/4/4) & \hmCell{4.67} (5/5/4) & \hmCell{3.00} (3/3/3) & \resizebox{0.8cm}{0.2cm}{\begin{sparkline}{2.5}
\sparkspike .00 .8
\sparkspike .33 .6
\sparkspike .67 .9
\sparkspike 1.0 .3
\end{sparkline}}
&{\scriptsize % Slightly smaller text so squares remain small
\begin{tabular}{%
  @{}c% no left gap
  @{\hskip 2pt}c%
  @{\hskip 2pt}c%
  @{\hskip 2pt}c%
  @{\hskip 2pt}c@{}}
  \ZeroCell{~} &
  \ZeroCell{~} &
  \ZeroCell{~} &
  \SquareCell{midblue1}{2} &
  \SquareCell{midblue2}{1} \\
\end{tabular}
}& \hmCell{4.67} (4/5/5) \\
        \bottomrule
    \end{tabular}
    \label{tab:evaluation-summary}
\end{table*}

\begin{table*}[ht]
    \caption{\textbf{Feedback evaluation per filter types.}
    We present the average feedback score for each filter type, based on iteration-level scoring. 
    \rev{We also provide a summary of the effectiveness and limitations of the filters used in \techname{} per topic, based on the comments provided by the expert evaluators.}
    The numbers in brackets indicate the corresponding count of feedback instances for each score. 
    Overall, designers improved when using the feedback, although novices found the CVD-related feedback more challenging to apply. }
    \centering
    \scalebox{0.96}{
    \begin{tabular}{lcccll}
    \toprule
     & \textbf{Novices} & \textbf{Intermediates} & \textbf{Experts} & \rev{\textbf{Effectiveness}} & \rev{\textbf{Limitations}} \\
    \midrule
    \textbf{Virtual eyetracker \Sal}    & \heatmap{3.06} (12) & \heatmap{3.79} (11) & \heatmap{4.00} (7) & \rev{Helps designers highlight key parts} & \rev{Task-specific saliency prediction} \\
    \textbf{Text \Text}   & \heatmap{3.28} (6) & \heatmap{4.00} (4) & \heatmap{4.08} (4) & \rev{Helps add titles and explanations} & \rev{Does not suggest text contents} \\
    \textbf{Visual representation \Chart} & \heatmap{3.44} (3) & \heatmap{4.00} (2) & \heatmap{3.00} (1) & \rev{Helps recommend simple charts} & \rev{Does not recommend complex charts} \\
    \textbf{Color perception \Col} & \heatmap{3.67} (4) & \heatmap{3.66} (3) & \heatmap{4.17} (2) & \rev{Guides proper color use by data type} & \rev{Color choice relies on the designer}\\ 
    \textbf{CVD \CVD}    & \heatmap{2.60} (5) & \heatmap{3.67} (2)  & \heatmap{3.00} (1) & \rev{Helps be aware of CVD} & \rev{Color choice relies on the designer }\\
    \bottomrule
    \end{tabular}
    }
    \label{tab:per-filtertype-scores}
    \end{table*}

\subsection{Design Evaluation}
\label{subsec:eval-design}

We here present the results of the design assessment by external reviewers, which is the final step of our study procedure (see \S~\ref{subsec:procedure}).
We recruited three senior visualization researchers who all have more than 7 years of experience to evaluate the designs conducted by the participants. 
Table~\ref{tab:evaluation-summary} shows various evaluations on the improvement of visualization designs conducted by our participants. 
Columns from second to fifth show the evaluation of visualization designer per iteration. 
The sixth column shows the increasing/decreasing trend of the scores per participant.
The seventh column shows which design the experts judged to be the best among each designer's iterations. 
Finally, the last column shows the overall improvement scores of visualization designs.
Note that the values in the second to fifth and the last columns are not on the quality of the design but on the improvement of the design.
\rev{Also, we analyze the evaluation scores, effectiveness, and limitations per feedback types.
This is shown in Table~\ref{tab:per-filtertype-scores}.}

On the whole, evaluators largely believed that the participants' designs improved from their starting point, averaging a score of 3.69 (3 is neutral).
N2, N4, and I1's designs were modestly below the neutral mark.
Evaluators observed consistency in N4's designs, with a slight decrease in color quality.
For I1, while adding a title was beneficial, the chart's readability declined.
\rev{N2} enlarged the chart for improved readability, but received mixed reviews about the choice of color.
We also compare the improvements per different expert levels. 
We find that the mean score of improvements from the novice group is $3.34$, from the intermediate group is $3.92$, and from the expert group is $4.11$. 
While we observe positive signs from all groups, we see more improvements from intermediate and expert groups.
Below, we provide detailed analyses of the tables. 

\noindent\textbf{When feedback is effective. }
We identified several ways in which the system’s feedback proved effective. 
First, as designers gained more expertise, they used the salience feature more effectively, which led to an improved data ink ratio. 
This was also noted by the evaluators. 
Most issues raised by the system were resolved within one or two iterations, although salience maps sometimes required additional effort. 
Even novice users managed salience related challenges by comparing designs across iterations and using the tracker interface (Fig.~\ref{fig:teaser} (D)) together with archives for more detailed heatmap comparisons. 
In addition, feedback on text placement and color usage led to tangible improvements.

These findings indicate that our ACG feedback framework, as well as feedback in textual form, was effective across various skill levels. 
We describe the role of the tracker interface (T) in \S~\ref{subsec:post-interview-study}.

\noindent\textbf{When feedback is not effective. }
Although the system demonstrated effectiveness in many situations, it was not universally successful. 
One reason is that novices did not receive feedback that was sufficiently direct or actionable. 
In practice, novices had difficulty adapting color palettes for color vision deficiencies, which indicates that selecting and applying a limited set of colors can be challenging for less experienced designers. 
Some also struggled with salience-related tasks, such as precisely shifting attention to specific chart elements. 
In the future, we plan to explore more accessible ways to recommend color palettes for CVD and to provide more direct salience-related feedback.

Furthermore, the system focuses on general communicative purposes, and thus does not yet account for specialized design intentions. 
For example, I1 prioritized data analysis rather than optimizing the chart for communicative goals. 
Although the expert evaluators were unaware of this, it may still have been appropriate for the designer’s specific objectives.

Moreover, we observe cases where the filters in the system occasionally malfunction. For example, salience may appear concentrated in a region with no chart elements (N6), or, if the chart is too complex (I1), Deplot might fail to read the data correctly and recommend an inappropriate chart. 
Although such cases are relatively not common, they reduce confidence in the filters.

\noindent\textbf{Best version from iterations. }
To determine which iteration yielded the most polished design, we asked evaluators to select the best version out of five. 
Our overview shows that when a design’s score is lower, evaluator opinions are more divided, whereas a higher score reflects consensus favoring the fifth version. 
However, the final design is not always the best for several reasons. 
First, if feedback-driven changes do not result in noticeable improvements, evaluators may choose different versions. 
For instance, in participant I1’s case, the design changes were minor and barely discernible, leading three evaluators to select three different versions. 
N4 exhibited a similar pattern. 
Second, designers’ goals can sometimes introduce unintended changes that degrade the overall design from an outside perspective. 
For example, near the end of participant E3’s process, an element became overly highlighted; evaluators, unaware of this intention, did not favor that final iteration. 
Third, when changes are trivial, observers may not detect a meaningful difference between one iteration and the next. 
For example, in participant N3’s final iteration, only a single red grid line was added to enhance clarity, which did not substantially shift evaluators’ perceptions.

Overall, when design changes are clear and demonstrably beneficial, evaluators’ judgments tend to converge. 
Conversely, if the changes are less distinct or their benefits uncertain, their evaluations diverge. 
We also observe that some designers have their own goals, and evaluators assessed them as a communicative visualization without knowing the designers' intents. 

\noindent\textbf{Variations in evaluators' scores. }
In some cases, the evaluators' scores converged, yet there were also instances of considerable divergence. 
Overall, the variance in final scores was generally minimal: most evaluators agreed or differed only slightly. 
However, participant E2 presented an exception, where two evaluators gave a score of 4 and one assigned a 1. 
The evaluator who gave a 1 stated, \textit{``I do not see how changing from a heatmap to a bar chart is an improvement. I might even prefer the heatmap.''} 
This discrepancy appears related to the communicative aspect of visualization, in which the evaluator’s limited understanding of the designer’s intentions played a role.

We observed even greater variations in the evaluators’ assessments across different iterations. 
Despite having solid expertise in visualization, each evaluator brought unique preferences and standards for communicative visualization. 
For example, moving from V2 to V3 in I2’s design, one evaluator was neutral about introducing multiple colors (assigning a 3), another greatly approved (giving a 5), and a third specifically praised the altered chart type (awarding a 4). 
Similar disagreements arose when some evaluators judged the overall positive effect of a design change, whereas others compared only the previous and current versions. 
One evaluator noted, \textit{`` To judge whether a change is good or bad, we need to consider the entire sequence of modifications. 
Design is not always a greedy process. 
Can we dismiss a change as having low value just because it looks unfavorable in one instance?}''

These findings underscore the complexity of evaluation when preferences vary. 
They also highlight that evaluators, unfamiliar with the designer's intention, may reach divergent conclusions. 
 
\subsection{Post-Study Interview}
\label{subsec:post-interview-study}

After the experiment, we conducted a post-interview session to examine closely the iterative improvements of the design and the role of \techname{} in making the changes. 
For every design refinement, we asked participants to mainly elaborate on the \textbf{role of \techname{}} in the changes.
We then transitioned to understand their experiences in using \techname{} during the design process, and their perspective on design feedback mechanisms. 
Detailed below are the summarized comments.

\smallskip

\noindent\textbf{Role of the feedback. }
\techname{} not only detects issues but also provides expert-level knowledge that many participants had not initially considered. 
In some cases, the system highlighted overlooked concerns, such as color deficiency filters. 
N4 noted, ``\textit{[the system] is capable of showing problems that I was not aware of, although I should have been,}'' while E2 remarked, \textit{``even experts may not be aware of those specific details about visualization principles.'' }
However, the system occasionally misdiagnosed the chart image, leading to feedback that was out of context. 

Moreover, \techname{} could provide a more detailed, actionable solutions, as some complex tasks still required deep human decisions. 
Although all participants attempted to address the system's feedback, some novice- and intermediate-level designers found it difficult to implement suggestions that were not specific enough, such as redistributing saliency attention. 
For instance, I2 devoted every iteration to balancing a chart’s salience, and I1 sought to focus salience on a particular area, underscoring the varying difficulty of applying automated feedback.

\smallskip

\noindent\textbf{Role of tracker.}
The consistent use of the tracker interface (T) for iterative refinements, especially when a problem could not be resolved in a single step, further underscores its supportive role in the design process.
We find that most participants that use virtual eyetracker have the experience of using the tracker interface to track their improvements (11/13). 
For people that use saliency maps, it functioned as a comparison system that measures whether there were improvements between the current and the past version.
I2, whose main goal throughout the study was to alleviate salience in textual areas, said, 
\textit{``I consistently checked the rate of salience in the tracking interface.
The drop in numbers reassured me that I was on the right path. 
The positive feedback felt like a reward for the adjustments I had made.''}
Furthermore, tracking changes per iteration allowed participants to catch up (3/13). 

There were also a few participants who did not use tracking. 
They uploaded the designs one after another without much delay. 
N4 was an example, noting \textit{``Because I remember what needs to be changed, I did not look into the system.''}

\smallskip

\noindent\textbf{Role of the textual report.}
Because the output of this report is generated by an LLM, it is predominantly textual.
The hierarchical layout of the design allowed participants to view the summary of each section, selectively pick the issues that they found relevant, and drill down to the important details. 
10 out 13 participants felt that this capability helped them be more organized, though some still felt overwhelmed at the volume of text.

\smallskip

\noindent\textbf{General assessment.}
Participants considered the system an assistant providing expert feedback. 
Many people originally thought that it would be difficult to provide useful insights, but were surprised at its capability to help the design. 
Furthermore, the use of an LLM created a sense of a ``human'' response, which made some participants feel less isolated during the design process. 

However, some areas could be enhanced.
One key observation was that, despite the hierarchical structure, participants still found the text volume overwhelming.

%% -------------------------------------------------------------
%% DISCUSSION 
%% -------------------------------------------------------------
\section{Discussion}
\label{sec:discussion}

The results from this study showed three things:
that (1) a ``vanilla'' LLM, when provided with appropriate symbolic metrics and a ``cheat sheet'' of visualization design knowledge, can provide meaningful and actionable feedback on flaws in a visualization and how to improve them; 
that (2) even experienced visualization designers can benefit from this feedback; and
that (3) our ACGT framework is effective for providing feedback for communicative visualizations.
Below we discuss issues on visualization design feedback, as well as the limitations of our work.

\subsection{Visualization Design Feedback}
\label{subsec:visualizationary-novice}

\noindent\textbf{Bridging design contexts.} 
Although we designed our feedback primarily to enhance clarity in visualizations, it is equally important to consider each designer’s unique goals and intentions. 
Some participants diverged from the system’s advice because it did not align with their specialized needs (see \S~\ref{subsec:eval-design}). 
This underscores that relying solely on general ``best practices'' can overlook the nuances of personal or domain-specific objectives.

Balancing clear, communicative design with each user’s specialized goals remains a significant challenge. 
On one hand, emphasizing universal guidelines can neglect individual contexts; on the other, tailoring feedback for every nuanced purpose requires more extensive modeling of the designer’s intent. 
We see this as an important avenue for future work, where feedback systems can be refined to address both communicative effectiveness and the broader range of a designer’s motivations.

\smallskip

\noindent\textbf{Human-like visualization feedback.}
AI models can help designers assess whether a visualization is ``good enough'' by applying best practices and accessibility guidelines. 
Systems like \techname{} use automated checklists to approximate human perception, for instance by evaluating color vision deficiency and visual saliency. 
One way to enhance this guidance is to add more filters tailored to designers’ needs. 
To further validate a design or create domain-specific conditions, designers can set benchmarks (e.g., speed of comprehension) to measure success.

Nevertheless, visualization design is context-dependent and subjective, as it involves aesthetic considerations and domain-specific requirements.
Can we develop a subjective design evaluator that declares a design ``complete'' and provides feedback in a style akin to a human visualization designer?
This would not only guide users to general feedback, but also provide design feedback that would be tailored to their owners’ styles. 
Achieving this would require extensive knowledge of best practices, accessibility standards, and real user feedback.
As a possible future direction, we plan to explore how effectively LLM-driven agents could contribute to building such models.

\smallskip

\noindent\textbf{Pitfalls of automated chart authoring.}
One important goal for \techname{} was to help novice designers create visualizations. 
Our results indeed show that our system is capable of detecting problems and providing relevant feedback according to existing design knowledge. 
However, it is also clear that participants with intermediate and expert knowledge used \techname{} more effectively than novices. 
Based on comments from novice participants, we speculate that automation and lack of fundamental design skills could be contributing factors.

By way of explanation, note that the tendency of novice users to rely on automated visualization suggestions, as offered by chart-authoring tools such as Microsoft Excel, contributes to this problem.
This reliance caused some novices to lose their manual chart design skills (or never to acquire them in the first place).
\rev{N2}, who obtained a professional certificate for Microsoft Excel nearly a decade ago, said, \textit{``the auto-generated charts are usually of good quality, so I don’t feel the need to alter them.''}
\rev{N1} shared similar comments. 
While automated tools have simplified visualization creation, their influence may have affected our results in Table~\ref{tab:evaluation-summary}.

\subsection{Limitations}
\label{subsec:limitations}

\topic{Avenues for future improvements. }
Our system is designed to enable a variety of future research avenues. 
To begin with, it can accommodate additional filters---for example, detecting charts that do not begin at zero (as in N2’s case) and introducing linting-like features to ensure consistency. 
\rev{Second, accurately predicting a saliency map for an overview task is challenging, as saliency varies by task and does not always align with overview-oriented goals. 
This presents the challenge of predicting saliency based on the user’s intent~\cite{wang24salcharqa}.
} 
Third, more advanced interpretability for each filter could yield more actionable feedback for designers. 
For instance, refining saliency-based methods may help pinpoint issues more precisely, and automating which part of the visualization the feedback refers to would streamline problem identification. 
Fourth, enhancing our system’s ability to deliver user-friendly and clear feedback further expands its potential by helping designers better understand the feedback. 
For example, by going beyond the ``if, then'' structure in our design guidelines (as noted in \S~\ref{sec:generating-feedback}) and exploring alternative tones, we can improve how feedback is delivered. 
We can also refine our experimental procedures by adopting more neutral language (e.g., ``What was your overall impression of the system?'') rather than “Did you like the system?” to mitigate bias. 
While we cannot address every challenge at once, we leave these avenues open for continued exploration.

\smallskip

\topic{Hallucination.}
Despite their capabilities, LLMs may produce nonsensical or unfaithful content, commonly referred to as \emph{hallucinations}~\cite{ji2023survey}.
Hallucinations may lead to feedback that contradicts the designer's original goals or fails to improve the visualizations.
This issue poses a potential risk to \techname{}, as users build trust in the models powering our system  when most of their outputs are consistently reliable.
Recent work has proposed defensive methods to mitigate hallucinations,  such as incorporating a user's feedback 
into the model's fine-tuning~\cite{lee2022factuality} and detecting self-contradiction in model-generated texts~\cite{mundler2023self}. %, manakul2023selfcheckgpt}.
Incorporating these methods remains an avenue for future work.

\smallskip

\topic{Privacy risks.}
Because our system are built on LLMs, one can be concerned about the associated privacy risks, 
such as membership inference or data extraction attacks~\cite{carlini2019secret}. %, carlini2021extracting, PI, carlini2022membership}.
User data may be intentionally (or unintentionally) leaked to an adversary  with access to these models~\cite{SamsungCase}.
The concerns primarily stem from two aspects:
(1) user data being shared with commercial chat-based services and 
(2) the fine-tuning of these models on user data.
However, the first concern can potentially be mitigated  by avoiding the use of third-party hosted LLMs. 
We can entirely prevent our system from accessing external services  by employing local open-source models. 
%such as Vicuna~\cite{chiang2023vicuna} and LLaMA~\cite{touvron2023llama}.
Second, we do \emph{not} fine-tune the models we use on user data, ensuring that user data cannot be memorized by our LLMs or leaked by malicious actors with access to the models.

%% -------------------------------------------------------------
%% CONCLUSION AND FUTURE WORK
%% -------------------------------------------------------------
\section{Conclusion}
\label{sec:conclusion}

We have presented \techname{}, a system that leverages LLMs and visualization guidelines to provide design feedback for visualization designers. 
\techname{} supports an automated visualization design workflow called ACGT (analyze-clarify-guide-track).
We created a web-based interface to evaluate the effectiveness of \techname{} and engaged 13 visualization designers of different seniority in a longitudinal design task. 
During this study, we conducted pre- and post-study interviews to further understand the designer's experience using the system. 
Finally, we asked three expert evaluators to assess their resulting designs.
Overall, our results show that an off-the-shelf LLM can indeed provide high-quality and actionable feedback for novice, intermediate, and expert users alike on how to design a visualization.

\section*{Acknowledgments}

We thank anonymous reviewers for their feedback.
This work was partly supported by grant IIS-1901485 from the U.S.\ National Science Foundation (Shin), a Google Faculty Research Award (Hong), and Villum Investigator grant VL-54492 by Villum Fonden (Elmqvist).
Any opinions, findings, and conclusions or recommendations expressed here are those of the authors and do not necessarily reflect the views of the funding agencies.

\bibliographystyle{abbrv-doi}
\bibliography{visualizationary}

\begin{IEEEbiography}[{\includegraphics[width=1in,height=1.25in,clip,keepaspectratio]{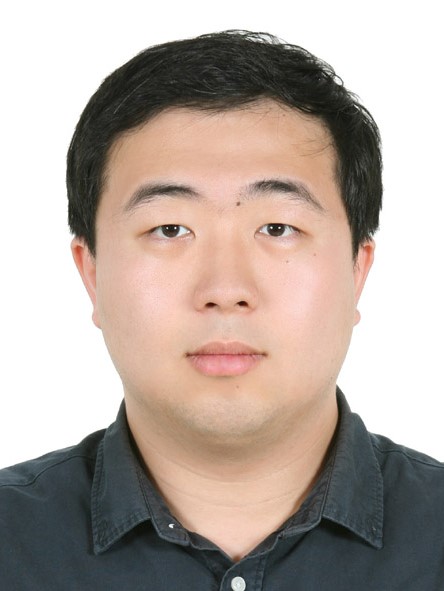}}]{Sungbok Shin} 
received the Ph.D. degree in 2024 from University of Maryland, College Park in College Park, MD, USA.
He is a postdoctoral researcher at Team Aviz in Inria-Saclay and Universit\'e Paris-Saclay in Saclay, France.
His research interest is in Human-Centered AI, visualization, and HCI. 
\end{IEEEbiography}

\begin{IEEEbiography}
[{\adjincludegraphics[width=1in,height=1.2in,trim={2.2cm 1.4cm 1.6cm 0},clip]{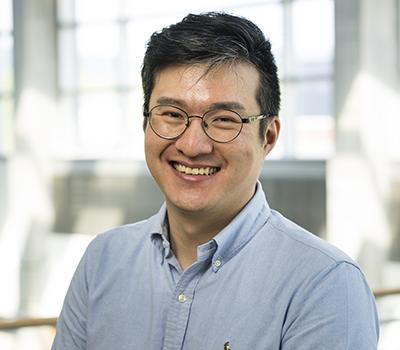}}]{Sanghyun Hong}
received the Ph.D. degree in 2021 from University of Maryland, College Park in College Park, MD, USA.
He is an assistant professor in the Department of Computer Science at Oregon State University in Corvallis, OR, USA.
His research interest is at the intersection of privacy, security and machine learning.
He is the recipient of the Google Faculty Research Award 2023 
and was also selected as a DARPA Riser 2022.
\end{IEEEbiography}

\begin{IEEEbiography}[{\includegraphics[width=1in,height=1.25in,clip,keepaspectratio]{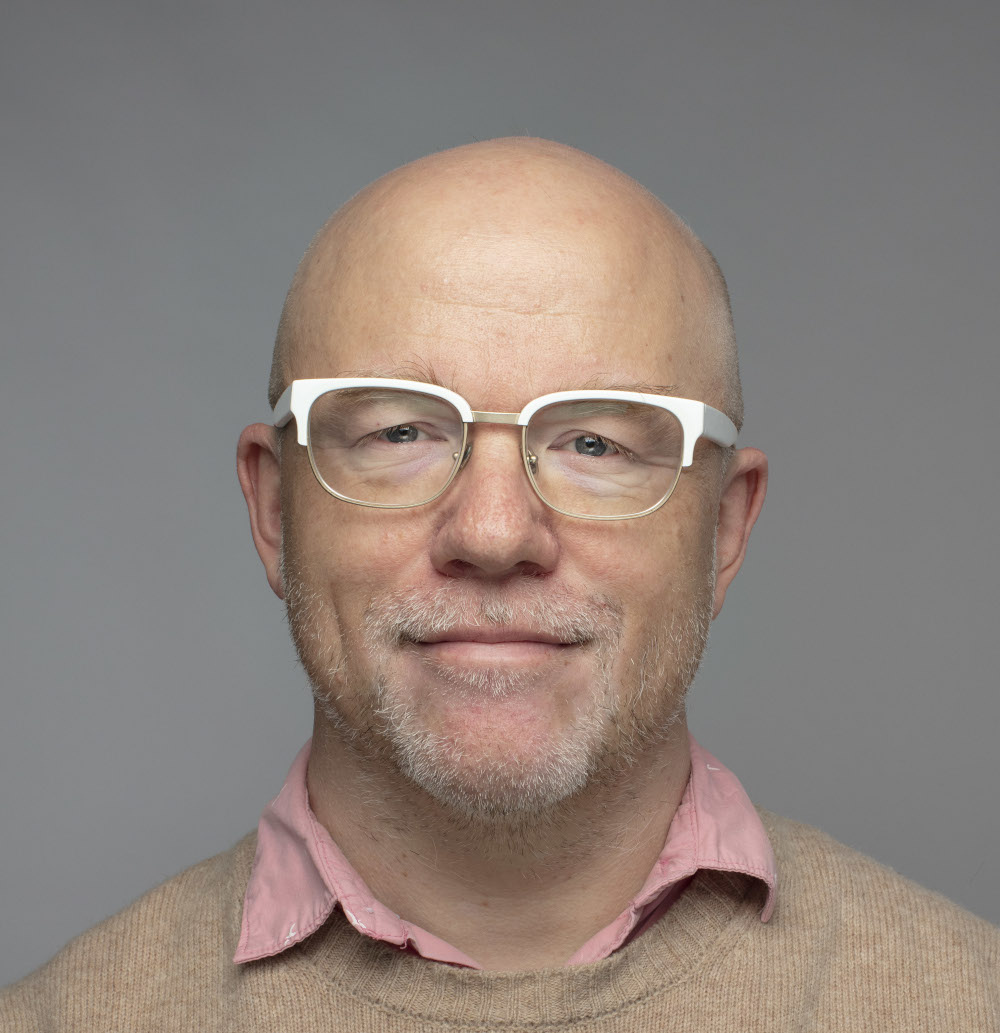}}]{Niklas Elmqvist}
received the Ph.D.\ degree in 2006 from Chalmers University of Technology in G\"{o}teborg, Sweden.
He is a Villum Investigator and professor in the Department of Computer Science at Aarhus University in Aarhus, Denmark.
He was previously faculty at University of Maryland, College Park from 2014 to 2023, and at Purdue University from 2008 to 2014. 
His research interests include visualization, HCI, and human-centered AI.
He is a Fellow of the IEEE and the ACM.
\end{IEEEbiography}

\vfill

\end{document}